\documentclass[12pt]{article}

\usepackage[height=8.85in,width=6.45in]{geometry}
\usepackage{amsmath,amssymb}
\usepackage{hyperref}
\usepackage{graphicx}

\usepackage{xcolor}
\usepackage{mdframed}
\newenvironment{claim}{  \begin{mdframed}[linecolor=black!0,backgroundcolor=black!10]\noindent\itshape\ignorespaces}{\end{mdframed}}

\renewenvironment{figure}[1][]{
  \begin{originalfigure}[#1]
    \begin{mdframed}[linecolor=black!0,backgroundcolor=black!2]
}{
    \end{mdframed}
  \end{originalfigure}
}

\usepackage{tikz}
\usetikzlibrary{arrows}
\usetikzlibrary{shapes.geometric,calc, positioning,shapes.misc}

\newcommand{\arrowpath}[3]{
	\draw[line width=1pt] #1--#2;
	\draw[->, >=stealth, line width=1pt, black] #1--($#1!#3!#2$);
}

\usepackage{times}
\usepackage{courier}
\numberwithin{equation}{section}
\numberwithin{figure}{section}

\newcommand{\bC}{\mathbb{C}}
\newcommand{\bF}{\mathbb{F}}
\newcommand{\bR}{\mathbb{R}}
\newcommand{\bZ}{\mathbb{Z}}
\newcommand{\tr}{\mathop{\mathrm{tr}}}
\newcommand{\Bimod}{\mathrm{Bimod}}

\newcommand{\Hom}{\mathrm{Hom}}
\newcommand{\uHom}{\underline{\mathrm{Hom}}}
\newcommand{\Mod}{\mathrm{Mod}}
\newcommand{\Rep}{\mathrm{Rep}}

\newcommand{\SO}{\mathrm{SO}}
\newcommand{\Sp}{\mathrm{Sp}}
\newcommand{\SU}{\mathrm{SU}}
\newcommand{\UU}{\mathrm{U}}
\newcommand{\id}{\mathrm{id}}
\newcommand{\co}[1]{{{}^\circ#1}}
\newcommand{\ev}{\epsilon}
\newcommand{\coev}{\co\ev}
\newcommand{\proj}{\pi}
\newcommand{\coproj}{\co\pi}
\renewcommand{\Vec}{\mathrm{Vec}}
\newcommand{\vev}[1]{\langle#1\rangle}
\newcommand{\inc}[1]{\vcenter{\hbox{\includegraphics[scale=.8]{#1}}}}
\newcommand{\incb}[1]{\vcenter{\hbox{\includegraphics[scale=.6]{#1}}}}
\newcommand{\incc}[1]{\vcenter{\hbox{\includegraphics[scale=.45]{#1}}}}
\newcommand{\cA}{\mathcal{A}}
\newcommand{\cC}{\mathcal{C}}
\newcommand{\cM}{\mathcal{M}}

\newcommand{\ot}{\otimes}

\newcommand{\be}{\begin{equation}}
\newcommand{\ee}{\end{equation}}
\newcommand{\bea}{\begin{eqnarray}}
\newcommand{\eea}{\end{eqnarray}}
\newcommand{\wh}{\widehat}

\newcommand{\ben}{\begin{enumerate}}
\newcommand{\een}{\end{enumerate}}

\begin{document}

\begin{titlepage}

\begin{flushright}
IPMU-17-0049
\end{flushright}

\vskip 4cm

\begin{center}

{\Large \bfseries On finite symmetries and their gauging in two dimensions}

\vskip 1cm
Lakshya Bhardwaj$^1$ and Yuji Tachikawa$^2$
\vskip 1cm

\begin{tabular}{ll}
$^1$  &Perimeter Institute for Theoretical Physics, \\
& Waterloo, Ontario, Canada N2L 2Y5\\
$^2$  & Kavli Institute for the Physics and Mathematics of the Universe, \\
& University of Tokyo,  Kashiwa, Chiba 277-8583, Japan\\
\end{tabular}

\vskip 1cm

\textbf{Abstract}  

\end{center}

\noindent 
It is well-known that if we gauge a $\mathbb{Z}_n$ symmetry in two dimensions, a dual $\mathbb{Z}_n$ symmetry appears, such that re-gauging this dual $\mathbb{Z}_n$ symmetry leads back to the original theory.
We describe how this can be generalized to non-Abelian groups, by enlarging the concept of symmetries from those defined by groups to those defined by unitary fusion categories. 
We will see that this generalization is also useful when studying what happens when  a non-anomalous subgroup of an anomalous finite group is gauged: for example, the gauged theory can have non-Abelian group symmetry even when the original symmetry is an Abelian group.
We then discuss the axiomatization of two-dimensional topological quantum field theories whose symmetry is given by a category. We see explicitly that the gauged version is a topological quantum field theory with a new symmetry given by a dual category.

\end{titlepage}

\setcounter{tocdepth}{2} 
\tableofcontents
\newpage

\section{Introduction}
Let us start by considering a two-dimensional theory $T$ with $\bZ_n$ symmetry.
We can  gauge it to get the gauged theory $T/\bZ_n$.
This gauged theory is known to have a new $\bZ_n$ symmetry,
and re-gauging it gives us back the original theory: $T/\bZ_n/\bZ_n =T$ \cite{Vafa:1989ih}. It is also well-known that this phenomenon generalizes to any arbitrary finite Abelian group $G$. That is, gauging a theory $T$ with $G$ symmetry results in a theory $T/G$ with a new finite Abelian group symmetry $\hat G$ such that gauging the new theory by the new symmetry takes us back to the original theory: $T/G/\hat G=T$.
One natural question arises: can it be generalized to higher dimensions?
Yes, according to \cite{Gaiotto:2014kfa}, where the generalized concept of $p$-form symmetries was introduced.
Our investigation starts from a related but different question: can it be generalized to non-Abelian finite groups?

The answer is again yes \cite{Brunner:2014lua}, and it again requires a generalization of the concept of symmetries, but in a direction different from that of \cite{Gaiotto:2014kfa}. To explain this, let us pose what we have said above in a different way. Traditionally, we say that $T$ has symmetry $G$ if we can find unitary operators $U_g$ labeled by elements $g\in G$ whose action on the Hilbert space commutes with the Hamiltonian. When $G$ is Abelian, the information about $G$ is captured in $T/G$ via unitary operators $U_{\hat g}$ labeled by elements $\hat g\in \hat G$ which commute with the Hamiltonian of $T/G$. When $G$ is non-Abelian, we will argue that the information of $G$ in $T/G$ is captured by operators $U_i$ which still commute with the Hamiltonian but these operators are now in general non-unitary. These operators can be constructed by wrapping a Wilson line for $G$ along the spatial circle. Hence, Wilson lines should be thought of as generalized symmetries for the theory $T/G$. In fact, we will also argue that there is a natural notion of gauging this symmetry formed by Wilson lines such that gauging $T/G$ results back in the original theory $T$.

This raises the following question: How do we specify a generalized symmetry that a theory can admit? In this paper, we give an answer to this question:
a general finite symmetry of a two-dimensional theory is specified by a structure which is known to mathematicians in the name of \emph{unitary fusion categories}. We prefer to call it \emph{symmetry categories}.\footnote{We do not claim that this is the ultimate concept for the 0-form finite symmetry in two dimensions; there still might be a generalization in the future.
For example, in other spacetime dimensions, a proposed generalization was to use the concept of $p$-groups, see e.g.~\cite{Sharpe:2015mja}, and one might want to unify the two approaches.
We also note that this extension of the concept of the symmetry in two dimensions from groups to categories was already proposed by many other authors in the past, and we are merely shedding a light to it from a slightly different direction. We will come back to this point later in the Introduction.}  
For the gauged theory $T/G$ for possibly non-Abelian group $G$,
the Wilson line operators form $\Rep(G)$, which is a symmetry category formed by the representations of $G$. Similarly, a general symmetry category $\cC$ physically corresponds to more general line operators of $T$.

We also discuss how a symmetry category $\cC$ can be gauged. It turns out that there is no canonical way of gauging a generic symmetry category. Pick one way $\cM$ of gauging the symmetry $\cC$ of a theory $T$.  Denote the gauged theory by $T/\cM$ and its symmetry category by $\cC'$. 
It then turns out that there exists a dual way $\cM'$ of gauging $\cC'$ such that $T/\cM/\cM'$ is equivalent to $T$. This generalizes the fact that regauging the gauged theory with symmetry $\Rep(G)$ gives back the original theory with symmetry $G$.

This generalization of the concept of symmetry allows us not only to perform the re-gauging of non-Abelian gauge theories, but also to answer various other questions.
First of all, we will see that symmetry categories $\cC$ capture symmetries together with their anomalies. 
Then, the machinery we spell out allows us to compute what is the symmetry of the gauge theory $T/H$ 
when we gauge a subgroup $H$ of a possibly anomalous flavor symmetry $G$.
For example, if we gauge a non-anomalous $\bZ_2$ subgroup of $\bZ_2\times \bZ_2\times \bZ_2$ with a suitable choice of the anomaly, we can get non-anomalous non-Abelian symmetry $D_8$ and $Q_8$, the dihedral group and the quaternion group of order 8.%
\footnote{Recently in \cite{Gaiotto:2017yup}, Gaiotto, Kapustin, Komargodski and Seiberg performed an impressive study of the phase structure of thermal 4d $\mathfrak{su}(2)$ Yang-Mills theory. 
One important step in the analysis is the symmetry structure of the thermal system, which is essentially three-dimensional. 
As a dimensional reduction from 4d, the system has  a $\bZ_2\times\bZ_2$ 0-form symmetry and a $\bZ_2$ 1-form symmetry, with a mixed anomaly.
Then the authors gauged the $\bZ_2$ 1-form symmetry, and found that the total 0-form symmetry is now $D_8$.
This $D_8$ was then used very effectively to study the phase diagram, but that part of their paper does not directly concern us here.
Their analysis of turning an anomalous Abelian symmetry by gauging a non-anomalous subgroup into a non-Abelian symmetry  is a 3d analogue of what we explain in 2d. 
See their Sec.~4.2, Appendix B and Appendix C. 
Clearly an important direction to pursue is to generalize their and our constructions to arbitrary combinations of possibly-higher-form symmetries in arbitrary spacetime dimensions, 
but that is outside of the scope of this paper.
}
In general, the symmetry $\cC$ of the gauged theory is neither a group nor $\Rep(G)$ for a finite group, 
and we need the concept of symmetry categories to describe it.

There are also vast number of symmetry categories not related to finite groups, formed by topological line operators of two-dimensional rational conformal field theories (RCFTs).
In particular, any unitary modular tensor category, or equivalently any Moore-Seiberg data, can be thought of as a symmetry category, by forgetting the braiding.

We should emphasize here that this generalization of the concept of symmetry from that defined by groups to that defined by categories was already done long ago by other authors, 
belonging to three somewhat independent lines of studies, namely
in the study of the subfactors and operator algebraic quantum field theories,
in the study of representation theory,
and in the study of RCFTs.
Each of the communities produced vast number of papers, and not all of them can be cited here.
We recommend to the readers textbooks by Bischoff, Longo, Kawahigashi and Rehren \cite{Bischoff:2014xea} and Etingof, Gelaki, Nikshych and Ostrik \cite{EGNO} from the first two communities
and the articles by Carqueville and Runkel \cite{Carqueville:2012dk} and by Brunner, Carqueville and Plencner \cite{Brunner:2013xna} from the third community as the starting points.

Our first aim in this paper is then to summarize the content of these past works in a way hopefully more accessible to other researchers of quantum field theory, including the authors of this paper themselves, 
emphasizing the point of view related to the modern study of  symmetry protected topological phases.
What we explain in this first part of the paper is not new, except possibly the way of the presentation, and can all be found in the literature in a scattered form. 

Our second aim is to  axiomatize two-dimensional topological quantum field theories (TFTs) whose symmetry is given by a symmetry category $\cC$.
This is a generalization of the work by Moore and Segal \cite{Moore:2006dw}, where two-dimensional TFTs with finite group symmetry were axiomatized.
We write down basic sets of linear maps between (tensor products of) Hilbert spaces on $S^1$
and basic consistency relations among them
which guarantee that a unique linear map is associated to any surface with $m$ incoming circles and $n$ outgoing circles together with arbitrary network of line operators from $\cC$.

The rest of the paper is organized as follows. 
First in Sec.~\ref{sec:regauging}, as a preliminary,  we recall how gauging of a finite Abelian symmetry $G$ can be undone by gauging the new finite Abelian symmetry $\hat G$, and then briefly discuss how this can be generalized to non-Abelian symmetries $G$, by regarding $\Rep(G)$ as a symmetry. This effort of generalizing the story to a non-Abelian group makes the possibility and the necessity of a further generalization to symmetry categories manifest. We exploit this possibility and describe the generalization in detail in subsequent sections.

Second, we have two sections that form the core of the paper.
In Sec.~\ref{sec:generalities}, we introduce the notion of  symmetry categories,
and discuss how we can regard as symmetry categories both a finite group $G$ with an anomaly $\alpha$ and the collection $\Rep(G)$ of representations of $G$.
We then explain in Sec.~\ref{sec:gauging}
that physically distinct gaugings of a given symmetry category $\cC$ correspond to indecomposable module categories $\cM$ of $\cC$,
and we describe how to obtain the new symmetry $\cC'$ of the theory $T/\cM$
for a given theory $T$ with a symmetry $\cC$.

Third, in  Sec.~\ref{sec:examples}, we give various examples illustrating the notions introduced up to this point. Examples include the form of new symmetry categories $\cC'$ when we gauge a non-anomalous subgroup $H$ of an anomalous finite group $G$, and the symmetry categories of RCFTs.

Fourth, in Sec.~\ref{sec:tqft}, we move on to the discussion of the axioms of two-dimensional TFTs whose symmetry is given by a symmetry category $\cC$.
We also construct the gauged TFTs $T/\cM$ given an original TFT $T$ with a symmetry category $\cC$ and a gauging specified by its module category $\cM$.
Sections~\ref{sec:examples} and \ref{sec:tqft} can be read independently.

Finally, we conclude with a brief discussion of what remains to be done in Sec.~\ref{sec:conclusions}.
We have an appendix~\ref{sec:cohomology} where we review basic notions of group cohomology used in the paper.

Before proceeding, we note that we assume that the space-time is oriented throughout the paper.
We also emphasize that all the arguments we give, except in Sec.~\ref{sec:tqft}, apply to non-topological non-conformal 2d theories.

\section{Re-gauging of finite group gauge theories}
\label{sec:regauging}
\subsection{Abelian case}
Let us start by reminding ourselves the following well-known fact \cite{Vafa:1989ih}:
\begin{claim}
Let $T$ be a 2d theory with flavor symmetry given by an Abelian group $G$.
Let us assume that $G$ is non-anomalous and can be gauged,
and denote the resulting theory by $T/G$. 
Then this theory  has the flavor symmetry $\hat G$,
which is the Pontrjagin dual of $G$, such that 
$T/G/\hat G=T$.
\end{claim}
Recall the definition of the Pontrjagin dual $\hat G$ of an Abelian group $G$.
As a set, it is given by \begin{equation}
\hat G=\{ \chi: G\to U(1) \mid  \text{$\chi$ is an irreducible representation }\}.
\end{equation} Note that $\chi$ is automatically one-dimensional.
Therefore the product of two irreducible representations is again an irreducible representation, which makes $\hat G$ into a group.
$G$ and $\hat G$ are isomorphic as a group but it is useful to keep the distinction because there is no canonical isomorphism between them.

In the literature on 2d theories, gauging of a finite group $G$ theory is more commonly called as orbifolding by $G$, and the fact above is often stated as follows: a $G$-orbifold has a dual $\hat G$ symmetry assigning charges to twisted sectors, and orbifolding again by this dual $\hat G$ symmetry we get the original theory back. 
This dual $\hat G$ symmetry is also known as the quantum symmetry in the literature.

This fact can be easily shown as follows. Let $Z_T[M,A]$ denote the partition function of $T$ on $M$ with the external background $G$ gauge field $A$. Here $A$ can be thought of as taking values in $H^1(M,G)$. 
Then the partition function of the gauged theory $T/G$ on $M$ is given by $Z_{T/G}[M]\propto\sum_A Z_T[M,A]$. Here and in the following we would be cavalier on the overall normalization of the partition functions. 
More generally, with  the background gauge field $B$ for the dual $\hat G$ symmetry, the partition function is given by 
 \begin{equation}
Z_{T/G}[M,B] \propto \sum_A e^{i (B,A)} Z_T[M,A] \label{orb}
\end{equation} where $B\in H^1(M,\hat G)$  and  $e^{i(B,A)}$ is obtained by the intersection pairing \begin{equation}
e^{i(-,-)} : H^1(M,\hat G) \times H^1(M,G) \to H^2(M,U(1)) \simeq U(1).
\end{equation} 
The equation \eqref{orb} says that the partition function of $T/G$ is essentially the discrete Fourier transform of that of $T$, and therefore we dually have $T=T/G/\hat G$: \begin{equation}
Z_{T}[M,A] \propto \sum_B e^{i (A,B)} Z_{T/G}[M,B]. \label{reorb}
\end{equation}

This statement was generalized to higher dimensions in e.g.~\cite{Gaiotto:2014kfa}: 
\begin{claim}
Let $T$ be a $d$-dimensional theory with $p$-form flavor symmetry given by an Abelian group $G$.
Let us assume that $G$ is non-anomalous and can be gauged,
and denote the resulting theory by $T/G$. 
Then this theory has the dual $(d{-}2{-}p)$-form flavor symmetry $\hat G$, such that 
$T/G/\hat G=T$.
\end{claim}
The derivation is entirely analogous to the 2d case, except that now $A\in H^{p+1}(M,G)$ and $B\in H^{d-1-p}(M,\hat G)$.

\subsection{Non-Abelian case}\label{sec:general-d-nonabelian}
The facts reviewed above means that the finite Abelian gauge theory $T/G$ still has the full information of the original theory $T$, which can be extracted by gauging the dual symmetry $\hat G$. 
It is natural to ask if this is also possible when we have a non-Abelian symmetry $G$, which we assume to be an ordinary 0-form symmetry.

This is indeed possible\footnote{%
That this is possible was already shown for two-dimensional theories in \cite{Brunner:2014lua}, as an example of a much more general story, which we will also review in the forthcoming sections. Here we describe the construction in an elementary language.  }
by suitably restating the derivation above, but we will see that we need to extend the concept of what we mean by \emph{symmetry}. To show this, we first massage (\ref{reorb}) in a suitable form which admits a straightforward generalization. Let us consider the case of $Z_T[M,A=0]$ for illustration. By Poincare duality, $B$ can also be represented as an element of $H_1(M,\hat G)$. Then, (\ref{reorb}) can be rewritten as
\be
Z_T[M]\propto \sum_{\hat{g}_1,\cdots,\hat{g}_n} Z_{T/G}[M,\hat{g}_1,\cdots,\hat{g}_n] \label{reorb2}
\ee
where $i\in\{1,\cdots,n\}$ labels generators of $H_1(M)$ and $\hat{g}_i$ is an element of $\hat G$ associated to the cycle labeled by $i$. 
Each summand on the right hand side, $Z_{T/G}[M,\hat{g}_1,\cdots,\hat{g}_n]$, is then the expectation value of Wilson loops in representations labeled by $\hat{g}_i$ placed along the cycle $i$. 

Now, we can sum the $\hat G$ elements for each $i$ separately to obtain
\be
Z_T[M]\propto Z_{T/G}[M,W^\text{reg}_1,\cdots,W^\text{reg}_n] \label{reorbwilson}
\ee
where $W^\text{reg}_i$ denotes the insertion of a Wilson line in the regular representation along the  cycle $i$. This is because, for an abelian $G$, the regular representation is just the sum of representations corresponding to elements $\hat g$ of $\hat G$. 

The relation (\ref{reorbwilson})  says that by inserting all possible Wilson lines on all possible cycles,
we are putting the delta function for the original gauge field $A$. 
We now note that the relation \eqref{reorbwilson}  holds for a non-Abelian $G$ as well, 
if we insert $W^\text{reg}_i$  not only for the generators of $H_1(M)$ but for the generators of $\pi_1(M)$.
This can be seen by the fact that $\tr g$ in the regular representation is nonzero if and only if $g$ is the identity.

The identity \eqref{reorbwilson} means that the ungauged theory $T$ can be recovered from the gauged theory $T/G$ by inserting line operators $W^\text{reg}_i$ in an appropriate manner.
This is  analogous to the construction of the gauged theory $T$ from the ungauged theory $T$ by inserting line operators representing the $G$ symmetry in an appropriate manner.
Given the importance of Wilson lines in recovering the information of the ungauged theory, we assign the status of dual \emph{symmetry} to Wilson lines. 

Let us phrase it another way. When $G$ is Abelian, the dual $(d-2)$-form $\hat G$ symmetries can be represented by 1-cycles labeled by elements of $\hat G$, forming a group.
When $G$ is non-Abelian, the dual \emph{symmetry} can  still be represented by 1-cycles labeled by representations $\Rep(G)$ of $G$. We can still multiply lines, corresponding to the tensor product of the representations, and this operation reduces to group multiplication of $\hat G$ in the abelian case.
But $\Rep(G)$ is not a group if $G$ is non-Abelian.
Therefore this is not a flavor symmetry \emph{group}, it is rather a flavor symmetry \emph{something}.
We summarize this observation as follows:
\begin{claim}
Let $T$ be a $d$-dimensional theory with $0$-form flavor symmetry given by a possibly non-Abelian group $G$.
Let us assume that $G$ is non-anomalous and can be gauged,
and denote the resulting theory by $T/G$. 
Then this theory has $\Rep(G)$ as the dual $(d-2)$-form flavor symmetry `something',
such that  $T/G/\Rep(G)=T$.
\end{claim}

\section{Symmetries as categories in two dimensions}
\label{sec:generalities}
Any finite group, possibly non-Abelian, can be the 0-form symmetry \emph{group} of a theory.
In addition to this, we saw in the last section that $\Rep(G)$, the representations of $G$, can also be the $(d-2)$-form symmetry \emph{something} of a $d$-dimensional theory.
We do not yet have a general understanding of what should be this \emph{something} in general $d$ dimensions for general combinations of various $p$-form symmetries. 
However, at least for $d{=}2$ and $p{=}0$, we already have a clear concept for this \emph{something} in the literature, which includes both groups $G$ and representations of groups $\Rep(G)$ and much more.
In this section we explain what it is.

\subsection{Basic notions of symmetry categories}
\label{sec:basicdefs}
In two dimensions, a 0-form finite symmetry element can be represented by a line operator with a label $a$. Inserting this line operator on a space-like slice $S$ corresponds to acting on the Hilbert space associated to $S$ by a possibly non-unitary operator $U_a$ corresponding to the symmetry element.
Moreover, $U_a$ commutes with the Hamiltonian $H$ associated to any foliation of the two-dimensional manifold.
In addition, $U_a$ cannot change under a continuous deformation of its path.
Therefore the line operator under consideration is automatically topological.

Topological line operators, in general, form a structure which mathematicians call a \emph{tensor category}. We want to restrict our attention to topological line operators describing a \emph{finite} symmetry. Such topological line operators form a structure which mathematicians call a \emph{unitary fusion category}.
We call it instead as a \emph{symmetry category}, to emphasize its role as the finite symmetry of unitary two-dimensional quantum field theory.
We start by stating the slogan, and then fill in the details:
\begin{claim}
Finite flavor symmetries of 2d theories are characterized by symmetry categories.
\end{claim}

\subsubsection{Objects}
The objects in a symmetry category $\cC$ correspond to topological line operators generating the symmetry. More precisely, a theory $T$ admitting the symmetry $\cC$ will admit topological line operators labeled by the objects of $\cC$. Henceforth, we will drop the adjective topological in front of line operators.

For any line operator labeled by an object $a$, we have a partition function of $T$ with the line inserted along an \emph{oriented} path $C$, which we can denote as \begin{equation}
\vev{\cdots a(C)\cdots}
\end{equation} where the dots stand for additional operators inserted away from $C$.

\subsubsection{Morphisms}
The morphisms in a symmetry category $\cC$ correspond to topological local point operators which can be inserted between two lines. More precisely, consider two labels $a$ and $b$, and a path $C$ such that up to a point $p\in C$ we have the label $a$ and from the point $p$ we have the label $b$. Then, we can insert a possibly-line-changing topological operator labeled by $m$ at $p$.
We call such $m$ a morphism from $a$ to $b$, and denote this statement interchangeably as either  \begin{equation}
m: a\to b 
\qquad \text{or}\qquad
m\in \Hom(a,b).
\end{equation}

The set $\Hom(a,b)$ is taken to be a complex vector space and it labels (a subspace of) topological local operators between line operators corresponding to $a$ and $b$ in $T$. From now on, we would drop the adjective topological in front of local operators.

\subsubsection{Existence of a trivial line}
$\cC$ contains an object $1$, which labels the trivial line of $T$. We have
\begin{equation}
\vev{\cdots 1(C)\cdots}
=\vev{\cdots \cdots}.
\end{equation}

\subsubsection{Additive structure}
Given two objects $a$ and $b$, there is a new object $a\oplus b$ in $\cC$. In terms of lines in $T$, we have
\begin{equation}
\vev{\cdots (a\oplus b)(C) \cdots}
=\vev{\cdots a(C)\cdots}+\vev{\cdots b(C)\cdots}.
\end{equation}
We abbreviate $a\oplus a$ by $2a$, $a\oplus a\oplus a$ by $3a$, etc.
The linear operator $U_{a\oplus b}$ acting on the Hilbert space obtained by wrapping $a\oplus b$ on a circle 
is then given by $U_a + U_b$.

\subsubsection{Tensor structure}
Given two objects $a$ and $b$, we have an object $a\otimes b$ in $\cC$.
This corresponds to considering two parallel-running line operators $a$ and $b$ as one line operator.
The linear operator $U_{a\otimes b}$ acting on the Hilbert space obtained by wrapping $a\otimes b$ on a circle is then given by $U_a U_b$.

The trivial object $1$ acts as an identity for this tensor operation. That is, there exist canonical isomorphisms $a\otimes 1\simeq a$ and $1\otimes a\simeq a$ for each object $a$. We can always find an equivalent category in which these isomorphisms are trivial, that is $a\otimes 1=1\otimes a =a$. Hence, we can assume that these isomorphisms have been made trivial in $\cC$. Henceforth, the unit object will also be referred to as the identity object.

\begin{figure}
\[
\inc{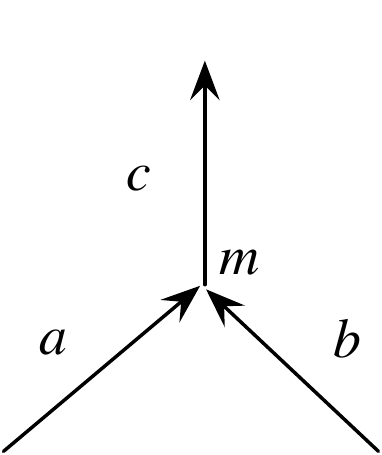}=\inc{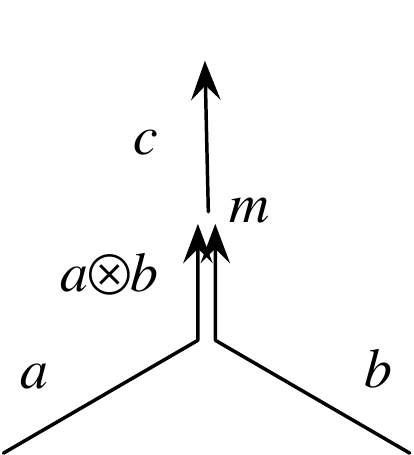}
\]
\caption{Two lines with the labels $a$ and $b$ can be fused to form a line with the label $a\otimes b$.\label{implicitjunction}}
\end{figure}

Consider three lines $C_{1,2,3}$ meeting at a point $p$,
with $C_{1,2}$ incoming and $C_3$ outgoing.
We can put the label $a$, $b$, $c$ on $C_{1,2,3}$, respectively.
We demand that the operators we can put at the junction point $p$
is given by $m\in \Hom(a\otimes b,c)$.
This label $a\otimes b$ corresponds to a composite line as can be seen by the following topological deformation shown in Fig.~\ref{implicitjunction}.

The definition of $a\otimes b$ here includes a choice of the implicit junction operator where the lines labeled by $a$, $b$ and $a\otimes b$ meet. In this paper, whenever we draw figures with such implicit junction operators, we always choose the operator to be the one labeled by the identity morphism $\mathrm{id}: a\otimes b\to a\otimes b$.

\subsubsection{Simplicity of the identity, semisimplicity, and finiteness}
The simple objects $a\in\cC$ are objects for which $\Hom(a,a)$ is one-dimensional. In general, for any object $x$, there is always a canonical \emph{identity morphism} from $x$ to $x$ which labels the identity operator on the line labeled by $x$. For a simple object $a$, the existence of the identity morphism implies that there is a natural isomorphism $\Hom(a,a)\simeq\bC$ as an algebra.
We assume for simplicity that the identity object 1 is simple.

We also assume that every object $x$ has a decomposition as a finite sum \begin{equation}
x=\bigoplus_a N_a a
\end{equation} where $N_a$ is a nonnegative integer and $a$ is \emph{simple}.
In other words, every object $x$ is \emph{semisimple}.

Finally, we assume \emph{finiteness}, that is the number of isomorphism classes of simple objects is finite.
Below, we will be somewhat cavalier on the distinction between simple objects and isomorphism classes of simple objects.

\subsubsection{Associativity structure}
The data in a symmetry category $\cC$ includes certain isomorphisms implementing associativity of objects \begin{equation}
\alpha_{a,b,c} \in \Hom((a\otimes b)\otimes c, a\otimes(b\otimes c))
\end{equation} which we call \emph{associators}.\footnote{%
Since we can and do choose the identity morphisms $1\otimes a \to a$ and $a\otimes 1\to a$ to be trivial, the associator $\alpha_{a,b,c}$ is also trivial when any of $a$, $b$, $c$ is trivial. \label{identitymorphisms}}
Fusion matrices $F$ for the Moore-Seiberg data,
and the (quantum) 6j symbols for the (quantum) groups are used in the literature to capture the data of associators.

\begin{figure}
\[
\inc{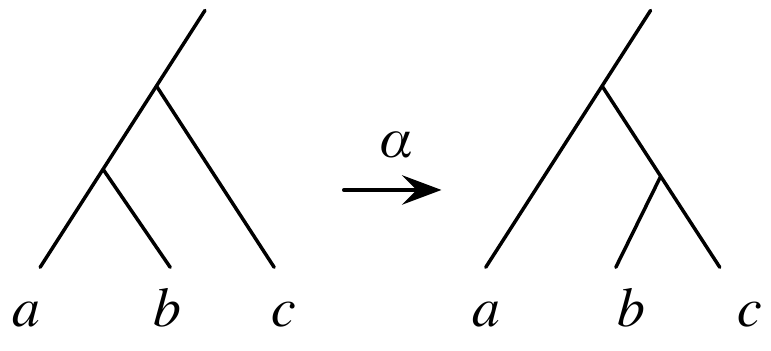}\qquad\qquad
\begin{tikzpicture}[line width=1pt,baseline=(b)]
\begin{scope}[every node/.style={sloped,allow upside down}]
\coordinate (023) at (0,0);
\coordinate (012) at ($(023)+(-1.5,1.2)$);
\coordinate (123) at ($(023)+(1.5,2.4)$);
\coordinate (013) at ($(023)+(0,3.6)$);
\coordinate (controlu) at ($(013)+(3,3)$);
\coordinate (controld) at ($(023)+(3,-3)$);
\coordinate (mid) at ($(controlu)!0.5!(controld)-(0.75,0)$);
\coordinate (up) at ($(013)+(0,1.5)$);
\coordinate (down) at ($(023)-(0,1.5)$);

\arrowpath{(023)}{(012)}{0.5};
\arrowpath{(023)}{(123)}{0.5};
\arrowpath{(012)}{(123)}{0.5};
\arrowpath{(012)}{(013)}{0.5};
\arrowpath{(123)}{(013)}{0.5};
\arrowpath{(013)}{(up)}{0.5};
\arrowpath{(down)}{(023)}{0.5};


\node[below left] at ($(023)!0.5!(012)$) {\scriptsize{$a\ot b$}};
\node[right] at ($(023)!0.5!(123)$) {\scriptsize{$c$}};
\node(b)[above] at ($(012)!0.5!(123)$) {\scriptsize{$b$}};
\node[left] at ($(012)!0.5!(013)$) {\scriptsize{$a$}};
\node[above right] at ($(123)!0.5!(013)$) {\scriptsize{$b\ot c$}};
\node[above] at (up) {\scriptsize{$a\ot(b\ot c)$}};
\node[below] at (down) {\scriptsize{$(a\ot b)\ot c$}};
\end{scope}
\end{tikzpicture}
\]
\caption{
Left: The associator relates two different orders to tensor three lines.
Right: Equivalently, the local operator labeled by the \emph{associator} morphism is obtained by squeezing the region between $(a\otimes b)\ot c$ and $a\ot(b\ot c)$ shown above to a point.} \label{fig:associator}
\end{figure}
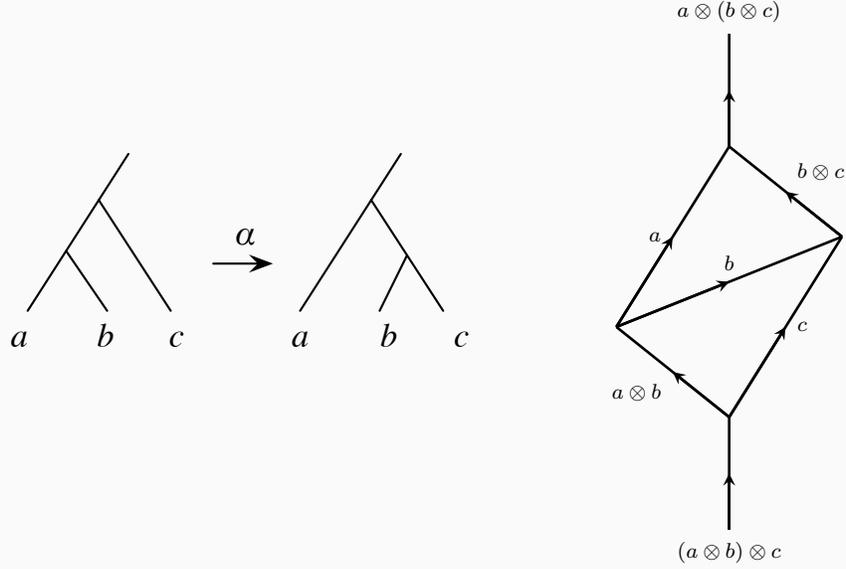

\begin{figure}
\[
\inc{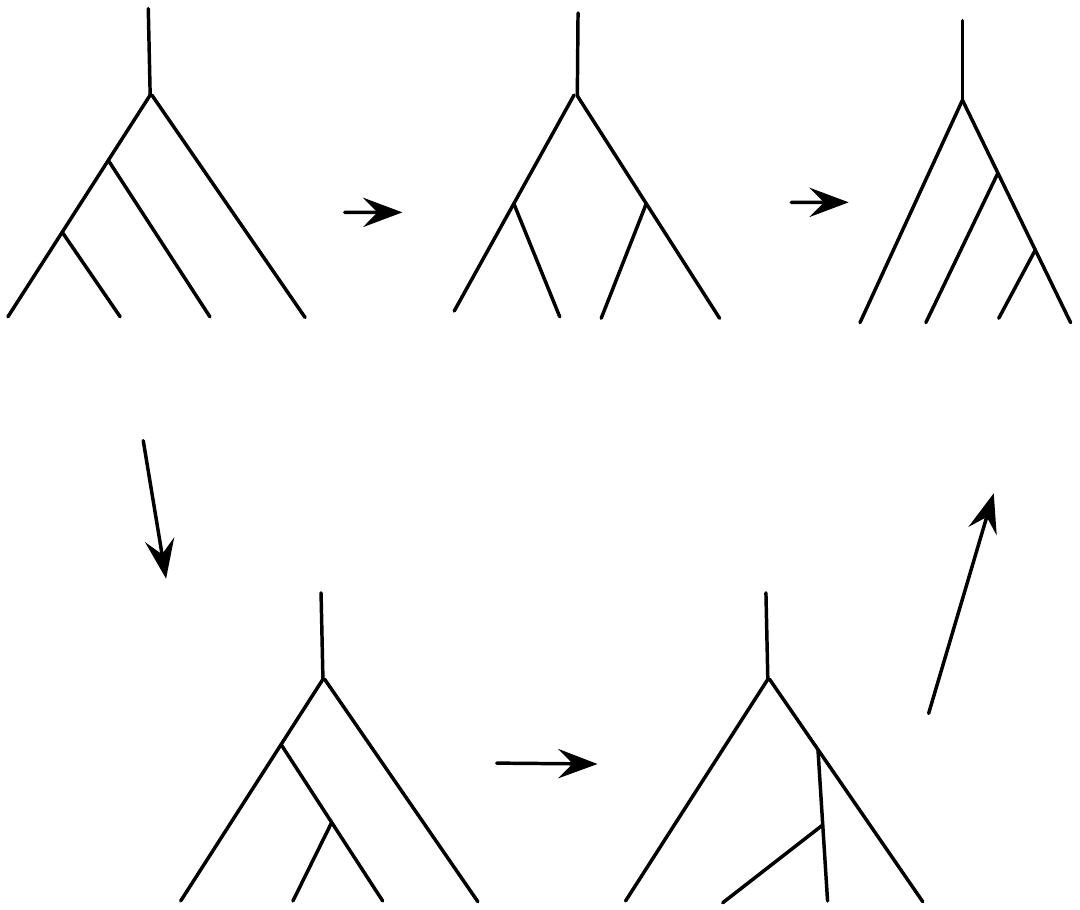}
\]
\caption{The pentagon identity guarantees that two distinct ways to rearrange the order of the tensoring of four lines lead to the same result.\label{fig:pentagon}}
\end{figure}

The associator $\alpha_{a,b,c}$ corresponds to a local operator which implements the process of exchanging line $b$ from the vicinity of $a$ to the vicinity of $c$. See Figure \ref{fig:associator}.
They satisfy the pentagon identity which states the equality of following two morphisms
\begin{multline}
((a\otimes b)\otimes c)\otimes d\to(a\otimes (b\otimes c))\otimes d\to a\otimes ((b\otimes c)\otimes d)\to a\otimes (b\otimes (c\otimes d))\\
=((a\otimes b)\otimes c)\otimes d\to(a\otimes b)\otimes (c\otimes d)\to a\otimes (b\otimes (c\otimes d))
\end{multline}
where each side of the equation stands for the composition of the  corresponding associators. The pentagon identity ensures that exchanging two middle lines $b$, $c$ between two outer lines $a$, $d$ in two different ways is the same, see Fig.~\ref{fig:pentagon}.

\subsubsection{Dual structure}
For every object $a$, $\cC$ contains a dual object $a^*$. The line labeled by the dual object has the property that \begin{equation}
\vev{\cdots a(C)\cdots}
=\vev{\cdots a^*(\tilde C)\cdots}.
\end{equation}
Here, $\tilde C$ denotes the same path $C$ but with a reverse orientation, and the morphisms attached at the junctions on $\tilde C$ need to be changed appropriately as we explain below at the end of this subsubsection. 

We require that the dual of the dual is naturally isomorphic to the original object: $(a^*)^*\simeq a$.
The dual operation also changes the order of the tensoring: \begin{equation}
(a\otimes b)^*=b^* \otimes a^*.
\end{equation}

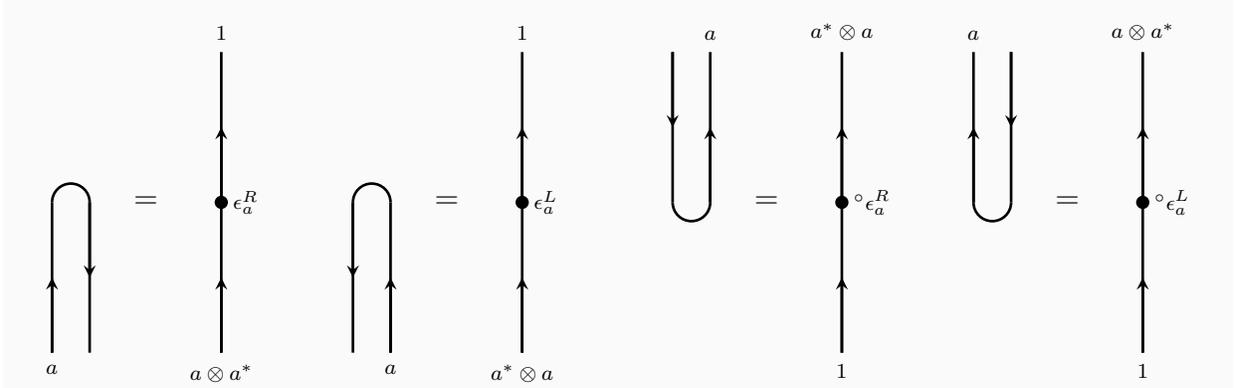
\begin{figure}
\centering
\begin{tikzpicture}[line width=1pt]
\begin{scope}[every node/.style={sloped,allow upside down}]
\coordinate (lowA1) at (0,0);
\coordinate (highA1) at ($(lowA1) + (0,2)$);
\coordinate (lowA2) at ($(lowA1)+(0.5,0)$);
\coordinate (highA2) at ($(lowA2) + (0,2)$);
\arrowpath{(lowA1)}{(highA1)}{0.5};
\arrowpath{(highA2)}{(lowA2)}{0.5};
\draw (highA1) arc[radius=0.25, start angle=180, end angle=0];
\node[below] at (lowA1) {\scriptsize{$a$}};

\coordinate (eqA) at ($(highA2)+(0.75,0)$);
\node at (eqA) {$=$};

\coordinate (lowA3) at ($(lowA2)+(1.75,0)$);
\coordinate (highA3) at ($(lowA3) + (0,4)$);
\coordinate (midA3) at ($(lowA3)!0.5!(highA3)$);
\arrowpath{(lowA3)}{(midA3)}{0.5};
\arrowpath{(midA3)}{(highA3)}{0.5};
\coordinate (halfboxA3) at (0.25, 0.2);
\draw[fill=black] (midA3) circle(2pt);
\node[right] at (midA3) {\scriptsize{$\ev_a^R$}};
\node[below] at (lowA3) {\scriptsize{$a \otimes a^*$}};
\node[above] at (highA3) {\scriptsize{$1$}};

\coordinate (lowA1X) at ($(lowA1)+(4,0)$);
\coordinate (highA1X) at ($(lowA1X) + (0,2)$);
\coordinate (lowA2X) at ($(lowA1X)+(0.5,0)$);
\coordinate (highA2X) at ($(lowA2X) + (0,2)$);
\arrowpath{(highA1X)}{(lowA1X)}{0.5};
\arrowpath{(lowA2X)}{(highA2X)}{0.5};
\draw (highA2X) arc[radius=0.25, start angle=0, end angle=180];
\node[below] at (lowA2X) {\scriptsize{$a$}};

\coordinate (eqAX) at ($(highA2X)+(0.75,0)$);
\node at (eqAX) {$=$};

\coordinate (lowA3X) at ($(lowA2X)+(1.75,0)$);
\coordinate (highA3X) at ($(lowA3X) + (0,4)$);
\coordinate (midA3X) at ($(lowA3X)!0.5!(highA3X)$);
\arrowpath{(lowA3X)}{(midA3X)}{0.5};
\arrowpath{(midA3X)}{(highA3X)}{0.5};
\coordinate (halfboxA3X) at (0.25, 0.2);
\draw[fill=black] (midA3X) circle(2pt);
\node[right] at (midA3X) {\scriptsize{$\ev_a^L$}};
\node[below] at (lowA3X) {\scriptsize{$a^* \otimes a$}};
\node[above] at (highA3X) {\scriptsize{$1$}};

\coordinate (lowB1) at ($(midA3)+(6,0)$);
\coordinate (highB1) at ($(lowB1) + (0,2)$);
\coordinate (lowB2) at ($(lowB1)+(0.5,0)$);
\coordinate (highB2) at ($(lowB2) + (0,2)$);
\arrowpath{(lowB2)}{(highB2)}{0.5};
\arrowpath{(highB1)}{(lowB1)}{0.5};
\draw (lowB1) arc[radius=0.25, start angle=180, end angle=360];
\node[above] at (highB2) {\scriptsize{$a$}};

\coordinate (eqB) at ($(lowB2)+(0.75,0)$);
\node at (eqB) {$=$};

\coordinate (lowB3) at ($(lowB2)+(1.75,-2)$);
\coordinate (highB3) at ($(lowB3) + (0,4)$);
\coordinate (midB3) at ($(lowB3)!0.5!(highB3)$);
\arrowpath{(lowB3)}{(midB3)}{0.5};
\arrowpath{(midB3)}{(highB3)}{0.5};
\coordinate (halfboxB3) at (0.25, 0.2);
\draw[fill=black] (midB3) circle(2pt);
\node[right] at (midB3) {\scriptsize{$\coev_a^R$}};
\node[above] at (highB3) {\scriptsize{$a^* \otimes a$}};
\node[below] at (lowB3) {\scriptsize{$1$}};

\coordinate (lowB1X) at ($(lowB1)+(4,0)$);
\coordinate (highB1X) at ($(lowB1X) + (0,2)$);
\coordinate (lowB2X) at ($(lowB1X)+(0.5,0)$);
\coordinate (highB2X) at ($(lowB2X) + (0,2)$);
\arrowpath{(lowB1X)}{(highB1X)}{0.5};
\arrowpath{(highB2X)}{(lowB2X)}{0.5};
\draw (lowB2X) arc[radius=0.25, start angle=360, end angle=180];
\node[above] at (highB1X) {\scriptsize{$a$}};

\coordinate (eqBX) at ($(lowB2X)+(0.75,0)$);
\node at (eqBX) {$=$};

\coordinate (lowB3X) at ($(lowB2X)+(1.75,-2)$);
\coordinate (highB3X) at ($(lowB3X) + (0,4)$);
\coordinate (midB3X) at ($(lowB3X)!0.5!(highB3X)$);
\arrowpath{(lowB3X)}{(midB3X)}{0.5};
\arrowpath{(midB3X)}{(highB3X)}{0.5};
\coordinate (halfboxB3X) at (0.25, 0.2);
\draw[fill=black] (midB3X) circle(2pt);
\node[right] at (midB3X) {\scriptsize{$\coev_a^L$}};
\node[above] at (highB3X) {\scriptsize{$a \otimes a^*$}};
\node[below] at (lowB3X) {\scriptsize{$1$}};

\end{scope}
\end{tikzpicture}
\caption{Folding a line and squeezing it gives rise to local operators labeled by evaluation and co-evaluation morphisms.}\label{fig:canonical}
\end{figure}

We demand that there are \emph{evaluation} morphisms
\be
\begin{aligned}
\ev^R_a&:a\otimes a^*\to 1,\label{eq:eval} &
\ev^L_a&:a^*\otimes a\to 1
\end{aligned}
\ee
and \emph{co-evaluation} morphisms\footnote{We use the convention that when something is denoted by $x$, co-something is denoted by $\co{x}$. This usage is unconventional, in particular for the case of coproduct for which $\Delta$ is definitely the standard notation, but it reduces the amount of notations that one has to remember. }
\be
\begin{aligned}
\coev_a^R&:1\to a^*\otimes a,\label{eq:coeval} &
\coev_a^L&:1\to a\otimes a^*.
\end{aligned}
\ee
These label local operators corresponding to the process of folding a line operator $a$. See Figure \ref{fig:canonical}.

We note that $\ev_a^R$ and $\ev_{a^*}^L$ are not necessarily equal. However, we require as part of definition of dual structure that they are related as follows
\be
\ev_a^R=\ev_{a^*}^L \circ (p_a\otimes 1)=\ev_{a^*}^L \circ (1\otimes p^{-1}_{a^*})\label{pev}
\ee
where $p_a$ is an isomorphism from $a$ to $a$ and $p_{a^*}$ is an isomorphism from $a^*$ to $a^*$. Similarly, we require
\be
\coev_a^R=(1\otimes p^{-1}_{a})\circ\coev_{a^*}^L= (p_{a^*}\otimes 1) \circ \coev_{a^*}^L\label{pcoev}
\ee
The data of $p_a$ and $p_{a^*}$ is referred to in the literature as a \emph{pivotal structure} on the fusion category $\cC$.

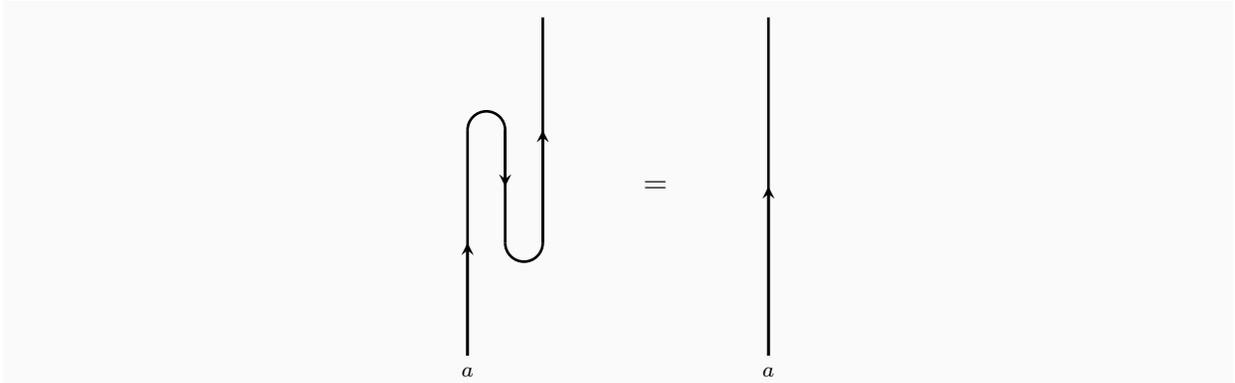
\begin{figure}
\centering
\begin{tikzpicture}[line width=1pt]
\begin{scope}[every node/.style={sloped,allow upside down}]
\coordinate (lowA1) at (0,0);
\coordinate (highA1) at ($(lowA1) + (0,3)$);
\coordinate (lowA2) at ($(lowA1)+(0.5,1.5)$);
\coordinate (highA2) at ($(highA1) + (0.5,0)$);
\arrowpath{(lowA1)}{(highA1)}{0.5};
\arrowpath{(highA2)}{(lowA2)}{0.5};
\draw (highA1) arc[radius=0.25, start angle=180, end angle=0];
\node[below] at (lowA1) {\scriptsize{$a$}};
\draw (lowA2) arc[radius=0.25, start angle=180, end angle=360];
\coordinate (lowA3) at ($(lowA2)+(0.5,0)$);
\coordinate (highA3) at ($(lowA3) + (0,3)$);
\arrowpath{(lowA3)}{(highA3)}{0.5};

\coordinate (eqA) at ($(lowA2)!0.5!(highA2)+(2,0)$);
\node at (eqA) {$=$};

\coordinate (lowA) at ($(lowA1)+(4,0)$);
\coordinate (highA) at ($(lowA)+(0,4.5)$);
\arrowpath{(lowA)}{(highA)}{0.5};
\node[below] at (lowA) {\scriptsize{$a$}};
\end{scope}
\end{tikzpicture}
\caption{Consistency condition on evaluation and co-evaluation morphisms resulting from a topological deformation.}\label{fig:unfold}
\end{figure}

The evaluation and co-evaluation morphisms have to satisfy the following consistency condition with the associator
\be
(\ev_a^R\otimes 1)\circ \alpha_{a,a^*,a}{}^{-1}\circ(1\otimes \coev_a^R)=1
\ee
as morphisms from $a$ to $a$. This ensures that a line with two opposite folds in the right direction can be unfolded as shown in Figure \ref{fig:unfold}. A similar identity is satisfied by $\ev_a^L$ and $\coev_a^L$ which ensures that two opposite folds in the left direction can be unfolded.

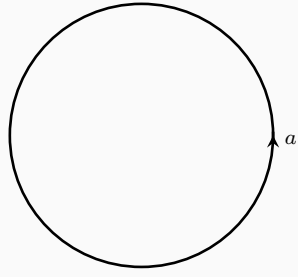
\begin{figure}
\centering
\begin{tikzpicture}[line width=1pt]
\begin{scope}[every node/.style={sloped,allow upside down}]
\coordinate (H) at ($(0,0)+(6,0)$);
\draw (H) circle(1.75cm);
\arrowpath{($(H)+(1.75,-0.05)$)}{($(H)+(1.75,0.05)$)}{0.5};
\node[right] at ($(H)+(1.75,-0.05)$) {\scriptsize{$a$}};
\end{scope}
\end{tikzpicture}
\caption{A loop of line $a$ constructed by composing evaluation and co-evalutaion morphisms. The loop, if it contains no other operators in it, can then be shrunk and the partition function with the loop is equal to $\dim_\text{CC} a=\dim a$ times the partition function without the loop, with all other insertions unchanged.}
\label{fig:dim}
\end{figure}

Using evaluation and co-evaluation morphisms of different parity we can construct loops of lines
\begin{align}
(\dim_\text{CC} a)\id &: 1\stackrel{\coev_{a}^R}{\longrightarrow} a^*\otimes a\stackrel{\ev_a^L}{\longrightarrow} 1,\\
(\dim_\text{C} a)\id &: 1\stackrel{\coev_{a}^L}{\longrightarrow} a\otimes a^*\stackrel{\ev_a^R}{\longrightarrow} 1.
\end{align}
These are morphisms from 1 to 1 and hence they are proportional to the identity morphism. The proportionality factors define two numbers: the counter-clockwise dimension $\dim_\text{CC} a$ of $a$ and the clockwise dimension $\dim_\text{C} a$ of $a$. See Figure \ref{fig:dim}.

Since we can replace the label $a$ by $a^*$ at the cost of flipping the orientation of line, we must have
\begin{align}
\dim_\text{CC} a&=\dim_\text{C} a^*,\\
\dim_\text{CC} a^*&=\dim_\text{C} a.
\end{align}
Indeed, this follows from (\ref{pev}) and (\ref{pcoev}).

In fact, it turns out that we can further argue that
\be
\dim_\text{CC} a=\dim_\text{C} a\equiv\dim a. \label{dimeq}
\ee
To see this, place a small counter-clockwise loop of line $a$ around the ``north pole'' on the sphere. Let there be no other insertions anywhere on the sphere. This evaluates to $\dim_\text{CC} a\times Z_{S^2}$ where $Z_{S^2}$ is the partition function on sphere. Now we can move the line such that it looks like a small clockwise loop around the ``south pole'' on the sphere. This evaluates to $\dim_\text{C} a\times Z_{S^2}$. Equating the two expressions we find (\ref{dimeq}) \footnote{The authors thank Shu-Heng Shao for discussion related to this point.}. This is a further constraint on $\cC$. If the fusion category $\cC$ satisfies (\ref{dimeq}), then $\cC$ is called a \emph{spherical fusion category} in the literature.

Since $\ev_a^R$ and $\ev_{a^*}^L$ are not necessarily equal, we have to specify whether a folding of line $a$ to the right should be read as the morphism $\ev_a^R$ or the morphism $\ev_{a^*}^L$. Similarly there is a specification of $\coev_a^R$ vs. $\coev_{a^*}^L$. 
This issue can be dealt with in two ways, which are technically equivalent but have a rather different flavor.

\paragraph{One method:}
One perspective is to regard that a line is always labeled by the pair (the local orientation, an object in $\cC$).
Then, a pair $(\uparrow,a)$ and $(\downarrow, a^*)$ are isomorphic but not actually the same.
We note that this distinction needs to be made even when $a\simeq a^*$.
Then we make the rule that when a vertical line is labeled by $(\uparrow,a)$ up to some point and then labeled by $(\downarrow, a^*)$ from that point, we insert the pivotal structure $p_a\in \Hom(a,a)$ at that point.
This approach would be preferred by those who have no trouble with adding local orientations as a new datum to a topological line operator.

\paragraph{Another method:}
Another perspective is to think that the change between $\ev_a^R$ and $\ev_{a^*}^L$
and between $\coev_a^R$ vs. $\coev_{a^*}^L$ 
is canceled by changing the nearby morphisms.
This method might be preferred by those who do not want to add local orientation as a new datum to a topological line operator.

We emphasize that, in this approach, the operation of exchanging $a$ by $a^*$ with a reversed orientation does not change the local operators at the junctions. 
Instead it changes the way the local operators at the junctions are read as morphisms in the associated symmetry category $\cC$.

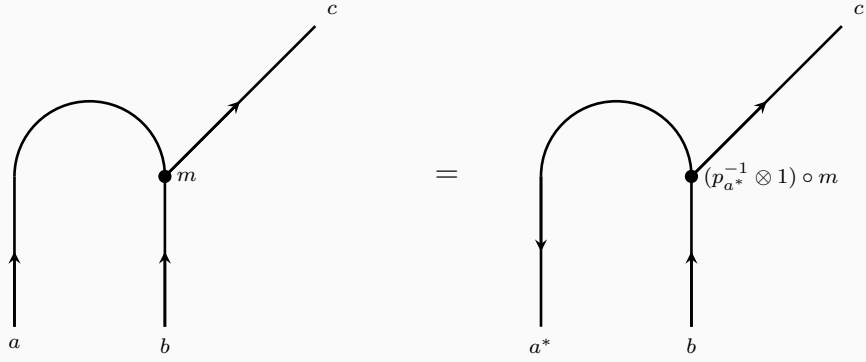
\begin{figure}
\centering
\begin{tikzpicture}[line width=1pt]
\begin{scope}[every node/.style={sloped,allow upside down}]
\coordinate (lowA1) at (0,0);
\coordinate (highA1) at ($(lowA1) + (0,2)$);
\coordinate (lowA2) at ($(lowA1)+(2,0)$);
\coordinate (highA2) at ($(lowA2) + (0,2)$);
\coordinate (high1) at ($(highA2) + (2,2)$);
\arrowpath{(lowA1)}{(highA1)}{0.5};
\arrowpath{(lowA2)}{(highA2)}{0.5};
\arrowpath{(highA2)}{(high1)}{0.5};
\draw (highA1) arc[radius=1, start angle=180, end angle=0];
\node[below] at (lowA1) {\scriptsize{$a$}};
\node[below] at (lowA2) {\scriptsize{$b$}};
\node[above right] at (high1) {\scriptsize{$c$}};
\node[right] at (highA2) {\scriptsize{$m$}};
\draw[fill=black] (highA2) circle(2pt);
\coordinate (eqA) at ($(highA2)+(3.75,0)$);
\node at (eqA) {$=$};

\coordinate (lowA1) at (7,0);
\coordinate (highA1) at ($(lowA1) + (0,2)$);
\coordinate (lowA2) at ($(lowA1)+(2,0)$);
\coordinate (highA2) at ($(lowA2) + (0,2)$);
\coordinate (high1) at ($(highA2) + (2,2)$);
\arrowpath{(highA1)}{(lowA1)}{0.5};
\arrowpath{(lowA2)}{(highA2)}{0.5};
\arrowpath{(highA2)}{(high1)}{0.5};
\draw (highA1) arc[radius=1, start angle=180, end angle=0];
\node[below] at (lowA1) {\scriptsize{$a^*$}};
\node[below] at (lowA2) {\scriptsize{$b$}};
\node[above right] at (high1) {\scriptsize{$c$}};
\node[right] at (highA2) {\scriptsize{$(p^{-1}_{a^*}\otimes 1)\circ m$}};
\draw[fill=black] (highA2) circle(2pt);
\end{scope}
\end{tikzpicture}
\caption{The fold in the diagram on left hand side is specified as $\ev_a^R$ and the fold on the right is specified as $\ev_{a^*}^L$. This changes the morphism from $m$ on the left side to $(p^{-1}_{a^*}\otimes 1)\circ m$ on the right side. These two diagrams provide two different categorical representations of the same physical configuration.}\label{move1}
\end{figure}

The following moves are sufficient to specify what happens in any situation:
\ben
\item Consider a morphism $\ev_a^R\circ\alpha^{-1}_{a,a^*,c}\circ(1\otimes m):a\otimes b\to c$ where $m:b\to a^*\otimes c$. This is equal to $\ev_{a^*}^L\circ\alpha^{-1}_{a,a^*,c}\circ(1\otimes n):a\otimes b\to c$ where $n=(p^{-1}_{a^*}\otimes 1)\circ m:b\to a^*\otimes c$.
\item Consider a morphism $\ev_{a^*}^R\circ\alpha_{c,a^*,a}\circ(m\otimes 1):b\otimes a\to c$ where $m:b\to c\otimes a^*$. This is equal to $\ev_{a}^R\circ\alpha_{c,a^*,a}\circ(n\otimes 1):b\otimes a\to c$ where $n=(p_{a^*}\otimes 1)\circ m:b\to c\otimes a^*$.
\item Consider a morphism $(1\otimes m)\circ\alpha_{a^*,a,b}\circ\coev_a^R:b\to a^*\otimes c$ where $m:a\otimes b\to c$. This is equal to $(1\otimes n)\circ\alpha_{a^*,a,b}\circ\coev_{a^*}^L:b\to a^*\otimes c$ where $n=m\circ(p_a^{-1}\otimes 1):a\otimes b\to c$.
\item Consider a morphism $(m\otimes 1)\circ\alpha^{-1}_{b,a,a^*}\circ\coev_{a^*}^R:b\to c\otimes a^*$ where $m:b\otimes a\to c$. This is equal to $(n\otimes 1)\circ\alpha^{-1}_{b,a,a^*}\circ\coev_{a}^L:b\to c\otimes a^*$ where $n=m\circ(1\otimes p_a):b\otimes a\to c$.
\een
These moves follow from (\ref{pev}) and (\ref{pcoev}). We draw a picture of the first move in Figure \ref{move1}. The other three moves are also described by similar pictures.

\subsubsection{Unitary structure}
The unitary structure requires an existence of a conjugate-linear involution sending $m\in \Hom(a,b)$ to $m^\dagger \in \Hom(b,a)$, generalizing the Hermitian conjugate in the standard linear algebra. 

We require that the evaluation and the coevaluation morphisms are related by this conjugate operation: \begin{equation}
\coev^R_{a^*} = (\ev^L_a)^\dagger,\qquad
\coev^L_{a^*} = (\ev^R_a)^\dagger.
\end{equation}

We further require $m^\dagger \circ m\in \Hom(a,a)$ to be positive semi-definite in the following sense:
Since we assumed the semisimplicity and the finiteness of the number of simple objects,
$\Hom(a,a)$ can naturally be identified with a direct sum of a matrix algebra.
Then we require $m^\dagger\circ m$ to have non-negative eigenvalues.

The above positivity condition requires $\dim a>0$ for all $a$. 
As we will see in Sec.~\ref{subsubsec:determinedim} below,
the unitarity implies sphericity.

\subsection{Comments}
We have several comments:
\begin{itemize}
\item Note that category theorists do not like unitary structures, since it is specific to the base field $\bC$ while they would like to keep everything usable for arbitrary base field. 
For this reason they often distinguish various concepts of $*$ operations and various structures satisfied by them, such as rigid structure, pivotal structure, spherical structure and pseudo-unitary structure.%
\footnote{The rigid structure posits the existence of the left dual $a^*$ and the right dual ${}^*a$, satisfying various conditions. It can be shown that $^{**}a\simeq a^{**}$, and $a^{****}\simeq a$. The pivotal structure is a collection of isomorphisms $a^{**}\simeq a$. 
In our description, the pivotal structure relates $\epsilon_a^L$ and $\epsilon_{a^*}^R$.
A pivotal structure is called spherical if $\dim a=\dim a^*$ for all $a$. A spherical structure is called pseudo-unitary if $|{\dim a}|$ is the largest eigenvalue of  $(N_a)^c_b$ for all simple $a$. For the definition of $N_{ab}^c$, see Sec.~\ref{subsubsec:fusionrule}.} 
If we consider unitary 2d quantum field theories (or more precisely its Wick-rotated versions which are reflection-positive), the unitary structure is the most natural one.

Operator algebraic quantum field theorists in fact work in this setting, since for them the existence of the positive-definite inner product on the Hilbert space is paramount.
Unfortunately their papers often phrase purely categorical results in the operator algebra theoretic language, which makes them somewhat harder for outsiders to digest.
From this point of view their review article \cite{Bischoff:2014xea} is very helpful, where a concise translation between terminologies of two different schools is given.
\item Every property given above, except the simplicity of identity, semisimplicity and finiteness, is a straightforward expression of how topological lines and the junction operators associated to symmetries should behave. 
We impose the simplicity of identity, semisimplicity and finiteness to make the situation tractable.
When the semisimplicity is dropped, the category is called a finite tensor category;
when the simplicity of identity is dropped, the category is called a finite multi-fusion category; when both are dropped, it is called a finite multi-tensor category. When finiteness is dropped, we simply drop the adjective ``finite".

Indeed, if we consider all topological lines in a given 2d theory and all topological operators on topological lines, they might not in general form a unitary fusion category.
Rather, our point of view is that we take a subset of topological lines and subspaces of topological operators on the lines so that they form a unitary fusion category, and then it can be thought of as a symmetry of the 2d theory.

\item A very similar categorical structure was introduced by Moore and Seiberg \cite{Moore:1989vd} in the analysis of 2d RCFTs and 3d TFTs. In the category theory they are now called unitary modular tensor categories. 
In fact, the unitary modular tensor categories \emph{are} also unitary fusion categories, where the latter description is obtained by forgetting the braiding.
\end{itemize}

\subsection{More notions of symmetry categories}
Before discussing examples, it is useful to set up a few more notions:

\subsubsection{`Homomorphisms' between symmetry categories}
In the case of two groups $G_1$ and $G_2$, we have the concept of homomorphisms $\varphi:G_1\to G_2$, preserving the group multiplication.
Similarly, we can talk about \emph{symmetry functors} $\varphi: \cC_1 \to \cC_2$ between two symmetry categories, together with the data specifying how the structures listed above are mapped. 
Among them are isomorphisms \begin{equation}
\epsilon_{a,b} \in \Hom(\varphi(a)\otimes \varphi(b) , \varphi(a\otimes b))\label{monoidalfunctor}
\end{equation}
which tell us how the tensor structure of $\cC_1$ is mapped into the tensor structure of $\cC_2$. 
For example, the morphisms $\epsilon_{a,b}$ map the associator of $\cC_1$ to the associator of $\cC_2$.

Two symmetry functors $\varphi,\varphi':\cC_1\to \cC_2$ are considered equivalent when there is 
a set of isomorphisms \begin{equation}
\eta_a \in \Hom(\varphi(a),\varphi'(a))
\end{equation} such that \begin{equation}
\eta_{a\otimes b} \epsilon_{a,b} = \epsilon'_{a,b}  (\eta_a \otimes \eta_b).\label{2coboundary}
\end{equation}
When a symmetry functor has an inverse, it is called an equivalence between symmetry categories.

\subsubsection{Products of symmetry categories}
In the case of two groups $G_1$ and $G_2$, their product $G_1\times G_2$ is also a group.
Similarly, given two symmetry categories $\cC_1$ and $\cC_2$, we denote their product as $\cC_1\boxtimes \cC_2$, whose simple objects are given by $a_1\boxtimes a_2$ where $a_{1,2}$ are simple objects of $\cC_{1,2}$, respectively. 
This product is called Deligne's tensor product of categories.

\subsubsection{Fusion rule of unitary fusion categories}
\label{subsubsec:fusionrule}
A symmetry category comes with a lot of structures.
Sometimes it is useful to forget about most of them as follows.
For each isomorphism class of simple objects $a$, introduce a symbol $[a]$,
and define their multiplication by $[a][b]:=\sum_c N_{ab}^c [c]$ when $a\otimes b=\bigoplus_c N_{ab}^c c$. 
This makes non-negative integral linear combinations of $[a]$'s into an algebra  over $\bZ_+$ with a specific given basis.
We call this algebra $R(\cC)$ 
the fusion ring of the symmetry category $\cC$. 
In the case of modular tensor categories, this algebra is also called the Verlinde algebra.
We would often call this algebra as just the fusion rule of $\cC$.

Let $n$ be the number of  isomorphism classes of simple objects. Then we can regard $(N_a)_b^c$ as $n\times n$ matrices and $[a]\mapsto N_a$ is the adjoint representation of the fusion ring.

\subsubsection{Determination of dimensions of objects}
\label{subsubsec:determinedim}
The dimensions are fixed by $N_{ab}^c$. To see this, consider the  $n$-dimensional vector $v:=(\dim a)_a$ where $a$ runs over the isomorphism classes of simple objects. 
Its entries are positive real numbers thanks to the unitarity.
Furthermore, $v$ is the simultaneous eigenvector of all $N_a$'s with eigenvalues $\dim a$.
Then by the Perron-Frobenius theorem, $\dim a$ is the largest eigenvalue of the matrix $N_a$, which is guaranteed to be positive.
The argument above applies both to $\dim_\text{C}$ and $\dim_\text{CC}$, and therefore the sphericity is implied by the unitarity.
Unitarity also guarantees $\dim a=\dim a^*$.

We define the total dimension of the symmetry category $\cC$ by the following formula: \begin{equation}
\dim \cC=\sum_a (\dim a)^2.
\end{equation}
Here  the sum runs over the isomorphism classes of simple objects.

\subsection{Groups and representations of groups as symmetry categories}
\subsubsection{Symmetry categories $\cC(G,\alpha)$}
As an example, let us recast an ordinary group $G$ as a symmetry category.
We first regard each element $g\in G$ as a simple object denoted by the same letter in the category.
We define \begin{equation}
g\otimes g':= gg', \qquad g^* :=g^{-1}.
\end{equation}
Taking $\alpha_{g_1,g_2,g_3}$ to be the identity maps, they clearly form a unitary fusion category, which we denote by $\cC(G)$.

\begin{figure}
\centering
\begin{tikzpicture}[line width=1pt]
\begin{scope}[every node/.style={sloped,allow upside down}]
\coordinate (l) at (3,0);
\coordinate (i) at (0,4);
\coordinate (j) at (2,4);
\coordinate (k) at (4,4);
\coordinate (ij) at ($(l)!0.67!(i)$);
\coordinate (jk) at ($(l)!0.33!(i)$);

\arrowpath{(l)}{(i)}{0.5};
\arrowpath{(l)}{(jk)}{0.5};
\arrowpath{(ij)}{(i)}{0.5};
\arrowpath{(jk)}{(k)}{0.5};
\arrowpath{(ij)}{(j)}{0.5};

\node[above] at (i) {\scriptsize{$g$}};
\node[above] at (j) {\scriptsize{$g'$}};
\node[above] at (k) {\scriptsize{$g''$}};
\node[below] at (l) {\scriptsize{$gg'g''$}};
\node[left] at ($(ij)!0.5!(jk)$) {\scriptsize{$gg'$}};

\coordinate (to) at ($(l)+(3.5,2)$);
\node at (to) {$\longrightarrow$};

\coordinate (L) at ($(l)+(9,0)$);
\coordinate (I) at ($(i)+(9,0)$);
\coordinate (J) at ($(j)+(9,0)$);
\coordinate (K) at ($(k)+(9,0)$);
\coordinate (JK) at ($(jk)+(9,0)$);
\coordinate (IJ) at ($(JK)!0.5!(K)$);

\arrowpath{(JK)}{(IJ)}{0.5};
\arrowpath{(L)}{(JK)}{0.5};
\arrowpath{(IJ)}{(K)}{0.5};
\arrowpath{(JK)}{(I)}{0.5};
\arrowpath{(IJ)}{(J)}{0.5};

\node[above] at (I) {\scriptsize{$g$}};
\node[above] at (J) {\scriptsize{$g'$}};
\node[above] at (K) {\scriptsize{$g''$}};
\node[below] at (L) {\scriptsize{$gg'g''$}};
\node[right] at ($(IJ)!0.5!(JK)$) {\scriptsize{$g'g''$}};

\end{scope}
\end{tikzpicture}
\caption{The same background connection represented in two different ways as networks of topological line operators. If the symmetry is anomalous, they lead to different partition functions.}\label{fig:F}
\end{figure}
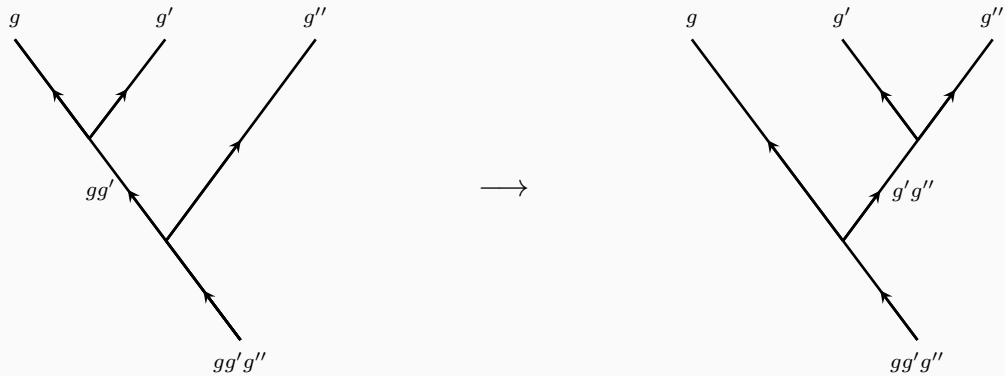

More generally, the pentagon identity among $\alpha_{g_1,g_2,g_3}$ says that $\alpha$ is a 3-cocycle on $G$ valued in $\UU(1)$.\footnote{%
As already noted in footnote \ref{identitymorphisms}, in our convention $\alpha_{g_1,g_2,g_3}$ is trivial whenever any of $g_{1,2,3}$ is the identity. Such a cocycle is called normalized. It is a well-known fact in group theory that group cohomology can be computed by restricting every cochains involved to be normalized.}
Denote the resulting fusion category by $\cC(G,\alpha)$. In the literature it is often denoted $\mathrm{Vec}_G^\alpha$.
We clearly have $\dim g=1$. Thus the total dimension of this symmetry category is the order of the group.

When $\alpha_1$ and $\alpha_2$ differ by a coboundary of a 2-cochain $\epsilon$, we can construct an equivalence of categories between $\cC(G,\alpha_1)$ and $\cC(G,\alpha_2)$ using the functor specified using the same $\epsilon$ in \eqref{monoidalfunctor}.
This means that in the definition of $\cC(G,\alpha)$, one can regard $\alpha\in H^3(G,\UU(1))$. 

It is also clear that any unitary fusion category whose simple objects are all invertible can be made to be of this form. Summarizing, 
\begin{claim}
A symmetry category $\cC$ whose simple lines are all invertible is equivalent to $\cC(G,\alpha)$ where $G$ is a finite group and $\alpha$ is an element in $H^3(G,\UU(1))$.
\end{claim}

\subsubsection{$\cC(G,\alpha)$ and the anomaly}
As is by now familiar, this cohomology class $\alpha\in H^3(G,\UU(1))$ specifies the anomaly of $G$ flavor symmetry in two dimensions \cite{Dijkgraaf:1989pz,Chen:2011pg}. 
One way to see it is as follows \cite{Gaiotto:2014kfa}: Insert a network of lines with trivalent junctions between them on the spacetime manifold $\Sigma$. Let the lines be labeled by simple objects of $\cC(G,\alpha)$, that is by group elements. And let every junction of the form $g\otimes g'\to gg'$ be labeled by the identity morphism $gg'\to gg'$. Such a configuration can also be thought of as reproducing the effects of a background connection on $\Sigma$ which has holonomies given by $g$ on crossing transversely a line labeled by $g$. Now, consider a local region looking like the left hand side of Figure \ref{fig:F}. Move the lines such that now it looks like the right hand side of Figure \ref{fig:F}. This changes the partition function by $\alpha(g,g',g'')$. The new background connection is just a gauge transform of the original background connection. Hence, we see that $\alpha$ precisely captures the anomaly in the flavor symmetry.
Morally, this means the following:
\begin{claim}
A symmetry category $\cC$ includes the specification of its anomaly.
\end{claim}

Fixing a group $G$, the set of its anomalies forms an Abelian group. 
Notice that, in our language, $\cC(G,\alpha)$ for different $\alpha$ have the same fusion ring $R=\bZ_+ G$.
Thus, we can ask the following more general question: does the set of symmetry categories $\cC$ with the same fusion ring $R$ form an Abelian group?
The answer is that we need a coproduct on $R$.
To see this, let us recall why the anomaly of a flavor symmetry forms an Abelian group from the perspective of quantum field theory.

In general, given two theories $T_{1,2}$, we can consider the product theory $T_1\times T_2$ which is just two decoupled theories considered as one.
When $T_i$  has flavor symmetry group $G_i$, the product $T_1\times T_2$ has flavor symmetry group $G_1\times G_2$. 
When $G_1=G_2=G$, we can take the diagonal $G$ subgroup of $G\times G$ and regard $T_1\times T_2$ to have flavor symmetry $G$.
Now, when $T_i$  has the anomaly $\alpha_i$, we define the anomaly of $T_1\times T_2$ to be the sum $\alpha_1+\alpha_2$. 
This abstractly defines the addition operation on the anomaly. 

The crucial step that does not directly generalize to symmetry categories is the existence of the diagonal subgroup $G\subset G\times G$. 
In order to define the addition operation on the set of fusion categories sharing the same fusion rule  $R$, similarly we need a coproduct $R\to R\otimes R$.

\subsubsection{$\cC(G,\alpha)$ and the $G$-SPT phases}

Next, fixing a 3-cocycle $\alpha$, let us ask what is the autoequivalence of $\cC(G,\alpha)$, that is, the self equivalence that preserves the structure as a symmetry category.

Pick an autoequivalence $\varphi:g\to \varphi(g)$ with the associated $\epsilon_{g,h} \in\Hom( \varphi(g)\varphi(h) ,\varphi(gh))$.
Clearly $\varphi$ is an automorphism of $G$. 
Fixing $\varphi$ to be the identity, $\epsilon$ needs to be a 2-cocycle so that it does not change $\alpha$.
Furthermore, two such $\epsilon$'s are considered equivalent when they differ by a 2-coboundary, due to \eqref{2coboundary}.
Therefore $\epsilon$ can be thought of as taking values in $H^2(G,\UU(1))$.
Summarizing,
\begin{claim}
Autoequivalence of $\cC(G,0)$ is the semidirect product $\mathrm{Aut}(G) \ltimes H^2(G,\UU(1))$.
\end{claim}

The $\mathrm{Aut}(G)$ part is clear: it just amounts to renaming the topological lines associated to the group operation. 
How should we think of $H^2(G,\UU(1))$? It is telling us that instead of choosing the identity operator as the implicit junction operator  for $g\otimes g'\to gg'$ as done in Fig.~\ref{implicitjunction}, we can choose $\epsilon_{g,g'}$ times the identity. This will not change the associator but will change the partition function associated to a background connection on $\Sigma$. 
This  corresponds to coupling a two-dimensional theory with $\cC(G,\alpha)$ symmetry with a two-dimensional bosonic symmetry protected topological (SPT) phase, which is specified by the 2-cocycle $\epsilon$ protected by the flavor symmetry $G$. 
The 2-cocycle $\epsilon$ is also known as a discrete torsion of $G$.

\begin{figure}
\[
\inc{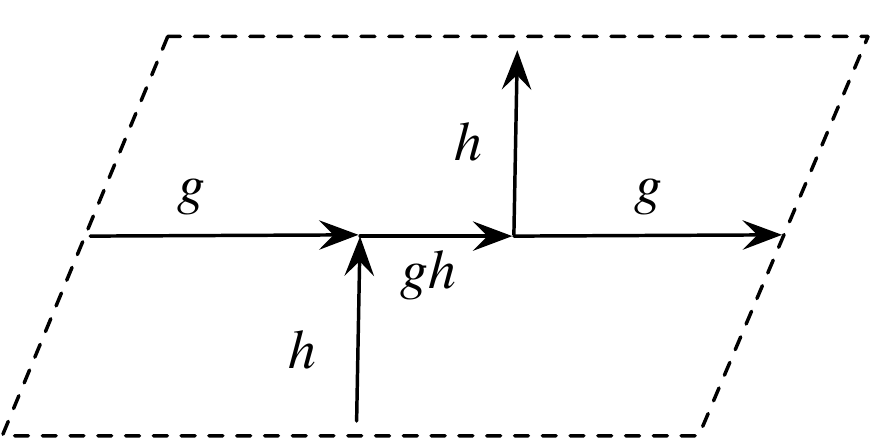}
\]
\caption{The lines representing a commuting holonomy $(g,h)$ on a torus. \label{fig:torusX}}
\end{figure}

As an example, consider a torus with holonomies $g,h$ around two 1-cycles. They can be represented using the topological lines as in Fig.~\ref{fig:torusX}.
There, we resolved the intersection to two trivalent junctions.

We now change the operators at the two junctions to $\epsilon_{g,h}$ and $\epsilon_{g^{-1},h^{-1}}$ given by the values of the 2-cocycle. 
In total the phase of the partition function changes by \begin{equation}
c_{g,h} = \epsilon_{g,h}/\epsilon_{h,g}
\end{equation} which is the standard relation between the discrete torsion phase $c$ on the torus and the 2-cocycle $\epsilon$ \cite{Douglas:1998xa}.
We can thus generalize as follows:
\begin{claim}
Autoequivalences of a symmetry category $\cC$ generalize the notion of  renaming and  multiplying by SPT-phases for a group symmetry.
\end{claim}
We need to keep in mind however that the phases introduced by $\epsilon$ in the general case do not have an interpretation of multiplying a SPT phase protected by $\cC$, since the product of two theories with symmetry $\cC$ has symmetry $\cC\boxtimes \cC$ but is not guaranteed to have symmetry $\cC$, as already discussed above.

\subsubsection{$\Rep(G)$ as symmetry category}\label{sec:repG}
Next, let us discuss $\Rep(G)$ for a finite group $G$.
Its structure as a symmetry category is straightforward: 
the objects are representations of $G$, the morphism space $\Hom(R,S)$ between two representations consists of intertwiners, as the tensor product $\otimes$ and the associator $\alpha_{R,S,T}$ we use the ordinary operations, the dual is the complex conjugate, etc.
The simple objects are the irreducible representations.
The dimension of a representation $R$ as defined above equals the ordinary definition.
Then, the total dimension of $\Rep(G)$ is the sum of the square of the dimensions of the irreducible representations, therefore \begin{equation}
\dim \Rep(G)=|G|.
\end{equation}

Clearly, when $G$ is an Abelian group, we have a natural equivalence $\cC(\hat G)=\Rep(G)$.
This means in particular that the symmetry categories $\Rep(G)$ for distinct Abelian $G$ are distinct.
In fact the fusion rings  $R(\Rep(G))$ are already distinct.

For non-Abelian $G$, the situation is more delicate. For example, take the dihedral group $D_8$ with eight elements (i.e.~the symmetry of a square in three-dimensional space) and the quaternion group $Q_8$ (i.e.~the group formed by eight quaternions $\pm 1$, $\pm i$, $\pm j$, $\pm k$).
Their character tables are the same. In particular, they both have four one-dimensional irreducible representations $1$, $a$, $b$ and $ab$ and a unique two-dimensional irreducible representation $m$, with $m\otimes m \simeq 1\oplus a \oplus b\oplus ab$.
Therefore, their fusion rings are the same.
In this case, however, the symmetry categories are different, since they are known to have distinct associator $\alpha_{m,m,m}$, see e.g.~\cite{TambaraYamagami}.
In this sense, the relation between $\Rep(D_8)$ and $\Rep(Q_8)$ is analogous to the relation between the symmetry categories $\cC(G,\alpha)$ associated to the same group but with a different anomaly.

It is also known that there are cases where two distinct groups lead to the same symmetry category \cite{EtingofGelaki,IzumiKosaki}. 
Let $\bF_2$ be the finite field with two elements.
The symplectic group $\Sp(2d,\bF_2)$ acts on $(\bF_2)^{2d}$. 
The extension of $(\bF_2)^{2d}$ by $\Sp(2d,\bF_2)$ with this natural action is 
classified by the cohomology $H^2(\Sp(2d,\bF_2),(\bF_2)^{2d})$, see Appendix A, in particular \eqref{crossed}.
It is known that when $d\ge 3$, 
there is a suitable nonzero $\nu\in H^2(\Sp(2d,\bF_2),(\bF_2)^{2d})$ 
such that  $G_d=\Sp(2d,\bF_2)\ltimes (\bF_2)^{2d}$. 
and $G_d'=\Sp(2d,\bF_2)\ltimes_\nu (\bF_2)^{2d}$
we have $\Rep(G_d)=\Rep(G_d')$ as a symmetry category.
The smallest example is when $d=3$, where $|G_3|=|G_3'|=(2^6-1)(2^4-1)(2^2-1)2^{3^2}\cdot 2^{3\cdot 2}=92897280$.

As already roughly discussed in Sec.~\ref{sec:general-d-nonabelian}, a theory $T/G$  where $T$ has flavor symmetry $G$ has a symmetry $\Rep(G)$, and re-gauging $\Rep(G)$ we get back the original theory $T$.
Equivalently, if a theory $T'$ has a symmetry $\Rep(G)$, 
gauging $\Rep(G)$ of $T'$ has the symmetry $G$.
In the example above, we said $\Rep(G_d)=\Rep(G_d')$ for different $G_d$ and $G_d'$.
This means that there are (at least) two distinct ways of gauging $\cC=\Rep(G_d)=\Rep(G_d')$, producing two different theories with different symmetries $G_d$ and $G_d'$.
Let us next study what is going on in detail.

\section{Gaugings and symmetry categories}
\label{sec:gauging}

Given a theory $T$ with non-anomalous $G$ symmetry, we can form the gauged theory $T'=T/G$, and $T'$ has $\Rep(G)$ as the symmetry category.
We would like to precisely state the process of gauging $\Rep(G)$ of $T'$ and getting back $T$.

To do this, it is helpful first to state two procedures, one of obtaining $\Rep(G)$ from $G$ and another of obtaining $G$ from $\Rep(G)$, in a manner that treats $G$ and $\Rep(G)$ on an equal footing.
This is done using the concept of module categories and bimodule categories.

Once we understand this process, it is straightforward to study how to gauge a more general symmetry category $\cC$. We will also be able to answer how the symmetry category $\cC$ changes under gauging in the general setting.

\subsection{Module and bimodule categories}
Readers would be familiar with the action of a group or an algebra on a vector space.
Then the vector space is called a representation or equivalently a module.
When there are two commuting actions, one from the left and another from the right, it is often called a bimodule. 
Module categories and bimodule categories are categorified versions of this construction. 
Our expositions will be brief; for more details, please consult Chapter 7 of the textbook \cite{EGNO}.

The categorified version of a linear space is an additive category. 
An additive category $\cM$ is a category with the additive structure whose morphism spaces are vector spaces.
Here and in the following,  we assume objects of $\cM$ to be of the form $\bigoplus_m N_m m $ where $m$ runs over simple objects and $N_m$ are non-negative integers. 
We also assume that the number of isomorphism classes of simple objects is finite.

The simplest additive category is $\Vec$, the category of finite dimensional vector spaces. There is only one simple object $\bC$, and every other object is isomorphic to $n\bC = \bC^n$.

Take $\cC$ to be a symmetry category. 
A left-module category over $\cC$ is an additive category $\cM$ such that for $a\in \cC$ and $m\in \cM$ the product $a \otimes m \in \cM$ is defined, together with the associativity morphisms \begin{equation}
\alpha_{a,b,m} \in \Hom((a\otimes b) \otimes m, a\otimes(b\otimes m))\label{module-associator}
\end{equation} satisfying the pentagon identity. Direct sums of module categories can be easily defined.
When $\cC$ is a symmetry category it is known that every module category can be decomposed into a direct sum of indecomposable module categories.

Physically, a module category specifies a topological boundary condition. Each object $m$ in the module category specifies the boundary condition and the ``type of flux" it carries. A line $a$ can end on $m$ at the cost of transforming $m$ to $a\otimes m$ which changes the flux at the boundary by $a$. If $a\otimes m\simeq m$ for all $a\in\cC$ then $m$, along with its direct sums with itself, describes an indecomposable module category by itself which corresponds to a boundary condition that absorbs all the flux and hence we can refer to it as the ``Dirichlet boundary condition''. In general, an indecomposable module category is a generalization of mixed Dirichlet-Neumann boundary conditions familiar from gauge theory. In the literature, the structure on $\Vec$ of a module category for $\cC$ is often called a fiber functor of $\cC$, but we try not to use this terminology.

We can similarly define a right-module category over $\cC$.
A $(\cC_1,\cC_2)$ bimodule category  is an additive category $\cM$ which has a left action of $\cC_1$ and a right action of $\cC_2$, together with further compatibility morphisms \begin{equation}
\alpha_{c_1,m,c_2} \in  \Hom((c_1\otimes m) \otimes c_2, c_1\otimes(m\otimes c_2)).
\end{equation} 
A bimodule category has the physical interpretation that it describes a topological interface where lines in the symmetry categories $\cC_{1,2}$ can end from the left and from the right, respectively.

Recall that  a representation of an algebra $A$ on a vector space $V$ can be thought of as a homomorphism from $A$ to the algebra of matrices $\Hom(V,V)$. 
Similarly, given an additive category $\cM$ we can form the symmetry category $\mathrm{Func}(\cM,\cM)$ of additive self functors, 
and then giving the structure of a module category on $\cM$ is the same as giving a homomorphism of symmetry categories from $\cC$ to $\mathrm{Func}(\cM,\cM)$.

The commutant of $\cC$ within $\mathrm{Func}(\cM,\cM)$ is denoted by $\cC_\cM^*$ and called the dual of $\cC$ with respect to $\cM$. When $\cM$ is indecomposable, it is known that $(\cC_\cM^*)_\cM^*\simeq \cC$.
$\cM$ is naturally a $(\cC,\cC')$ bimodule category where $\cC'$ is $\cC_\cM^*$ with the order of the tensor product reversed. 
Since we will only use $\cC'$ and not $\cC_\cM^*$ in the following,
 we will call $\cC'$ the dual of $\cC$ with respect to $\cM$.
Two symmetry categories $\cC$ and $\cC'$ such that $\cC'$ is the dual of $\cC$ acting on $\cM$ are called \emph{categorically Morita equivalent}. 
We will come back to this notion in Sec.~\ref{sec:moreregauging}.

\subsection{Duality of $\cC(G)$ and $\Rep(G)$}

Let us discuss examples of module categories.
We can give $\Vec$ a structure of a module category over $\cC(G)$: for $g\in \cC(G)$ and $\bC\in \Vec$, we just define $g \otimes \bC=\bC$, and let the associator be trivial.
Note that this construction does not work for $\cC(G,\alpha)$ with nonzero anomaly $\alpha\in H^3(G,\UU(1))$, since the pentagon will not be satisfied. 

We can also give $\Vec$ a structure of a module category over $\Rep(G)$: for $R \in \Rep(G)$ and $\bC\in \Vec$, we just define $R\otimes \bC=R \in \Vec$, and let the associator be the natural one induced from the tensor product structure of $G$.
We note that for $\cC=\Rep(G_d)=\Rep(G_d')$ at the end of Sec.~\ref{sec:repG},
the module category structures on $\Vec$ as $\Rep(G_d)$ and $\Rep(G_d')$ are different,
since the associators \eqref{module-associator} as morphisms in $\Vec$ turn out to be distinct.

Furthermore, we note that $\Vec$ is a $(\cC(G),\Rep(G))$ bimodule category. The only additional ingredient is the associator 
\begin{equation}
\alpha_{g,\bC,R} \in \Hom((g\otimes \bC)\otimes R,g\otimes(\bC\otimes R))=\Hom(R,R).
\end{equation} 
The compatibility condition just says that $\alpha_{g,\bC,R}$ are representation matrices of $g$ on $R$, chosen compatible with the tensor product.
Physically, we can regard $\Vec$ as the boundary between a system with $G$ symmetry on the left and a gauged system with $\Rep(G)$ symmetry on the right.

In fact, $\Rep(G)$ is characterized as the maximal set of right actions  compatible with the left action of $\cC(G)$, and $\cC(G)$ as the maximal set of left actions compatible with the right action of $\Rep(G)$. 
As we will explicitly confirm in Sec.~\ref{sec:example},
 $\cC(G)$ and $\Rep(G)$ are dual to each other with respect to $\Vec$: 
 \begin{claim}
 The symmetry category $\cC(G)$ and $\Rep(G)$ are dual with respect to the module category $\Vec$.
\end{claim}

\subsection{Gauging by an algebra object} \label{sec:algebra}

\begin{figure}
\centering
\begin{tikzpicture}[line width=1pt]
\begin{scope}[every node/.style={sloped,allow upside down}]
\coordinate (l) at (3,0);
\coordinate (i) at (0,-4);
\coordinate (j) at (2,-4);
\coordinate (k) at (4,-4);
\coordinate (ij) at ($(l)!0.67!(i)$);
\coordinate (jk) at ($(l)!0.33!(i)$);

\arrowpath{(i)}{(l)}{0.5};
\arrowpath{(jk)}{(l)}{0.5};
\arrowpath{(i)}{(ij)}{0.5};
\arrowpath{(k)}{(jk)}{0.5};
\arrowpath{(j)}{(ij)}{0.5};

\node[below] at (i) {\scriptsize{$A$}};
\node[below] at (j) {\scriptsize{$A$}};
\node[below] at (k) {\scriptsize{$A$}};
\node[above] at (l) {\scriptsize{$A$}};
\node[left] at (ij) {\scriptsize{$\mu$}};
\node[left] at (jk) {\scriptsize{$\mu$}};
\node[below right] at ($(ij)!0.5!(jk)$) {\scriptsize{$A$}};

\coordinate (to) at ($(l)+(4,-2)$);
\node at (to) {$=$};

\coordinate (L) at ($(l)+(9,0)$);
\coordinate (I) at ($(i)+(9,0)$);
\coordinate (J) at ($(j)+(9,0)$);
\coordinate (K) at ($(k)+(9,0)$);
\coordinate (JK) at ($(jk)+(9,0)$);
\coordinate (IJ) at ($(JK)!0.5!(K)$);

\arrowpath{(IJ)}{(JK)}{0.5};
\arrowpath{(JK)}{(L)}{0.5};
\arrowpath{(K)}{(IJ)}{0.5};
\arrowpath{(I)}{(JK)}{0.5};
\arrowpath{(J)}{(IJ)}{0.5};

\node[below] at (I) {\scriptsize{$A$}};
\node[below] at (J) {\scriptsize{$A$}};
\node[below] at (K) {\scriptsize{$A$}};
\node[above] at (L) {\scriptsize{$A$}};
\node[right] at (IJ) {\scriptsize{$\mu$}};
\node[left] at (JK) {\scriptsize{$\mu$}};
\node[below left] at ($(IJ)!0.5!(JK)$) {\scriptsize{$A$}};

\end{scope}
\end{tikzpicture}
\caption{Associativity axiom for an algebra object $A\in\cC$}\label{fig:ass}
\end{figure}

\begin{figure}
\centering
\begin{tikzpicture}[line width=1pt]
\begin{scope}[every node/.style={sloped,allow upside down}]
\coordinate (C1) at (0,0);
\coordinate (mid1) at ($(C1)-(0,2)$);
\coordinate (A1) at ($(mid1)+(-1.5,-1.25)$);
\coordinate (B1) at ($(mid1)+(0,-2)$);

\arrowpath{(mid1)}{(C1)}{0.5};
\arrowpath{(A1)}{(mid1)}{0.5};
\arrowpath{(B1)}{(mid1)}{0.5};

\draw[fill=black] (A1) circle(1.5pt);

\node[above] at (C1) {\scriptsize{$A$}};
\node[below] at (A1) {\scriptsize{$u$}};
\node[below] at (B1) {\scriptsize{$A$}};
\node[right] at (mid1) {\scriptsize{$\mu$}};
\node[above left] at ($(A1)!0.5!(mid1)$) {\scriptsize{$A$}};

\coordinate (to) at ($(mid1)+(2,0)$);
\node at (to) {$=$};

\coordinate (C11) at ($(C1)+(4,0)$);
\coordinate (mid11) at ($(C11)-(0,2)$);
\coordinate (A11) at ($(mid11)+(1.5,-1.25)$);
\coordinate (B11) at ($(mid11)+(0,-2)$);

\arrowpath{(mid11)}{(C11)}{0.5};
\arrowpath{(A11)}{(mid11)}{0.5};
\arrowpath{(B11)}{(mid11)}{0.5};

\draw[fill=black] (A11) circle(1.5pt);

\node[above] at (C11) {\scriptsize{$A$}};
\node[below] at (A11) {\scriptsize{$u$}};
\node[below] at (B11) {\scriptsize{$A$}};
\node[left] at (mid11) {\scriptsize{$\mu$}};
\node[above right] at ($(A11)!0.5!(mid11)$) {\scriptsize{$A$}};

\coordinate (to) at ($(mid11)+(3,0)$);
\node at (to) {$=$};

\coordinate (C2) at ($(C11)+(5,0)$);
\coordinate (B2) at ($(B11)+(5,0)$);

\arrowpath{(B2)}{(C2)}{0.5};

\node[below] at (B2) {\scriptsize{$A$}};

\end{scope}
\end{tikzpicture}
\caption{Unit axiom for an algebra object $A\in\cC$.}\label{fig:unit}
\end{figure}

How is this duality of $\cC(G)$ and $\Rep(G)$ related to gauging?
To understand this, we need a digression.
Take $A=\bigoplus_g g$ in $\cC(G)$. 
The group multiplication gives a multiplication morphism $\mu:A\otimes A\to A$, and and the map $1\mapsto\bigoplus_g\delta_{1g}g$ defines a unit morphism $u:1\to A$. These two operations satisfy the properties in Figures \ref{fig:ass} and \ref{fig:unit}. In general, if we have an object $A$ in a general symmetry category $\cC$ along with morphisms $\mu$ and $u$ which satisfy these properties, then $A$ is called an \emph{algebra} in $\cC$. 

We also have corresponding co-morphisms. The map $g\mapsto\frac{1}{|G|}\bigoplus_h(h\otimes h^{-1}g)$ defines a co-multiplication $\co\mu:A\to A\otimes A$ and the projection $g\mapsto\delta_{1g}1$ defines a co-unit morphism $\co{u}:A\to 1$. $\co\mu$ satisfies a co-associativity axiom which is given by turning the graphs in Figure \ref{fig:ass} upside down and changing the directions of all the arrows. Similarly, $\co\mu$ and $\co{u}$ together satisfy a co-unit axiom which is given by turning the graphs in Figure \ref{fig:unit} upside down and changing the directions of all the arrows. In general, if we have an object $A\in\cC$ with morphisms $\co\mu$ and $\co{u}$ satisfying these axioms, we say that $A$ is a \emph{co-algebra} in $\cC$.

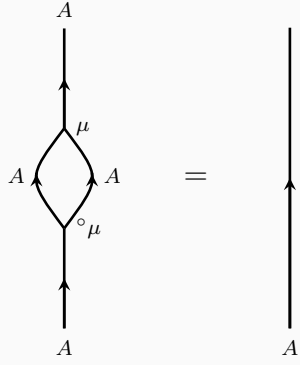
\begin{figure}
\centering
\begin{tikzpicture}[line width=1pt]
\begin{scope}[every node/.style={sloped,allow upside down}]
\coordinate (C2) at (0,0);
\coordinate (mid2) at ($(C2)+(0,1.33)$);
\coordinate (mid3) at ($(mid2)+(0,1.33)$);
\coordinate (C3) at ($(mid3)+(0,1.33)$);

\arrowpath{(C2)}{(mid2)}{0.5};
\arrowpath{(mid3)}{(C3)}{0.5};
\draw (mid2) .. controls ($(mid2)+(-0.5,0.67)$) .. (mid3);
\draw (mid2) .. controls ($(mid2)+(0.5,0.67)$) .. (mid3);
\arrowpath{($(mid2)+(-0.37,0.7)$)}{($(mid2)+(-0.37,0.74)$)}{0.5};
\arrowpath{($(mid2)+(0.37,0.7)$)}{($(mid2)+(0.37,0.74)$)}{0.5};

\node[below] at (C2) {\scriptsize{$A$}};
\node[above] at (C3) {\scriptsize{$A$}};
\node[left] at ($(mid2)+(-0.37,0.7)$) {\scriptsize{$A$}};
\node[right] at ($(mid2)+(0.37,0.7)$) {\scriptsize{$A$}};
\node[right] at (mid2) {\scriptsize{$\co\mu$}};
\node[right] at (mid3) {\scriptsize{$\mu$}};

\coordinate (eq) at ($(mid2)+(1.75,0.67)$);
\node at (eq) {$=$};

\coordinate (C4) at ($(C2)+(3,0)$);
\coordinate (C5) at ($(C4)+(0,4)$);
\arrowpath{(C4)}{(C5)}{0.5};

\node[below] at (C4) {\scriptsize{$A$}};
\end{scope}
\end{tikzpicture}
\caption{Separability axiom for an algebra object $A\in\cC$.}\label{fig:sep}
\end{figure}

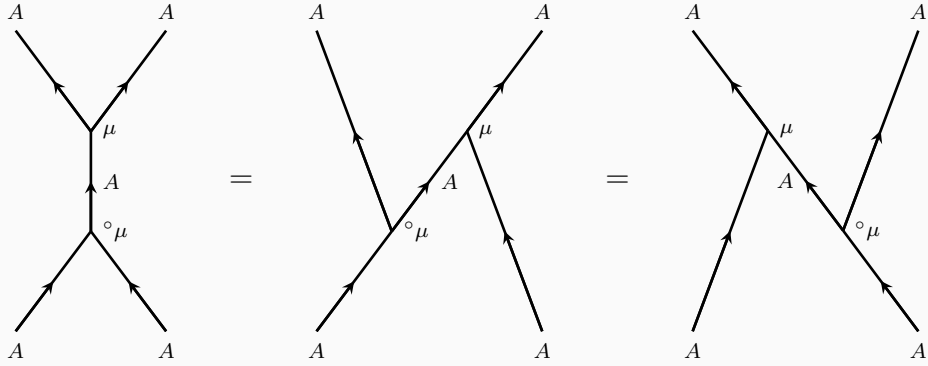
\begin{figure}
\centering
\begin{tikzpicture}[line width=1pt]
\begin{scope}[every node/.style={sloped,allow upside down}]
\coordinate (i0) at (0,0);
\coordinate (j0) at ($(i0)+(2,0)$);
\coordinate (k0) at ($(i0)+(1,1.33)$);
\coordinate (k1) at ($(k0)+(0,1.33)$);
\coordinate (i1) at ($(i0)+(0,4)$);
\coordinate (j1) at ($(j0)+(0,4)$);

\arrowpath{(i0)}{(k0)}{0.5};
\arrowpath{(j0)}{(k0)}{0.5};
\arrowpath{(k0)}{(k1)}{0.5};
\arrowpath{(k1)}{(i1)}{0.5};
\arrowpath{(k1)}{(j1)}{0.5};

\node[above] at (i1) {\scriptsize{$A$}};
\node[above] at (j1) {\scriptsize{$A$}};
\node[below] at (i0) {\scriptsize{$A$}};
\node[below] at (j0) {\scriptsize{$A$}};
\node[right] at (k0) {\scriptsize{$\co\mu$}};
\node[right] at (k1) {\scriptsize{$\mu$}};
\node[right] at ($(k0)!0.5!(k1)$) {\scriptsize{$A$}};

\coordinate (eq1) at ($(j0)+(1,2)$);
\node at (eq1) {$=$};

\coordinate (i2) at ($(j0)+(2,0)$);
\coordinate (j2) at ($(i2)+(3,0)$);
\coordinate (k2) at ($(i2)+(1,1.33)$);
\coordinate (k3) at ($(k2)+(1,1.33)$);
\coordinate (i3) at ($(i2)+(0,4)$);
\coordinate (j3) at ($(j2)+(0,4)$);

\arrowpath{(i2)}{(k2)}{0.5};
\arrowpath{(j2)}{(k3)}{0.5};
\arrowpath{(k2)}{(k3)}{0.5};
\arrowpath{(k2)}{(i3)}{0.5};
\arrowpath{(k3)}{(j3)}{0.5};

\node[above] at (i3) {\scriptsize{$A$}};
\node[above] at (j3) {\scriptsize{$A$}};
\node[below] at (i2) {\scriptsize{$A$}};
\node[below] at (j2) {\scriptsize{$A$}};
\node[right] at (k2) {\scriptsize{$\co\mu$}};
\node[right] at (k3) {\scriptsize{$\mu$}};
\node[right] at ($(k2)!0.5!(k3)$) {\scriptsize{$A$}};

\coordinate (eq2) at ($(j2)+(1,2)$);
\node at (eq2) {$=$};

\coordinate (i4) at ($(j2)+(2,0)$);
\coordinate (j4) at ($(i4)+(3,0)$);
\coordinate (k4) at ($(i4)+(1,2.67)$);
\coordinate (k5) at ($(k4)+(1,-1.33)$);
\coordinate (i5) at ($(i4)+(0,4)$);
\coordinate (j5) at ($(j4)+(0,4)$);

\arrowpath{(i4)}{(k4)}{0.5};
\arrowpath{(j4)}{(k5)}{0.5};
\arrowpath{(k5)}{(k4)}{0.5};
\arrowpath{(k4)}{(i5)}{0.5};
\arrowpath{(k5)}{(j5)}{0.5};

\node[above] at (i5) {\scriptsize{$A$}};
\node[above] at (j5) {\scriptsize{$A$}};
\node[below] at (i4) {\scriptsize{$A$}};
\node[below] at (j4) {\scriptsize{$A$}};
\node[right] at (k4) {\scriptsize{$\mu$}};
\node[right] at (k5) {\scriptsize{$\co\mu$}};
\node[left] at ($(k4)!0.5!(k5)$) {\scriptsize{$A$}};
\end{scope}
\end{tikzpicture}
\caption{Frobenius axiom for an algebra object $A\in\cC$.}\label{fig:fro}
\end{figure}

\begin{figure}
\centering
\begin{tikzpicture}[line width=1pt]
\begin{scope}[every node/.style={sloped,allow upside down}]
\coordinate (i4) at (0,0);
\coordinate (j4) at ($(i4)+(2,0.67)$);
\coordinate (k4) at ($(i4)+(1,2.67)$);
\coordinate (k5) at ($(k4)+(1,-1.33)$);
\coordinate (i5) at ($(i4)+(1,3.33)$);
\coordinate (j5) at ($(i4)+(3,4)$);

\arrowpath{(i4)}{(k4)}{0.5};
\arrowpath{(k5)}{(k4)}{0.5};
\arrowpath{(k4)}{(i5)}{0.5};
\arrowpath{(j5)}{(k5)}{0.5};

\draw[fill=black] (i5) circle(1.5pt);

\node[above] at (i5) {\scriptsize{$\co{u}$}};
\node[above] at (j5) {\scriptsize{$A^*$}};
\node[below] at (i4) {\scriptsize{$A$}};
\node[right] at (k4) {\scriptsize{$\mu$}};
\node[below] at (k5) {\scriptsize{$\coev_{A}^L$}};
\node[left] at ($(k4)!0.5!(k5)$) {\scriptsize{$A$}};

\coordinate (eq2) at ($(i4)+(4,2)$);
\node at (eq2) {$=$};

\coordinate (i2) at ($(i4)+(5,0)$);
\coordinate (j2) at ($(i2)+(3,0)$);
\coordinate (k2) at ($(i2)+(1,1.33)$);
\coordinate (k3) at ($(k2)+(1,1.33)$);
\coordinate (i3) at ($(i2)+(0,4)$);
\coordinate (j3) at ($(j2)+(-1,3.33)$);

\arrowpath{(j2)}{(k3)}{0.5};
\arrowpath{(k2)}{(k3)}{0.5};
\arrowpath{(i3)}{(k2)}{0.5};
\arrowpath{(k3)}{(j3)}{0.5};

\draw[fill=black] (j3) circle(1.5pt);

\node[above] at (i3) {\scriptsize{$A^*$}};
\node[above] at (j3) {\scriptsize{$\co{u}$}};
\node[below] at (j2) {\scriptsize{$A$}};
\node[below] at (k2) {\scriptsize{$\coev_A^R$}};
\node[right] at (k3) {\scriptsize{$\mu$}};
\node[right] at ($(k2)!0.5!(k3)$) {\scriptsize{$A$}};
\end{scope}
\end{tikzpicture}
\caption{Symmetricity axiom for an algebra object $A\in\cC$. Here $\coev_a$ denotes the co-evaluation map for line $a$.}\label{fig:sym}
\end{figure}

Moreover, in the group case we can explicitly check that $A$ satisfies additional properties which relate the algebra and co-algebra structure on $A$. These are shown in Figures \ref{fig:sep}, \ref{fig:fro} and \ref{fig:sym}. An object $A\in\cC$ satisfying  these properties is called as a \emph{symmetric Frobenius algebra object}  in $\cC$.

Now, given a two-dimensional theory $T$ with flavor symmetry $G$, the partition function of the gauged theory $T/G$ can be described using this object $A$.
Namely, $Z_{T/G}(M)$ on a manifold $M$ is defined by the partition function of $T$ on $M$ with a \emph{fine-enough} trivalent mesh of topological lines all labeled by $A$. By fine-enough, we mean a mesh which can be obtained as the dual graph of a triangulation of the 2d manifold $M$.
The relations above guarantee that the result does not depend on the choice of the mesh as long as it is fine enough. Dually, these relations guarantee invariance under changes of triangulations by Pachner moves.
Furthermore, by decomposing $A=\bigoplus_g g$, one can see that this trivalent mesh is a superposition of various $G$ bundles over $M$,
giving us back the standard definition of the gauged theory.

The crucial idea of \cite{Fuchs:2002cm,Carqueville:2012dk} is to take this as the definition of gauging in the generalized sense. 
Namely, given a theory $T$ with a symmetry category $\cC$, pick a symmetric Frobenius algebra object $A\in \cC$. 
Then, the gauged theory $T/A$ is defined exactly as in the previous paragraph.

Note that there can be multiple possible choices of $A$ for a given $\cC$.
For example, when $\cC=\cC(G)$, we can pick any subgroup $H\subset G$ and
take $A_H:=\bigoplus_{h\in H} h$.
When $\cC=\cC(G,\alpha)$, we can check that $A_H$ is an algebra only when the anomaly $\alpha\in H^3(G,\UU(1))$ when restricted to $H$ is trivial.
We can also twist the multiplication morphism $m: A\otimes A\to A$ by using \begin{equation}
\epsilon_{g,h}\in \Hom(g\otimes h,gh)
\end{equation} where $\epsilon_{g,h}\in H^2(G,\UU(1))$.
The choice of $A$ then can be thought of as the choice of a gauge-able subsymmetry together with the choice of the discrete torsion \cite{Brunner:2014lua}. 
We summarize:
\begin{claim}
Gauging a gauge-able subpart of $\cC$ is done by inserting a fine-enough mesh of an algebra object $A$ in $\cC$.
The choice of $A$ in $\cC$ generalizes the notion of choosing a non-anomalous subgroup and the discrete torsion in the case of group symmetry.
\end{claim}

We have some comments:
\begin{itemize}
\item There is in general no canonical maximal $A$ in $\cC$. 
For example, given $\cC(G,\alpha)$ for a nontrivial anomaly, there is not necessarily a unique maximal non-anomalous subgroup.
\item In the case of $\cC(G,\alpha)$, finding possible algebras $A$ is equivalent to finding a symmetry subcategory of the form $\cC(H,0)$. This is however not the general situation. Algebras $A$ specifying possible gaugings do not necessarily correspond to symmetry subcategories.
\item Note also that two distinct $A$ and $A'$ might give rise to the same gauged theory. We will come back to this question in Sec.~\ref{sec:distinctgaugings}.
\end{itemize}

\subsection{Symmetries of the gauged theory from bimodules for the algebra object} \label{sec:bimod}
What is the symmetry category of the gauged theory $T/A$? 
Since $T/A$ is just the original theory $T$ with a fine mesh of $A$, we can still consider  topological lines $p\in \cC$ of the original theory. 
But for $p$ to be topological in the presence of an arbitrary mesh of $A$, the lines of $A$ need to be able to end on $p$ both on the left and the right consistently.

\begin{figure}
\centering
\begin{tikzpicture}[line width=1pt]
\begin{scope}[every node/.style={sloped,allow upside down}]
\coordinate (l) at (3,0);
\coordinate (i) at (0,-4);
\coordinate (j) at (2,-4);
\coordinate (k) at (3,-4);
\coordinate (jk) at ($(l)!0.33!(k)$);
\coordinate (ij) at ($(i)!0.33!(jk)$);

\arrowpath{(ij)}{(jk)}{0.5};
\arrowpath{(jk)}{(l)}{0.5};
\arrowpath{(i)}{(ij)}{0.5};
\arrowpath{(k)}{(jk)}{0.5};
\arrowpath{(j)}{(ij)}{0.5};

\node[below] at (i) {\scriptsize{$A$}};
\node[below] at (j) {\scriptsize{$A$}};
\node[below] at (k) {\scriptsize{$p$}};
\node[above] at (l) {\scriptsize{$p$}};
\node[left] at (ij) {\scriptsize{$\mu$}};
\node[right] at (jk) {\scriptsize{$x_L$}};
\node[below right] at ($(ij)!0.5!(jk)$) {\scriptsize{$A$}};

\coordinate (to) at ($(l)+(2,-2)$);
\node at (to) {$=$};

\coordinate (L) at ($(l)+(6,0)$);
\coordinate (I) at ($(i)+(6,0)$);
\coordinate (J) at ($(j)+(6,0)$);
\coordinate (K) at ($(k)+(6,0)$);
\coordinate (JK) at ($(jk)+(6,0)$);
\coordinate (IJ) at ($(JK)!0.5!(K)$);

\arrowpath{(IJ)}{(JK)}{0.5};
\arrowpath{(JK)}{(L)}{0.5};
\arrowpath{(K)}{(IJ)}{0.5};
\arrowpath{(I)}{(JK)}{0.5};
\arrowpath{(J)}{(IJ)}{0.5};

\node[below] at (I) {\scriptsize{$A$}};
\node[below] at (J) {\scriptsize{$A$}};
\node[below] at (K) {\scriptsize{$p$}};
\node[above] at (L) {\scriptsize{$p$}};
\node[right] at (IJ) {\scriptsize{$x_L$}};
\node[right] at (JK) {\scriptsize{$x_L$}};
\node[below right] at ($(IJ)!0.5!(JK)$) {\scriptsize{$p$}};

\end{scope}
\end{tikzpicture}
\caption{Associativity axiom for a left $A$-module $(p,x_L)$.}\label{fig:modass}
\end{figure}
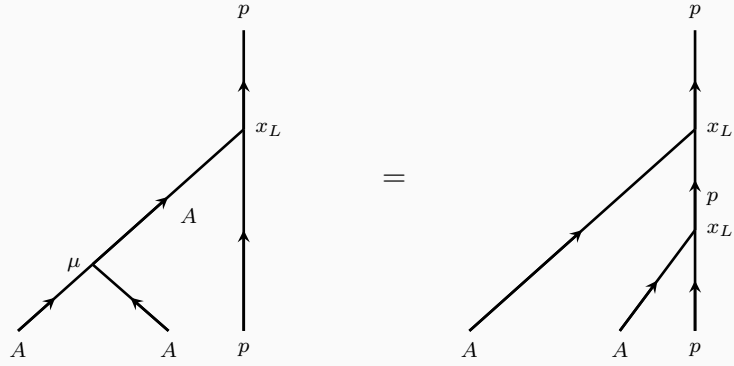

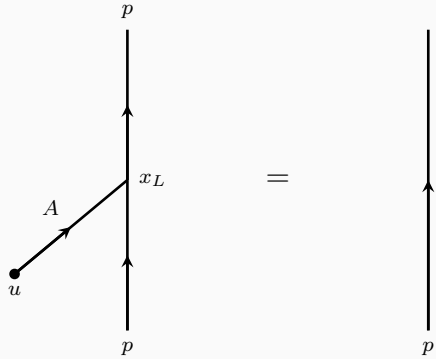
\begin{figure}
\centering
\begin{tikzpicture}[line width=1pt]
\begin{scope}[every node/.style={sloped,allow upside down}]
\coordinate (C1) at (0,0);
\coordinate (mid1) at ($(C1)-(0,2)$);
\coordinate (A1) at ($(mid1)+(-1.5,-1.25)$);
\coordinate (B1) at ($(mid1)+(0,-2)$);

\arrowpath{(mid1)}{(C1)}{0.5};
\arrowpath{(A1)}{(mid1)}{0.5};
\arrowpath{(B1)}{(mid1)}{0.5};

\draw[fill=black] (A1) circle(1.5pt);

\node[above] at (C1) {\scriptsize{$p$}};
\node[below] at (A1) {\scriptsize{$u$}};
\node[below] at (B1) {\scriptsize{$p$}};
\node[right] at (mid1) {\scriptsize{$x_L$}};
\node[above left] at ($(A1)!0.5!(mid1)$) {\scriptsize{$A$}};

\coordinate (to) at ($(mid1)+(2,0)$);
\node at (to) {$=$};

\coordinate (C2) at ($(C1)+(4,0)$);
\coordinate (B2) at ($(B1)+(4,0)$);

\arrowpath{(B2)}{(C2)}{0.5};

\node[below] at (B2) {\scriptsize{$p$}};

\end{scope}
\end{tikzpicture}
\caption{Unit axiom for a left $A$-module $(p,x_L)$.}\label{fig:modunit}
\end{figure}

\begin{figure}
\centering
\begin{tikzpicture}[line width=1pt]
\begin{scope}[every node/.style={sloped,allow upside down}]
\coordinate (L) at (2,0);
\coordinate (I) at (0,-4);
\coordinate (J) at (3,-4);
\coordinate (K) at (2,-4);
\coordinate (JK) at ($(L)!0.33!(K)$);
\coordinate (IJ) at ($(L)!0.67!(K)$);

\arrowpath{(IJ)}{(JK)}{0.5};
\arrowpath{(JK)}{(L)}{0.5};
\arrowpath{(K)}{(IJ)}{0.5};
\arrowpath{(I)}{(JK)}{0.5};
\arrowpath{(J)}{(IJ)}{0.5};

\node[below] at (I) {\scriptsize{$A$}};
\node[below] at (J) {\scriptsize{$A$}};
\node[below] at (K) {\scriptsize{$p$}};
\node[above] at (L) {\scriptsize{$p$}};
\node[right] at (IJ) {\scriptsize{$x_R$}};
\node[right] at (JK) {\scriptsize{$x_L$}};
\node[below right] at ($(IJ)!0.5!(JK)$) {\scriptsize{$p$}};

\coordinate (to) at ($(l)+(2,-2)$);
\node at (to) {$=$};

\coordinate (l) at ($(L)+(5,0)$);
\coordinate (i) at ($(I)+(6,0)$);
\coordinate (j) at ($(J)+(6,0)$);
\coordinate (k) at ($(K)+(5,0)$);
\coordinate (jk) at ($(JK)+(5,0)$);
\coordinate (ij) at ($(IJ)+(5,0)$);

\arrowpath{(ij)}{(jk)}{0.5};
\arrowpath{(jk)}{(l)}{0.5};
\arrowpath{(k)}{(ij)}{0.5};
\arrowpath{(i)}{(ij)}{0.5};
\arrowpath{(j)}{(jk)}{0.5};

\node[below] at (i) {\scriptsize{$A$}};
\node[below] at (j) {\scriptsize{$A$}};
\node[below] at (k) {\scriptsize{$p$}};
\node[above] at (l) {\scriptsize{$p$}};
\node[right] at (ij) {\scriptsize{$x_L$}};
\node[right] at (jk) {\scriptsize{$x_R$}};
\node[below right] at ($(ij)!0.5!(jk)$) {\scriptsize{$p$}};

\end{scope}
\end{tikzpicture}
\caption{Definition of $(A,A)$-bimodule $(p,x_L,x_R)$.}\label{fig:bimod}
\end{figure}
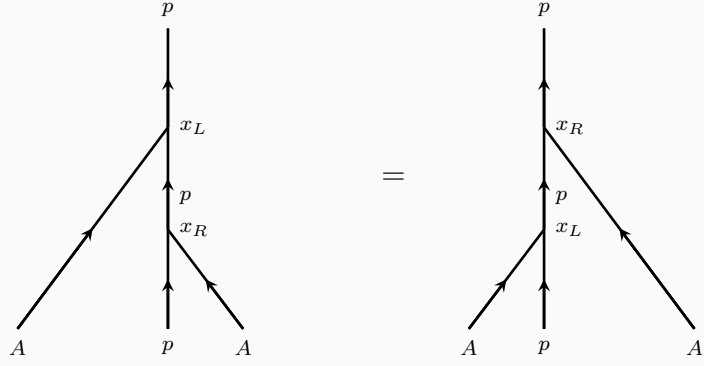

First of all, there must be a morphism $x_L:A\otimes p\to p$ such that conditions in Figures \ref{fig:modass} and \ref{fig:modunit} are satisfied. These properties make $p$ into a left $A$-module in $\cC$. Similarly, there must be a morphism $x_R:p\otimes A\to p$ which satisfies similar conditions and makes $p$ into a right $A$-module as well. Moreover, there must be morphisms $\co{x}_L:p\to A\otimes p$ and $\co{x}_R:p\to p\otimes A$ which satisfy co-conditions obtained by reflecting the graphs in Figures \ref{fig:modass} and \ref{fig:modunit} upside down and reversing all the arrows. This would make $p$ into a left and right $A$-comodule. It turns out that for a symmetry category, we can always get the co-module structure by combining the module structure with the co-evaluation map for $A$ or $A^*$. Hence, from now on we restrict our attention only to the module structure. On top of all these conditions, $p$ must also satisfy conditions which allow us to commute the left and right actions of $A$ on $p$. This is the condition shown in Figure \ref{fig:bimod} and it makes $p$ into an $(A,A)$-bimodule in $\cC$.

The topological operators between two lines $p$ and $q$, both of which are bimodules, need to be compatible with the action of $A$ from both sides. That is, the insertion of the bimodule changing operator and the action of $A$ must commute. They give rise to the category $\Bimod_\cC(A)$ of $(A,A)$ bimodules of $A$ within $\cC$. 
Note that the concept of the category of $(A,A)$ bimodules is different from the concept of bimodule categories over $\cC$ that we encountered above.

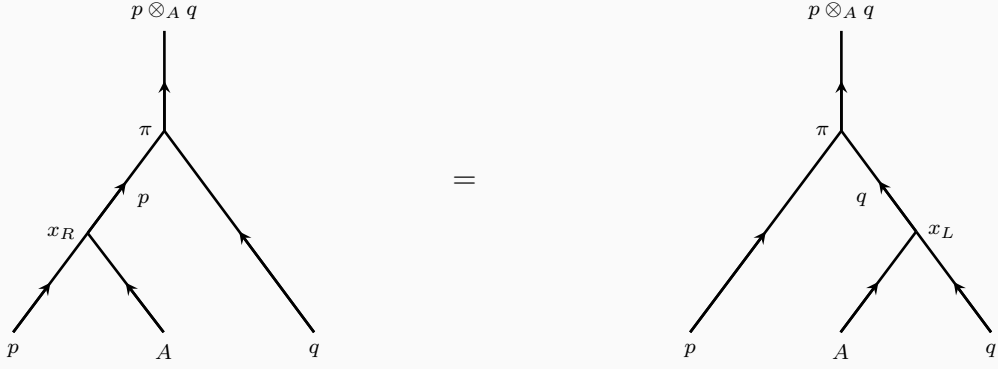
\begin{figure}
\centering
\begin{tikzpicture}[line width=1pt]
\begin{scope}[every node/.style={sloped,allow upside down}]
\coordinate (l) at (3,0);
\coordinate (i) at (0,-4);
\coordinate (j) at (2,-4);
\coordinate (k) at (4,-4);
\coordinate (ij) at ($(l)!0.67!(i)$);
\coordinate (jk) at ($(l)!0.33!(i)$);
\coordinate (l') at ($(jk)+(0,1.33)$);

\arrowpath{(ij)}{(jk)}{0.5};
\arrowpath{(jk)}{(l')}{0.5};
\arrowpath{(i)}{(ij)}{0.5};
\arrowpath{(k)}{(jk)}{0.5};
\arrowpath{(j)}{(ij)}{0.5};

\node[below] at (i) {\scriptsize{$p$}};
\node[below] at (j) {\scriptsize{$A$}};
\node[below] at (k) {\scriptsize{$q$}};
\node[above] at (l') {\scriptsize{$p\ot_A q$}};
\node[left] at (ij) {\scriptsize{$x_R$}};
\node[left] at (jk) {\scriptsize{$\proj $}};
\node[below right] at ($(ij)!0.5!(jk)$) {\scriptsize{$p$}};

\coordinate (to) at ($(l)+(3,-2)$);
\node at (to) {$=$};

\coordinate (L) at ($(l)+(9,0)$);
\coordinate (I) at ($(i)+(9,0)$);
\coordinate (J) at ($(j)+(9,0)$);
\coordinate (K) at ($(k)+(9,0)$);
\coordinate (JK) at ($(jk)+(9,0)$);
\coordinate (IJ) at ($(JK)!0.5!(K)$);
\coordinate (L') at ($(JK)+(0,1.33)$);

\arrowpath{(IJ)}{(JK)}{0.5};
\arrowpath{(JK)}{(L')}{0.5};
\arrowpath{(K)}{(IJ)}{0.5};
\arrowpath{(I)}{(JK)}{0.5};
\arrowpath{(J)}{(IJ)}{0.5};

\node[below] at (I) {\scriptsize{$p$}};
\node[below] at (J) {\scriptsize{$A$}};
\node[below] at (K) {\scriptsize{$q$}};
\node[above] at (L') {\scriptsize{$p\ot_A q$}};
\node[right] at (IJ) {\scriptsize{$x_L$}};
\node[left] at (JK) {\scriptsize{$\proj $}};
\node[below left] at ($(IJ)!0.5!(JK)$) {\scriptsize{$q$}};

\end{scope}
\end{tikzpicture}
\caption{The compatibility condition defining balanced tensor product of two $A$-bimodules $p$ and $q$ via $\proj $.}\label{fig:balanced}
\end{figure}

The tensor product in the category $\Bimod_\cC(A)$ is written as $\otimes_A$ and it ensures that any insertions of $A$ between $p$ and $q$ in $p\otimes_A q$ can be removed. The product $p\otimes_A q$ is given as a subobject of $p\otimes q$ defined  by the most general projection $\proj $ \begin{equation}
\proj: p\otimes q\to p\otimes_A q \label{eq:projection}
\end{equation}
 which satisfies the equation
\be
(p\otimes A)\otimes q\stackrel{x_R}{\to} p\otimes q\stackrel{\proj }{\to} p\otimes_A q=(p\otimes A)\otimes q\stackrel{\alpha}{\to} p\otimes (A\otimes q)\stackrel{x_L}{\to} p\otimes q\stackrel{\proj }{\to} p\ot_A q \label{bal}
\ee
where each side of the equation stands for the composition of the morphisms shown, which are the associators $\alpha$, the projection $\proj $ and the morphisms $x_L$ and $x_R$ defining the action of $A$ on $p$ and $q$. See Figure \ref{fig:balanced}. 
The equation tells us that the right action of $A$ on $p$ is \emph{balanced} against the left action of $A$ on $q$. 

The left action of $A$ on $p\ot_A q$ is defined by the compatibility condition
\be
A\ot(p\ot q)\stackrel{\proj}{\to} A\ot(p\ot_A q)\stackrel{x_L}{\to} p\ot_A q=A\ot(p\ot q)\stackrel{\alpha}{\to}(A\ot p)\ot q\stackrel{x_L}{\to} p\ot q\stackrel{\proj}{\to} p\ot_A q \label{comp}
\ee
where each side of the equation means the composition of the appropriate morphisms, namely the associators, the projection, and the left action of $A$ on $p$. The reader can draw a figure for \eqref{comp} in a similar fashion as to Figure \ref{fig:balanced} for \eqref{bal}. Similarly, we define the right action of $A$ on $p\ot_A q$. These actions manifestly commute and convert $p\ot_A q$ into an $A$-bimodule.

\begin{figure}
\centering
\begin{tikzpicture}[line width=1pt]
\begin{scope}[every node/.style={sloped,allow upside down}]
\coordinate (023) at (0,0);
\coordinate (012) at ($(023)+(-1.5,1.2)$);
\coordinate (123) at ($(023)+(1.5,2.4)$);
\coordinate (013) at ($(023)+(0,3.6)$);
\coordinate (controlu) at ($(013)+(3,3)$);
\coordinate (controld) at ($(023)+(3,-3)$);
\coordinate (mid) at ($(controlu)!0.5!(controld)-(0.75,0)$);
\coordinate (up) at ($(013)+(0,1.5)$);
\coordinate (down) at ($(023)-(0,1.5)$);

\arrowpath{(023)}{(012)}{0.5};
\arrowpath{(023)}{(123)}{0.5};
\arrowpath{(012)}{(123)}{0.5};
\arrowpath{(012)}{(013)}{0.5};
\arrowpath{(123)}{(013)}{0.5};
\arrowpath{(013)}{(up)}{0.5};
\arrowpath{(down)}{(023)}{0.5};

\node[left] at (012) {\scriptsize{$\coproj _{p,q}$}};
\node[right] at (023) {\scriptsize{$\coproj _{p\ot_A q,r}$}};
\node[left] at (013) {\scriptsize{$\proj _{p,q\ot_A r}$}};
\node[right] at (123) {\scriptsize{$\proj _{q,r}$}};

\node[below left] at ($(023)!0.5!(012)$) {\scriptsize{$p\ot_A q$}};
\node[right] at ($(023)!0.5!(123)$) {\scriptsize{$r$}};
\node[above] at ($(012)!0.5!(123)$) {\scriptsize{$q$}};
\node[left] at ($(012)!0.5!(013)$) {\scriptsize{$p$}};
\node[above right] at ($(123)!0.5!(013)$) {\scriptsize{$q\ot_A r$}};
\node[above] at (up) {\scriptsize{$p\ot_A(q\ot_A r)$}};
\node[below] at (down) {\scriptsize{$(p\ot_A q)\ot_A r$}};
\end{scope}
\end{tikzpicture}
\caption{The associator in $\Bimod_\cC(A)$.} \label{fig:bimodass}
\end{figure}

The dual of $\proj $ is $\coproj : p\otimes_A q\to p\otimes q$ and we have $\proj \circ \coproj =1$.  The associator in $\Bimod_\cC(A)$ is then defined as 
\be
\tilde \alpha_{p,q,r}=\proj _{p,q\otimes_A r}\circ(1\otimes \proj _{q,r})\circ \alpha_{p,q,r}\circ(\coproj _{p,q}\otimes 1)\circ(\coproj _{p\otimes_A q, r}),
\ee
see Figure \ref{fig:bimodass}. This ensures that any insertions of $A$ can be removed in the diagram defining the associator. 

Similarly, the evaluation and co-evaluation maps in $\Bimod_\cC(A)$ are defined as $\tilde \ev^{L,R}_p=u\circ \ev^{L,R}_p\circ \coproj $ and $\widetilde \coev^{L,R}_p=\proj \circ \coev^{L,R}_p\circ \co{u}$ where $u$ and $\co{u}$ are unit and co-unit morphisms defining $A$.
The argument above suggests that we should take $\Bimod_\cC(A)$ to be the symmetry category of the gauged theory $T/A$.

Now, consider two half-spaces, and put the mesh only on the right half. 
We now have a domain wall between the original theory $T$ on the left and the gauged theory $T/A$ on the right.
On this domain wall, the lines of the theory $T$, in particular those from $\cC$, should be able to end on the left
and the lines of the theory $T/A$, namely those from $\Bimod_\cC(A)$, should be able to end on the right. 
This suggests that the domain wall is described by the category $\Mod_{\cC}(A)$, the category of right $A$-modules.
Note that $\Mod_{\cC}(A)$ has a natural left action of $\cC$ and a natural right action of $\Bimod_\cC(A)$, making it naturally a $(\cC,\Bimod_\cC(A))$ bimodule category.
Almost by definition, $\Bimod_\cC(A)$ is the dual of $\cC$ with respect to the module category $\Mod_\cC(A)$.

\subsection{Gauging of $\cC(G)$ to get $\Rep(G)$ and vice versa} \label{sec:example}
As an exercise,  let us describe the process explicitly when $\cC=\cC(G)$ and $\Rep(G)$.

\paragraph{From $\cC(G)$ to $\Rep(G)$:}
We have already defined an algebra $A$ in $\cC(G)$ in Sec. \ref{sec:algebra}.
Let us  determine $\Mod_{\cC(G)}(A)$, the category of right modules of $A$.

Consider an object $M=\bigoplus N_g g$ where $N_g$ are non-negative integers. 
We  constrain $M$ by demanding that it forms a right-module for $A$.
First of all, we need a morphism $x:M\otimes A\to M$, which is  characterized by  morphisms $x_{g,g'}:N_g g\otimes {g'}\to N_{gg'}{gg'}$, each of which can be though of as an $N_{gg'}\times N_{g}$ matrix.
These matrices have to satisfy two equations. The first one involves the product axiom of $A$ and it tells us that
\begin{equation}
x_{gg',g''}x_{g,g'}=x_{g,g'g''}.
\end{equation}
The second one involves the unit axiom of $A$ and it tells us that
\begin{equation}
x_{g,e}=1.
\end{equation}
Combining these two, we find that
\begin{equation}
x_{g,g'}x_{gg',g'^{-1}}=1
\end{equation}
which implies that $N_g=N$ for all $g$ and
\begin{equation}
x_{g,g'}=x_{e,gg'}(x_{e,g})^{-1}. \label{e,g}
\end{equation}
So, we find that the right modules are $M=N\bigoplus_g g=NA$ with arbitrary invertible $N\times N$ matrices $x_{e,g}$.

Now, we find the morphisms in $\Mod_{\cC(G)}(A)$. The first observation is that all $(NA,x_{e,g})$ are isomorphic to $(NA,1)$. A morphism from right to left is provided by sending $Ng$ inside $(NA,1)$ to $Ng$ inside $(NA,x_{e,g})$ by the matrix $x_{e,g}$. The condition that this morphism commutes with the action of $A$ turns out to be \eqref{e,g} and hence this morphism is indeed a module morphism. The morphism from left to right via the inverse matrix is the inverse of this module morphism. So, we can restrict our attention to objects $NA\equiv(NA,1)$. 

Let us then find module morphisms from $NA$ to $N'A$. Such a morphism $\varphi$ is specified by an $N\times N'$ matrix $\varphi_g$ sending $Ng$ inside $NA$ to $N'g$ inside $N'A$. The condition that $\varphi$ be a module morphism implies that $\varphi$ is a constant $N\times n'$ matrix  independent of $g$. Thus, we identify $\Mod_{\cC(G)}(A)$ as the category $\Vec$.

Next, let us determine $\Bimod_{\cC(G)}(A)$. 
By a similar analysis as above, we can write such a bimodule as $(NA,(x_L)_{g,e},(x_R)_{e,g})$ where $(x_R)_{e,g}$  and $(x_L)_{g,e}$  encode the morphisms $M\otimes A\to M$ and $A\otimes M\to M$. As before we can restrict our attention to $(NA,(x_L)_{g,e},1)$. Demanding that the left and right actions of $A$ are compatible, we find that
\begin{equation}
(x_L)_{g,e}(x_L)_{g',e}=(x_L)_{gg',e}
\end{equation}
which means that the objects of $\Bimod_{\cC(G)}(A)$ are identified as representations of $G$. One can check that the morphisms of $\Bimod_{\cC(G)}(A)$ are precisely the intertwiners between the representations. Thus $\Bimod_{\cC(G)}(A)$ is equivalent to $\Rep(G)$.
We can also work out $\otimes_A$ and it agrees with the tensor product on $\Rep(G)$.

\paragraph{From $\Rep(G)$ to $\cC(G)$:}
Now the algebra object $A\in \Rep(G)$ is the regular representation. The regular representation is spanned by basis vectors $\wh g$ in one-to-one correspondence with the group elements $g$. If we denote the action of $g$ by $\rho(g)$ then $\rho(g)\wh h=\wh{gh}$. The algebra multiplication takes $\wh g\otimes \wh h$ to $\delta_{gh}\wh g$ which is an intertwiner. The unit morphism $1\to A$ corresponds to choosing $\sum_g\wh g$ in $A$. 

First, we look for right modules of $A$. Choose some representation $R$ of $G$. We denote the action of $g$ on $\vec q\in R$ as $g\vec q$. Note that $g\vec{q}\neq 0$ for any $g$ and any non-zero $\vec q$ because otherwise this would imply that $\vec q=g^{|G|}\vec q =0$.
The right-action of $A$ on $R$ must satisfy $g(\vec{p}\:\wh h)=(g\vec{p}\,)\wh h$. We must also have that $(\vec p\:\wh h)\:\wh k=\delta_{hk}\vec p \:\wh h$ and $\vec p(\sum_g\wh g)=\vec p$. 
Now, start with some arbitrary vector $\vec q\ne 0$ such that $\vec q\:\wh g\ne 0$ for some $g$. Then, it must be true that $\vec p:=(g^{-1}\vec q)\:\wh 1\ne 0$. Then, $\vec p\:\wh 1=\vec p$ and $\vec p\:\wh g=0$ for all $g\ne 1$. This implies that for every $g$ we obtain a vector $\vec p_g=g\vec p$ which has the following properties: $\vec p_g\:\wh h=\delta_{gh}\vec p_g$. Also, all $\vec p_g$ must be linearly independent because $\sum\alpha_g\vec p_g=0$ can be hit by $\wh h$ from the right to yield $\alpha_h=0$. Thus the set formed by $\vec p_g$ for all $g$ is the regular representation $A$ of $G$ and $R$ breaks up as a sum of regular representations. Thus, the objects of $\Mod_{\Rep(G)}(A)$ are all isomorphic to $NA$ for some non-negative integer $N$. One can easily work out the module morphisms as well and then one finds that $\Mod_{\Rep(G)}(A)$ is equivalent to $\Vec$.

Next, we look for bi-modules of $A$. Pick $A$ as a particular right module for $A$. It can be converted into a bimodule $V_g$ by simply stating that $\wh h\:\vec p_1=\delta_{gh}\vec p_1$. The $G$-equivariance then determines that $\wh h\:\vec p_k=\delta_{kg,h}\vec p_k$. Thus objects of $\Bimod_{\Rep(G)}(A)$ are labeled by non-negative integers $n_g$ for each $g$. One can similarly figure out bimodule morphisms leading to the result that $\Bimod_{\Rep(G)}(A)$ is equivalent to $\cC(G)$.
Again, the tensor product $\otimes_A$ can be worked out and it equals the group multiplication.

\subsection{Gaugings and module categories}\label{sec:distinctgaugings}
Let us come back to the general question.
We said that given a theory $T$ with a symmetry $\cC$ and a symmetric Frobenius algebra object $A\in \cC$, we can put a fine mesh of $A$ on the two-dimensional manifold to define the gauged theory $T/A$.
We now ask the question: when do two algebra objects $A$ and $A'$ give rise to the same gauged theory $T/A\simeq T/A'$? 
For this, the  symmetries of the gauged theories should be the same, $\Bimod_\cC(A)\simeq \Bimod_\cC(A')$ and the properties of the domain walls should also be the same, $\Mod_\cC(A)\simeq \Mod_\cC(A')$. 

Let us see that conversely, if $\Mod_\cC(A)\simeq \Mod_\cC(A')$, the gauged theories are the same $T/A\simeq T/A'$. Indeed, the equivalence is captured by an $(A,A')$ bimodule $m$ in $\cC$ which sends an object $M'$ of $\Mod_\cC(A')$ to an object $M$ of $\Mod_\cC(A)$ via $M=m\otimes_{A'}M'$ where $\otimes_{A'}$ denotes a tensor product in the category $\Bimod_\cC(A')$. There is an inverse $(A',A)$ bimodule $n$ providing the inverse map. Physically, $m$ corresponds to a topological interface between $T/A$ and $T/A'$ which can be fused with boundary conditions of $T/A'$ to yield boundary conditions of $T/A$. $n$ is the inverse interface. Algebras $A$ and $A'$ such that $\Mod_\cC(A)\simeq \Mod_\cC(A')$ are called \emph{Morita equivalent}.

We claim that the existence of such an invertible interface guarantees that $T/A$ and $T/A'$ are isomorphic theories. For instance, consider the Hilbert space $V$  of $T/A$ on $S^1$. 
It can be mapped to the Hilbert space $V'$ of $T/A'$ on $S^1$ by considering a cylinder geometry of infinitesimal time with the insertion of a wrapped domain wall $n$ in between. Similarly, we have an inverse map from $V'$ to $V$ constructed similarly from $m$. Now, we can compose these maps by putting one cylinder on top of the other. As the interfaces are topological, they can be moved towards each other and ultimately fused away as they are inverses of each other. Thus, we are left with a unitary evolution on an infinitesimal length cylinder geometry. We can now take the infinitesimal to zero in all the steps above to obtain an isomorphism between $V$ and $V'$. 
Similar arguments can be used to show that any kind of data of $T/A$ is isomorphic to the same kind of data of $T/A'$. 

We would like to stress that this equivalence of $T/A$ and $T/A'$ is different from our claim that $T$ and $T/A$ contain the same amount of information. 
For example, in the latter case, the Hilbert spaces would be different, as we will detail in Sec.~\ref{sec:Hil}.

All of this means that what really characterizes different gaugings are not the algebra objects used in the gaugings themselves but the associated module categories $\Mod_\cC(A)$ over $\cC$.
We can then ask the question: does every module category $\cM$ over $\cC$ come from a symmetric Frobenius algebra object $A$ in $\cC$, so that $\cM$ arises as the domain wall between the original theory $T$ and the gauged theory $T/A$?
The answer is yes \cite{EGNO,Schaumann}. 

The construction goes roughy as follows. 
Given a module category $\cM$ over $\cC$, consider the morphism spaces $\Hom(c\otimes m,n)$ for $c\in \cC$ and $m,n\in \cM$. 
There is an object called the internal hom and denoted  by $\uHom(m,n)$ in $\cC$ which satisfies \begin{equation}
\Hom_\cM(c\otimes m,n) \simeq \Hom_\cC(c,\uHom(m,n)).
\end{equation} At the level of objects, this is given by \begin{equation}
\uHom(m,n) = \bigoplus_c (\dim \Hom_\cM(c\otimes m,n) ) c
\end{equation} where $c$ runs over the isomorphism classes of simple objects in $\cC$ as can be simply verified by computing $\Hom_\cC(c,\uHom(m,n))$ from this equation.

Now, the internal homs can be concatenated naturally, in the sense that there is a natural morphism \begin{equation}
\mu: \uHom(m,n)\otimes \uHom(n,o) \to \uHom(m,o)
\end{equation}
and in particular $\uHom(m,m)$ has the multiplication. 
It can be shown \cite{EGNO} that $A=\uHom(m,m)$ for a simple object $m\in \cM$ is an algebra object such that $\Mod_A(C)\simeq \cM$. 
Furthermore, $A$ is automatically symmetric Frobenius if $\cC$ is a symmetry category \cite{Schaumann}.
Summarizing, we have 
\begin{claim}
Distinct choices of gaugings of a theory with symmetry category $\cC$ are in one to one correspondence with choices of  indecomposable module categories $\cM$ over $\cC$ characterizing the domain wall between the original theory and the gauged theory. The gauged theory has the symmetry $\cC'$ which is the dual of $\cC$ with respect to $\cM$.
\end{claim}
We denote the gauged theory corresponding to a module category $\cM$ by $T/\cM$. We would like to note that a decomposable module category corresponds to a direct sum of algebras $A=A_1\oplus A_2$ where the algebra structure of $A$ comes from algebra structure of $A_1$ and the algebra structure of $A_2$. Gauging a theory $T$ using $A$ produces $T/A=T/A_1\oplus T/A_2$ because a mesh of $A$ is the same as a mesh of $A_1$ plus a mesh of $A_2$.

\subsection{(Re-)gauging and its effect on the symmetry category}
\label{sec:moreregauging}
Now we can give a unified description of the (re-)gauging process. 
Consider a theory $T$ with a symmetry $\cC$.
Pick an indecomposable module category $\cM$ of $\cC$.
There is an algebra object $A\in \cC$ such that $\cM\simeq \Mod_\cC(A)$,
and then the gauged  theory $T/\cM:=T/A$ has the symmetry $\cC'=\Bimod_\cC(A)$ such that 
$\cM$ is a natural $(\cC,\cC')$ bimodule category.

Now, pick an indecomposable module category $\cM'$ of $\cC'$ and repeat the same process. We get the gauged theory $T/\cM/\cM'$ which has the symmetry $\cC''$  such that $\cM'$ is a $(\cC',\cC'')$ bimodule category. 
We now see that the `set' of all symmetry categories can be subdivided into `subsets' consisting of symmetry categories that can be converted into each other by a sequence of gauging.

Note that this double gauging can be done in one step, since from two bimodule categories $\cM$ and $\cM'$ we can form the tensor product $\cM\boxtimes_{\cC'}\cM'$ which is a $(\cC,\cC'')$ bimodule category. 
Notationally it is convenient to write a $(\cC,\cC')$ bimodule category $\cM$ as a morphism $\cM:\cC\to \cC'$.
Then we can regard $\cM\boxtimes_{\cC'}\cM'$ as $\cM'\circ \cM$.
Then $T/\cM/\cM'\simeq T/ (\cM'\circ\cM)$.
Any multiple gauging can be done in one step.

In particular, $\cM^\text{op}$  given by reversing the order of the tensor product is naturally a $(\cC',\cC)$ bimodule category, and $\cM^\text{op}\circ \cM$ is the identity. Therefore we have $T/\cM/\cM^\text{op}=T$.
The example we started this paper with, $T/G/\hat G=T$ for an Abelian group $G$, is a special instance of this construction.

It is instructive to phrase the re-gauging process in terms of the algebra $A$ as well, whose detail can be found in \cite{Carqueville:2012dk} with plenty of helpful drawings.
The category of bimodules $\Bimod_\cC(A)$ have an algebra object $B=A^*\otimes A$. $B$ is also trivially an algebra in $\cC$ and its algebra multiplication is given by the evaluation map. Moreover, we can embed $A$ into $B$ as $A\to A^*\otimes A\otimes A\to A^*\otimes A=B$ where the first map is the co-evaluation map and the second map is the multiplication for $A$. We are looking for $B$-bimodules in $\Bimod_\cC(A)$ which can be identified with $B$-bimodules in $\cC$ because $A$ embeds inside $B$ and hence a $B$-bimodule structure carries an $A$-bimodule structure. Now, $B$-bimodules in $\cC$ are trivially obtained from any object $a\in\cC$ by the mapping $A^*\otimes a\otimes A$. This mapping gives us an equivalence of $\cC$ and the category of $B$-bimodules in $\Bimod_\cC(A)$. 

Physically, this means that the lines of $T/\cM/\cM^\text{op}$ come as lines of $T$ dressed by $A^*$ on one side and by $A$ on the other. The local operators in $T/\cM/\cM^\text{op}$ also correspond to a local operator of $T$ which appears at the junction of ``middle" lines and the $A$-lines on the side are just joined smoothly. Let's insert a network of lines of $T/\cM/\cM^\text{op}$ on a surface $\Sigma$ and compute the partition function. The network will look like a network of lines of $T$ with additional $A$ loops, one for each ``plaquette". These $A$ loops can be shrunk away leaving the partition function of a network of lines of $T$. Thus we see explicitly how we obtain the same theory on regauging. Later on, we will discuss 2d TFTs with $\cC$ symmetry and we will give a prescription for constructing the gauged TFT which uses heavily the algebra $A$ rather than $\cM$. From this argument, it is clear that regauging the gauged TFT will give back the original TFT.

Summarizing, we have the following:
\begin{claim}
Denote by $\cM:\cC\to \cC'$ when the symmetry category $\cC$ has a gauging $\cM$ under which the dual symmetry is $\cC'$. 
Two consecutive gaugings $\cM:\cC\to \cC'$ and $\cM':\cC'\to \cC''$ can be composed to $\cM'\circ \cM:\cC \to \cC''$, and any gauging $\cM:\cC\to \cC'$ has an inverse $\cM^{-1}:\cC'\to \cC$
such that $\cM^{-1}\circ \cM$ is the identity.
\end{claim}

Now, let us ask when are two symmetry categories $\cC$ and $\cC'$ dual with respect to some indecomposable module category $\cM$. 
It is known \cite{ENO} that this happens if and only if  $Z(\cC)\simeq Z(\cC')$, where $Z(\cC)$ is the Drinfeld center of $\cC$. 
This has the following physical interpretation. 

Recall first that any symmetry category where every simple object is invertible is of the form $\cC=\cC(G,\alpha)$.
From the same data of $G$ and $\alpha\in H^3(G,\UU(1))$,
we can construct a 3d TFT called the Dijkgraaf-Witten theory \cite{Dijkgraaf:1989pz}.
On the boundary of the Dijkgraaf-Witten theory, we can put a 2d theory with symmetry $G$ with an anomaly $\alpha$.

There is a generalization of this construction that gives a 3d TFT starting from any symmetry category $\cC$, sometimes called the generalized Turaev-Viro construction.
We will just call them the Dijkgraaf-Witten theory associated to $\cC$.
The Drinfeld center $Z(\cC)$ is the  braided tensor category which captures the  properties of the line operators of this 3d TFT,
and this 3d TFT can have a boundary where a 2d theory with symmetry $\cC$ lives.
For a recent exposition, see e.g.~\cite{Bhardwaj:2016clt}. 

Since $T/A$ is obtained by just adding a mesh of $A$ on $T$, $T/A$ and $T$ can be put on boundaries of the same 3d TFT. 
Therefore their Drinfeld centers should be the same.
Conversely, specifying an isomorphism $Z(\cC)\simeq Z(\cC')$ corresponds to specifying a 1d domain wall on a 2d boundary of a single 3d TFT, such that we have lines from $\cC$ on the left and lines from $\cC'$ on the right of the domain wall. 
This corresponds to specifying a $(\cC,\cC')$ bimodule implementing the gauging. 
Summarizing,
\begin{claim}
Two symmetry categories $\cC$ and $\cC'$ can be transformed to each other by an appropriate gauging if and only if the 3d TFTs associated to them are equivalent, i.e.~$Z(\cC)\simeq Z(\cC')$.
\end{claim}

The fact $Z(\cC)\simeq Z(\cC')$ also implies that the total dimension of $\cC$ and $\cC'$ are the same, \begin{equation}
\dim \cC=\dim \cC'.
\end{equation} This is because it is known that $\dim Z(\cC)=|\dim \cC|^2$.

Another related point is the following. This mathematical process of choosing a module category or an algebra object corresponding to it for a symmetry category and taking the dual category can be performed for a modular tensor category. 
A modular tensor category describes topological line operators in a three-dimensional theory and therefore this operation should be thought of as creating a new three-dimensional theory from an old one.
This operation is in fact known under the name of \emph{anyon condensation} \cite{Kong:2013aya}.

\subsection{The effect of the gauging on Hilbert space on $S^1$}
Let us now discuss the Hilbert space on $S^1$ of a theory with symmetry $\cC$,
and how it is modified by the gauging.

\subsubsection{Backgrounds for $\cC$ symmetry} \label{sec:background}

\begin{figure}
\centering
\[
\incb{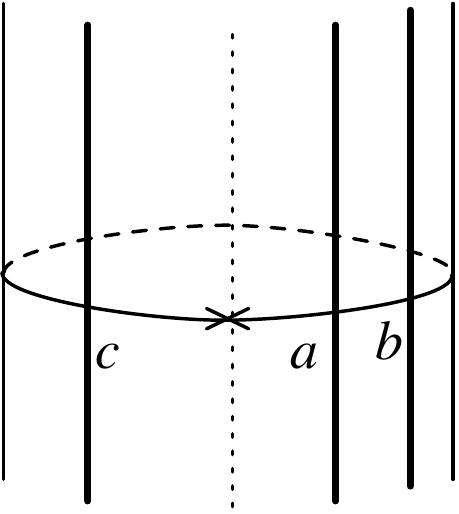}
\]
\caption{An example of a circle with  transverse line operators. The point marked with an X labels the choice of the base point.  In the diagrams here and below, the time flows upwards, and the lines also carry arrows in the upward direction unless otherwise mentioned.
}
\label{fig:aux}
\end{figure}

Consider a cylinder of the form $S^1\times \bR_t$, with  lines $a_i$ in $\cC$ transverse to the constant time slice $t=0$. 
Let us assume that other data about the circle like its size, spin-structure etc. have been fixed so that we can associate a Hilbert space of states to this circle.
See Figure \ref{fig:aux} for an example.
In the figures here and below, the time flows from the bottom to the top.

We would like to identify this configuration of lines with an object $a$ of $\cC$ and say that the circle carries the \emph{background} $a$ for the symmetry $\cC$. When $\cC=\cC(G)$, the backgrounds are labeled by group elements and correspond to holonomy of the $G$-connection around the circle. As the lines are topological, we can try to define such an $a$ by fusing them together. Clearly, there are choices in the precise way one performs this procedure: there is  a choice of the \emph{base-point} from which we start taking the tensor product, and there is also a choice in the relative order of fusion of the lines, which is related to the associators in $\cC$. 
To spell this out more fully, we need some notations and operations.

\begin{figure}
\[
\incb{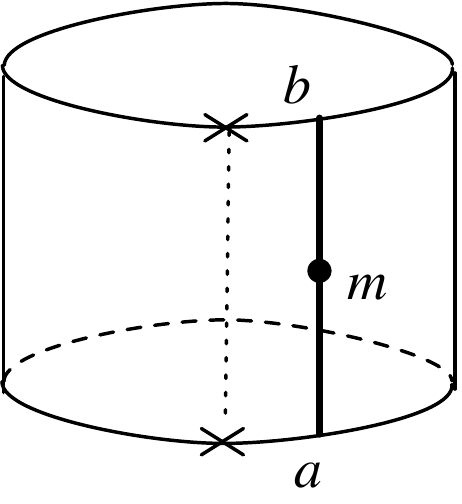}
\]
\caption{A morphism $m:a\to b$ defines an operator $Z(m):V_a\to V_b$. \label{fig:Z(m)}}
\end{figure}

Let us denote the Hilbert space for a cylinder with just one line operator $a\in \cC$ by $V_a$. 
We require $V_{a\oplus b}=V_a\oplus V_b$.
A morphism $m:a\to b$ defines an operator $Z(m):V_a\to V_b$. 
This is given by the geometry shown in Fig.~\ref{fig:Z(m)}.
Note that in a non-topological theory, the vertical height of the cylinder containing the topological line and/or the topological operator needs to be taken to zero.
The same comment applies to all the figures discussed below in this section.
Now, two choices of fusing $a$, $b$ and $c$ in Fig.~\ref{fig:aux} can be taken care of by using the associator $\alpha$ and $Z$: \begin{equation}
Z(\alpha_{a,b,c}) : V_{(a\otimes b)\otimes c} \to V_{a\otimes(b\otimes c)}.
\end{equation}

\begin{figure}
\[
\begin{array}{c@{\qquad}c}
X_{a,b} & X_{a,b}^{-1}\\
\incb{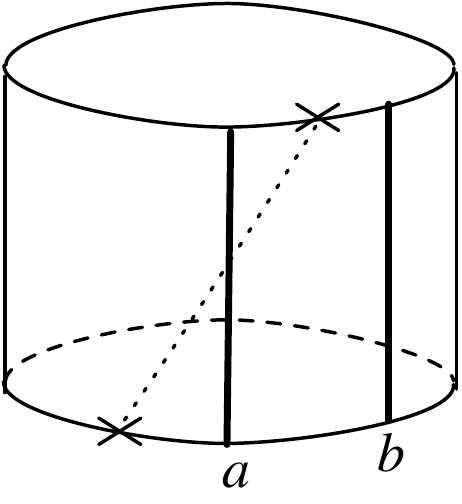} & \incb{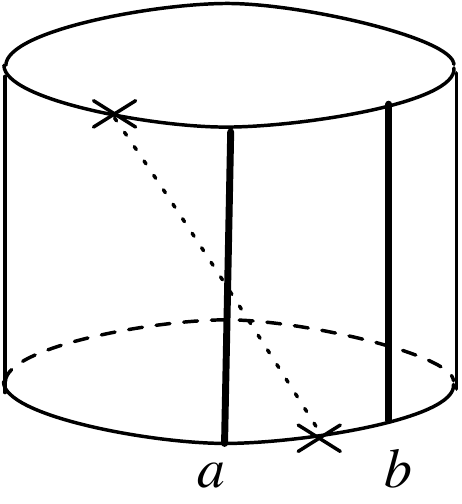} 
\end{array}
\]
\caption{$X_{a,b}$ and its inverse tracks the change of the base points. \label{fig:X}}
\end{figure}

To track the change of the base point, we need to introduce the map $X_{a,b}: V_{a\otimes b}\to V_{b\otimes a}$ defined by Fig.~\ref{fig:X},
together with its inverse.
They are associated to the intersection of the topological line and the \emph{auxiliary line}, which is the trajectory of the base point.

The operations $X_{a,b}$ and $Z(\alpha)$ satisfy many consistency relations,
which we spell out in full in Sec.~\ref{sec:TFTcyl}.
With them, we can keep track of the change of the base point and the change of the order of fusing the lines in a consistent manner.
This allows us to associate a Hilbert space to a circle with multiple transverse line operators $a_1$, \ldots, $a_k$.

\begin{figure}
\[
\incb{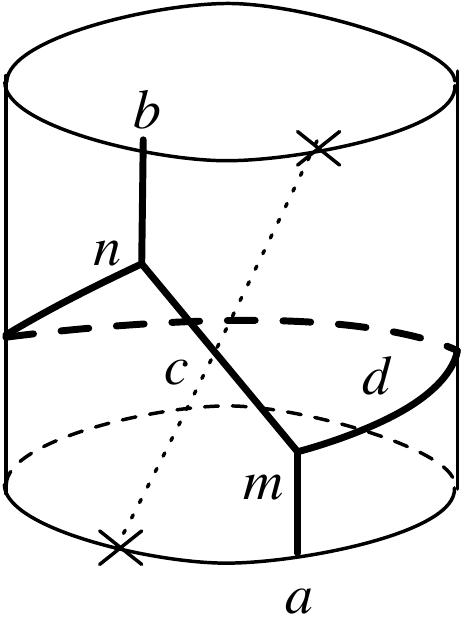}
\]
\caption{The action $U_{c,d}(m,n)$ of a line operator wrapped around $S^1$.\label{fig:U}}
\end{figure}

The operators $X$ and $Z$ can also be used to express how a line operator wrapped around $S^1$ acts on the Hilbert space.
Indeed, given two morphisms $m: a\to c\otimes d$ and $n:d\otimes c \to b$, we can define the operator $U_{c,d}(m,n)$ using the network drawn in Fig.~\ref{fig:U}.
Explicitly, this is given by \begin{equation}
U_{c,d}(m,n)=Z(n) X_{c,d} Z(m).
\end{equation}
The action $U_a$ of a line operator labeled by $a$ wrapped around on $S^1$ on the Hilbert space $V_1$ on $S^1$ without any transverse line, referred to in Sec.~\ref{sec:basicdefs}, is a special example of this construction: $U_a=U_{a^*,a}(\coev_a,\ev_a)$ where $\coev_a$, $\ev_a$ are the co-evaluation map and the evaluation map introduced in \eqref{eq:eval}, \eqref{eq:coeval}.

\subsubsection{The Hilbert space of the gauged theory} \label{sec:Hil}

So far we discussed the Hilbert spaces $V_a$ of the original theory $T$ with the symmetry $\cC$, where $a\in \cC$ is the label of the line operator transverse to the constant time slice $S^1$.
Let us now discuss how Hilbert spaces $W_p$ of the gauged theory $T/A$ can be found, where $p\in \cC'$.

A line in the gauged theory is given by an $(A,A)$ bimodule in $\cC$. 
Each such bimodule $p$ can be viewed as an object in $\cC$ and hence has a Hilbert space $V_p$ associated to it by the ungauged theory $T$. 
We construct  the Hilbert space $W_p$ associated to $p$ by the gauged theory $T/A$ as a subspace of $V_p$ in the following manner.

\begin{figure}
\[
\incb{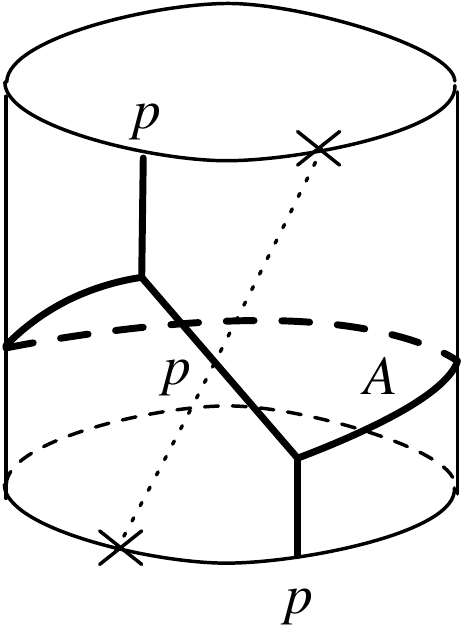}
\]
\caption{The projector defining the Hilbert space of the gauged theory is a specific instance of an action of a wrapped line operator. The morphisms used at the trivalent vertices are those specifying the bimodule structure of $p$.\label{fig:proj}}
\end{figure}

In the case of $\cC(G)$, one is traditionally instructed to project this space to the subspace left invariant by the action of the group $G$. 
For a general symmetry category $\cC$ and its gauging $A$, we need a projector $P:V_p \to W_p$. 
Such a projector is naturally given by $P:=U_{p,A}(\co{x}_R,x_L)$ where $\co{x}_R:p\to p\otimes A$ and $x_L:A\otimes p\to p$ are the morphisms defining the module and comodule structures on $p$, see Fig.~\ref{fig:proj}. This projector was written down in \cite{Brunner:2013ota}.

Let us now show that this projector to $W_p$ agrees with the traditional definition in the case of $\cC(G)$, when $p$ is the identity object in the category of bimodules.
This means that $p=A=\bigoplus_g g$ as an object in $\cC(G)$.
Therefore we are going to project $V_A=\bigoplus_g V_g$.
The operator $P$ restricted to $V_0$ is given by $|G|^{-1}\sum_g g$, and this is indeed a projector to the $G$-invariant subspace. 
The action of $P$ on the whole $V_A$ can be found similarly, and we find that \begin{equation}
W_1 = PV_A \simeq \bigoplus_{C} (\text{$G_c$-invariant subspace of $V_c$})
\end{equation} where $C$ runs over conjugacy classes of $G$, $c\in C$ is a representative element, and $G_c$ is the commutant of $c$ in $G$.
This is as it should be.

Summarizing, we have the following statement:
\begin{claim}
The Hilbert space of the gauged theory $T/A$ is obtained by taking the invariant part of the ungauged theory $T$ under the projector naturally defined by wrapping $A$ around $S^1$.
\end{claim}

\section{More examples of symmetry categories and their gauging}
\label{sec:examples}
In the previous sections we reviewed the general theories of symmetry categories and their gauging.
So far, however, we only saw basic examples where symmetry categories are either of the form $\cC(G,\alpha)$ where $G$ is a finite group and $\alpha$ its anomaly, or of the form $\Rep(G)$ where $G$ is a group.
In this section we discuss many other examples and their gauging. 

\subsection{Symmetry category with two simple lines}
The simplest symmetry category consists of just one simple object $1$ and its multiples. 
This can be thought of as a symmetry of any 2d theory, but it is not very interesting. 

Let us then consider the simplest nontrivial symmetry category, consisting of simple objects $1$ and $x$.
The dual of $x$ can only be $x$, and then $x\otimes x$ can only contain one copy of $1$.
Therefore the tensor product can only be of the form $x\otimes x=nx\oplus 1$, where $n$ is a non-negative integer.

When $n=0$, then the simple objects $1$ and $x$ form the group $\bZ_2$.
As we already discussed,
there are two possible symmetry categories $\cC(\bZ_2,\alpha)$ where $\alpha\in H^3(\bZ_2,\UU(1))=\bZ_2$ determines the anomaly.

When $n=1$, the dimension of $x$ is easily determined to be $(1+\sqrt{5})/2$.
The only nontrivial condition is the pentagon identity of the associator $a_{x,x,x}$,
and can be solved uniquely \cite{Moore:1988qv}.
This symmetry category in fact has a braiding which turns it into a modular tensor category describing an anyon system.
Recently this is known under the name of Fibonacci anyons because fusing $n$ copies of $x$ generates the $F_{n-1}1\oplus F_n x$ where $F_n$ is the $n$th Fibonacci number.
For this reason we denote this category $\mathrm{Fib}$.

It is known that we cannot have $n>1$, as shown by Ostrik \cite{OstrikRank2}.
Therefore, possible symmetry categories with two simple lines are just three, $\cC(\bZ_2,0)$, $\cC(\bZ_2,1)$ and $\mathrm{Fib}$.
Ostrik also classified all possible symmetry categories with three simple objects \cite{OstrikRank3}.

\subsection{Symmetry category of $\SU(2)$ WZW models and other RCFTs}
Next, we review the symmetry category of RCFTs, following the construction of \cite{Frohlich:2009gb}.
Let us take a rational chiral algebra $\cA$ in two dimensions and consider a conformal field theory $T$ which corresponds to the diagonal modular invariant of this algebra $\cA$. 
As is well-known, from a chiral vertex operator $a$ corresponding to an irreducible representation of $\cA$, we can construct a topological line operator $a$ of this theory $T$.
Because the theory $T$ corresponds to the diagonal modular invariant, chiral and antichiral vertex operators generate the same line operators.
Therefore the theory has topological lines generated by irreducible representations of a single copy of $\cA$.
They of course are specified by the Moore-Seiberg data, or equivalently they form a unitary modular tensor category $\cC$.
We can forget the braiding of $\cC$ and regard it as the symmetry category of this theory $T$.

The essential observation of \cite{Frohlich:2009gb} is that the choice of gauge-able subpart of $\cC$, or equivalently the choice of the module category $\cM$ over $\cC$, is in one to one correspondence with the choice of modular invariants of the chiral algebra $\cA$.
In particular, all modular invariants, including the exceptional ones, arise as the result of a generalized gauging.

Let us describe them in more detail in the case of $\SU(2)$ WZW models, following \cite{Ostrik}.
The chiral algebra is $\widehat\SU(2)_k$, which has $k+1$ irreducible representations \begin{equation}
V_j,\qquad j=0,1/2,\cdots, k/2
\end{equation} with the fusion rule \begin{equation}
V_j \otimes V_{j'} = V_{|j-j'|}\oplus V_{|j-j'|+1}\oplus \cdots \oplus V_{m}
\end{equation} where $m=\mathrm{min}(j+j',k-(j+j'))$.
We have \begin{equation}
\dim V_j= \frac{q^{j+1/2}-q^{-j-1/2}}{q^{1/2}-q^{-1/2}}\qquad\text{where}\quad q=e^{2\pi i /(k+2)}.
\end{equation}
They form a symmetry category we denote by $\Rep(\widehat\SU(2)_k)$.
The object $V_0$ is the identity.
It is clear that the objects $V_j$ with integral $j$ form a symmetry subcategory,
and can be denoted by $\Rep(\widehat\SO(3)_k)$.
In particular, when $k=3$, this is equivalent to the symmetry category $\mathrm{Fib}$ discussed above.

Since $V_{k/2}\otimes V_{k/2} = V_0$, the simple lines $V_0$ and $V_{k/2}$ form a sub-symmetry category.
From our general discussion above, this is equivalent to $\cC(\bZ_2,\alpha)$ where $\alpha\in H^3(\bZ_2,\UU(1))=\bZ_2$. 
This $\alpha$ is determined in terms of the associator, or equivalently the fusion matrix or the quantum 6j symbol involving four $V_{k/2}$, and is known to be $\alpha=k  \in \bZ_2$.  
In particular $\Rep(\widehat\SU(2)_1)=\cC(\bZ_2,1)$.
This also means that the subsymmetry formed by $V_0$ and $V_{k/2}$ is gauge-able when $k$ is even. Gauging it we obtain the modular invariant of type $D_{k/2+2}$.

The $E_{6,7,8}$ modular invariants correspond to algebra objects \begin{equation}
A=\begin{cases}
V_0\oplus V_3, & k=10 \quad (E_6),\\
V_0\oplus V_4\oplus V_8, & k=16 \quad (E_7),\\
V_0\oplus V_{5}\oplus V_{9} \oplus V_{14}, & k=28 \quad (E_8).
\end{cases}
\end{equation}

The type $X_n$ of the modular invariants, or equivalently of the possible gauging, specifies the corresponding module category structure as follows: the isomorphism classes of simple objects in the module category are labeled by the nodes of the Dynkin diagram of $X_n$, and the edges describe how $V_{1/2}$ acts on the simple objects.

\subsection{Gauging a subgroup of a possibly-anomalous group}
\subsubsection{Generalities}
We now turn our attention to a more traditional setup of gauging a subgroup $H$ of a bigger group $G$.
We will soon see that already in this traditional-looking setup we encounter various surprises. 

We start by specifying the anomaly of the bigger group; we start from a symmetry category $\cC(G,\alpha)$ where $\alpha\in H^3(G,\UU(1))$.
Possible gaugings are classified by their module categories, as already discussed.
They turn out to be in one-to-one correspondence with a pair $(H,\psi)$ 
where $H$ is a subgroup such that the restriction of $\alpha$ to $H$ is trivial, and $\psi$ is an element of $H^2(H,\UU(1))$.\footnote{More precisely, when $\alpha$ is nontrivial, $\psi$ is an element of a torsor over $H^2(H,\UU(1))$.}
This result agrees with the more traditional viewpoint: we choose a non-anomalous subgroup $H$ and then choose the discrete torsion $\psi$.

From the general machinery described above, the gauged theory has a symmetry category which is the dual of $\cC(G,\alpha)$ with respect to $(H,\psi)$.
Let us denote the resulting symmetry category by $\cC':=\cC(G,\alpha;H,\psi)$.
When $\alpha=0$, $H=G$, $\psi=0$, we already know that $\cC'=\Rep(G)$.
Furthermore,  when $G$ is abelian, $\cC'=\cC(\hat G,0)$.
The explicit structure of $\cC'$ in the general case can be determined by realizing it as a category of bimodules $\Bimod_\cC(A)$ for the algebra object $A$ corresponding to $(H,\psi)$.

Let us see what we can say generally.
Firstly, there are two general facts:
\begin{itemize}
\item the dimensions of lines in $\cC'$ are all integral; such symmetry categories  are called as integral symmetry categories.
\item the total dimension $\dim \cC':= \sum_a (\dim a)^2$ is the same as the original one: $\dim \cC'=|G|$.
\end{itemize}

Secondly, there are cases where the dual symmetry $\cC(G,\alpha;H,\psi)$ itself is of the form $\cC(G',\alpha')$ for some group $G'$ and the anomaly $\alpha'$.
There is a  theorem by Naidu, Nikshych \cite{Naidu,NaiduNikshych} and Uribe \cite{Uribe} determining exactly when this happens, and if so, explicitly the form of $G'$ and $\alpha'$. 
The general formula is too complicated to reproduce in full here. 
A necessary condition is that $H$ is an Abelian normal subgroup. When $\alpha$ is trivial this in fact suffices. 
In the next subsubsection we describe its explicit structure.

\subsubsection{Gauging a normal Abelian subgroup of a non-anomalous group}
\paragraph{Statement:}
Let us choose a group $G$ and its normal Abelian subgroup $H$.
We then gauge $H$. The gauged theory then has a symmetry group $G'$ with an anomaly $\alpha'\in H^3(G',\UU(1))$, given as follows.

The fact that $H$ is a normal Abelian subgroup means that
$G$ is an extension \begin{equation}
0\to H\to G\to K=G/H\to 0 
\end{equation} and as such it is determined by an an action of $K$ on $H$ by inner automorphisms in $G$, and an element $\kappa\in H^2(K,H)$ defined using the group action. The group $G$ is a crossed product, $G= H \rtimes_\kappa K$. 
Let us identity $G=H\times K$ \emph{as a set}. Then the group structure is given as follows: \begin{equation}
(h,k) (h',k')=(h(k\triangleright h') \kappa(k,k'), kk')
\end{equation} where $k\triangleright h'$ is the action of $k$ on $h'$ and $\kappa(k,k')$ is an $H$-valued 2-cocycle of $K$.

Denote by $\hat H$ the dual group of $H$. 
There is a natural action of $K$ on $\hat H$ given by $k\triangleright\rho(h)=\rho(k^{-1}\triangleright h)$ for arbitrary elements $k\in K$, $h\in H$ and $\rho\in\hat H$. Under this action, 
\begin{equation}
G'=\hat H\rtimes K\label{G'}
\end{equation} with the trivial two-cocycle in $H^2(K,\hat H)$,
and $\alpha'$ is given by 
\begin{equation}
\alpha'=\alpha_\kappa\qquad \text{where}\quad \alpha_\kappa((\rho,k),(\rho',k'),(\rho'',k'')) = \rho''(kk'\triangleright\kappa(k,k')) \label{GKKS}
\end{equation}
Note that the nontriviality $\kappa$ of the crossed product in the original $G$ side is traded for the nontriviality of the anomaly $\alpha_\kappa$ on the $G'$ side.
Summarizing, we have \begin{claim}
Let us gauge a normal Abelian subgroup $H$ of a symmetry group $G$. 
$G$ is then necessarily of the form $G=H\rtimes_\kappa K$, where $\kappa\in H^2(K,H)$.
When $G$ has no anomaly, the gauged theory has the symmetry group $G'=\hat H\rtimes K$, and the resulting anomaly $\alpha'$ is given in terms of $\kappa$ as in \eqref{GKKS}.
\end{claim}

\paragraph{Derivation:}
Let us now derive the description of $G'$ given in the last paragraph. Our starting category is $\cC(G)$ and we want to gauge it by the algebra object $A=\bigoplus h$ where $h\in H$. The simple objects of gauged category $\Bimod_{\cC(G)}(A)$ are bimodules which can be seen to form the set $\hat H\times K$ using arguments very similar to those in Sec.~\ref{sec:example}. An object $(\rho,k)$ in $\Bimod_{\cC(G)}(A)$ is built from the object $\bigoplus_h (h,k)$ in $\cC(G)$. Our choice of the bimodule structure on $(\rho,k)$ is that the right action by $A$ is trivial and the left action by $A$ is given in terms of morphisms $(x_L)_{h,h'}:(h,e)\otimes (h',k)\to (hh',k)$ satisfying the familiar condition
\be
(x_L)_{h,h'}=(x_L)_{hh',e}((x_L)_{h',e})^{-1}
\ee
with
\be
(x_L)_{h,e}=\rho(h).
\ee

The balanced tensor product of $(\rho,k)$ and $(\rho',k')$ in $\Bimod_{\cC(G)}(A)$ is given in terms of projectors $\proj _{h,h'}:(h,k)\ot (h',k')\to (h\kappa(k,k')(k\triangleright h'),kk')$. 
The equation \eqref{bal} tells us that
\be
\proj _{h(k\cdot h'),e}=\proj _{h,h'}\rho'(h') \label{balance}
\ee
Demanding the right action on $(\rho,k)\ot_A (\rho',k')$ to be trivial leads us to the condition that
\be
\proj _{h,h'}=\proj _{h,e}
\ee
which can be substituted into \eqref{balance} to simplify it to
\be
\proj _{h(k\triangleright h'),e}=\proj _{h,e}\rho'(h') \label{rebalance}.
\ee
Via \eqref{comp}, the left action $\rho''$ on $(\rho,k)\ot_A (\rho',k')$ satisfies
\be
\rho''(h)\proj _{e,e}=\proj _{h,e}\rho(h)
\ee
which can be combined with \eqref{rebalance} to yield
\be
\rho''(h)=\rho(h)\rho'(k^{-1}\triangleright h).\label{groupstructure}
\ee
In particular we have \begin{equation}
\proj _{h,e}=\rho'(k^{-1}\triangleright h).
\end{equation}

The equation \eqref{groupstructure} means that $\Bimod_{\cC(G)}(A)$ is equivalent to $\cC(G',\alpha')$ 
where $G'=\hat H\rtimes K$ for some yet to be determined $\alpha'$.

The associator can be computed from the graph in Figure \ref{fig:bimodass}. Let the objects $p,q,r$ be $(k,\rho),(k',\rho'),(k'',\rho'')$. It suffices to restrict each object to the sub-object $(k,e)$ in $\cC(G)$. Without loss of generality, we can assume $\proj _{e,e}=1$ because factors of $\proj_{e,e}$ are canceled by factors of $\coproj _{e,e}$. Then the only contribution comes from what is denoted as $\coproj_{p\ot_A q,r}$ in Figure \ref{fig:bimodass} and we find the anomaly $\alpha'$ given in \eqref{GKKS}.

\paragraph{Examples:}
As an example, consider $G=\bZ_{2n}$ generated by $x$ with $x^{2n}=1$, and gauge the $\bZ_2$ subgroup generated by $x^n$.
When $n$ is odd, $G=\bZ_2\times \bZ_n$, and the dual symmetry is clearly just $G'=\hat\bZ_2\times \bZ_n$ without any anomaly, since the part $\bZ_n$ does not matter.
When $n$ is even, $G$ is a nontrivial extension $0\to \bZ_2 \to G=\bZ_{2n} \to \bZ_n \to 0$, corresponding to a nonzero $\kappa \in H^2(\bZ_{n},\bZ_2)$.
This means that the dual is $G'=\hat \bZ_2\times \bZ_n$, with a nontrivial anomaly $\alpha_\kappa$ as given above.

As another example, consider $G=D_{2n}$, the dihedral group of $2n$ elements, generated by two elements $r,s$ such that $r^n=s^2=1$, $srs^{-1}=r^{-1}$.
In particular, let $n=2m$. Then $x:=r^m$ generates the center $\bZ_2=\vev{x}$ of $D_{2n}$.
Let us then gauge the center.  
Since the extension $0\to \bZ_2 \to D_{2n} \to D_n \to 0$ is nontrivial, the dual group $G'=\hat{\bZ_2}\times D_n$ has a nontrivial anomaly $\alpha_{\kappa_D}$,
given in terms of  a nonzero $ \kappa_D\in H^2(D_n,\bZ_2)$ describing the extension.
In particular, for $D_{2n}=D_8$, the dual group $G'=\hat{\bZ_2} \times \bZ_2\times \bZ_2$ is Abelian.
Dually, this means that by gauging $\hat{\bZ_2}$ of the Abelian group $G'$ with an anomaly $\alpha_{\kappa_D}$ turns the symmetry into a non-Abelian group $D_8$.

As a final example in this subsection, consider $G=Q_8$, the quaternion group of eight elements, formed by eight quaternions $\pm1$, $\pm i$, $\pm j$, $\pm k$. 
This is naturally a subgroup of $\SU(2)$ since  quaternions of absolute value 1 form the group $\SU(2)$,
and as such the lift to $\SO(3)$ of a finite subgroup of $\SO(3)$, this case $D_4=\bZ_2\times \bZ_2$.
This means that we have a nontrivial extension \begin{equation}
0\to \bZ_2\to Q_8\to D_4\to 0.
\end{equation} 
This extension is again nontrivial, whose class $\kappa_Q\in H^2(\bZ_2\times \bZ_2,\bZ_2)$ is different from $\kappa_D$ in the case of $D_8$.
The dual group is then $G'=\hat{\bZ_2} \times \bZ_2\times \bZ_2$ but with a different anomaly $\alpha_{\kappa_Q}$. 

Dually, we can say as follows. 
The same Abelian group, $\hat{\bZ_2} \times \bZ_2\times \bZ_2$ with two different anomalies $\kappa_D$ and $\kappa_Q$ dualizes, under gauging of $\hat{\bZ_2}$, into two different non-Abelian groups $D_8$ and $Q_8$.

\subsection{Integral symmetry categories of total dimension 6}
Let us study the symmetry categories of total dimension 6 in detail.
We already know a few such symmetry categories, $\cC(\bZ_2\times \bZ_3, \alpha)$, $\cC(S_3,\alpha)$ and $\Rep(S_3)$, where $S_3$ is the symmetric group acing on three objects.
Let us study what the gauging of their subgroups leads to.
We will see that there are in fact two more integral symmetry categories of total dimension 6.

\paragraph{From $\cC(\bZ_2\times \bZ_3,\alpha)$:}
Here the anomaly is determined by $\alpha\in H^3(\bZ_2\times \bZ_3,\UU(1))=\bZ_2\times \bZ_3$.
	\begin{itemize}
	\item $\bZ_1$ is always gaugeable,
	\item $\bZ_2$ is gaugeable only when $\alpha$ is from $\bZ_3$ and then the dual is itself,
	\item $\bZ_3$ is gaugeable only when $\alpha$ is from $\bZ_2$ and then the dual is itself,
	\item $\bZ_6$ is gaugeable only when $\alpha$ is trivial.
	\end{itemize}
So there is nothing particularly interesting going on here.

\paragraph{From $\Rep(S_3)$:}
Any possible gauging of $\Rep(S_3)$ can always be done by first gauging $\Rep(S_3)$ back to $\cC(S_3)$ and then gauge one of its subgroup. 
Therefore we do not have to study it separately.

\paragraph{From $\cC(S_3,\alpha)$:}
Here the anomaly is determined by $\alpha\in H^3(S_3,\UU(1))=\bZ_2\times \bZ_3$.
Let us denote by $a$ and $b$ the generators of $\bZ_2$ and of $\bZ_3$, respectively.
	\begin{itemize}
	\item $\bZ_1$ is always gaugeable and the dual is itself.
	\item The  subgroup $\bZ_2$ is gaugeable only when $\alpha=b^i$ with $i=0,1,2$.
	The dual is \emph{not} of the form $\cC(G',\alpha')$ because this subgroup is not normal.
	When $\alpha=0$  the dual turns out to be $\Rep(S_3)$.
	When $i=1,2$, the duals \emph{cannot} be $\Rep(S_3)$, since if so, a further gauging will produce $\cC(S_3,b^0)$ from $\cC(S_3,b^{1,2})$. 
	But this is impossible, since these two symmetry categories have different number of possible gaugings.	
	\item The normal subgroup $\bZ_3$ is gaugeable only when $\alpha=a^{0,1}$. Gauging it leads back to itself, with the same anomaly.
	\item $S_3$ is gaugeable only when $\alpha$ is trivial. The dual is $\Rep(S_3)$.
\end{itemize}

From the analysis above, we find that symmetry categories $\cC(S_3,b^{1,2};\bZ_2, 0)$ obtained by gauging the $\bZ_2$ subgroup of $S_3$ with a nontrivial anomaly $\alpha=b^{1,2}$ is neither of the form $\cC(G,\alpha)$ nor of the form $\Rep(S_3)$.
It turns out that they have the same fusion rule as $\Rep(S_3)$, namely there are simple objects $1,x$ of dimension 1 and $a$ of dimension 2, such that $x^2=1$, $ax=a$, $a^2=1+x+a$.
These are the smallest  integral symmetry categories which is neither $\cC(G,\alpha)$ nor $\Rep(G)$.

It is known that the symmetry categories we listed so far exhaust all possible integral symmetry categories of dimension 6. This was shown in  \cite{EGOclassification}.

\subsection{Integral symmetry categories of total dimension 8}

Let us next have a look at symmetry categories of dimension 8.
There are five finite groups $G$ of order 8, namely the three Abelian ones $\bZ_8$, $\bZ_2\times \bZ_4$, $(\bZ_2)^3$ and two non-Abelian ones $D_8$ and $Q_8$.
Correspondingly, we already see that there are symmetry categories $\cC(G,\alpha)$ constructed from these group, where the possible anomalies are given as follows: \begin{equation}
\begin{array}{c||c|c|c|c|c}
G& \bZ_8 & \bZ_2\times \bZ_4 & \bZ_2^3 & D_8 & Q_8 \\
\hline
H^3(G,\UU(1))& \bZ_8&\bZ_2^2\times \bZ_4 & \bZ_2^7 & \bZ_2^2\times \bZ_4 & \bZ_8
\end{array}\,.
\end{equation}
We also know two other symmetry categories of total dimension 8, namely the representation categories $\Rep(D_8)$ and $\Rep(Q_8)$.

These two representation categories have the same fusion rules: there are four dimension-1 simple objects $1$, $a$, $b$, $ab$ forming an Abelian group $\cA=\bZ_2\times \bZ_2$, and one dimension-2 simple object $m$, such that the fusion rule is commutative, $a\otimes m=b\otimes m=m$, and  \begin{equation}
m\otimes m=1\oplus a\oplus b\oplus ab.
\end{equation}
There are in fact two more symmetry categories, known as $\mathrm{KP}$ and $\mathrm{TY}$ with this fusion rule \cite{TambaraYamagami}.  
For all these four cases, it is known that  we can gauge the subsymmetry $\bZ_2=\{1,a\}$ and obtain $\cC(D_8,\alpha)$ where $\alpha\in H^3(D_8,\UU(1))$  is chosen depending on the four cases. 

The four symmetry categories with the above fusion rule  have a nice uniform description due to Tambara and Yamagami, which is applicable to a more general case based on any Abelian group $\cA$. 
The gauging of $\bZ_2=\{1,a\}$ leading to $\cC(D_8,\alpha)$ also has an explanation
in the larger context of Tambara-Yamagami categories. 
We will study them in more detail in the next subection.

The only remaining choice of the fusion rule of an integral symmetry category of total dimension 8 has the following form \cite{Blau}: there are four dimension-1 simple objects $1$, $c$, $c^2$, $c^3$ forming an Abelian group $\cA=\bZ_4$, and one dimension-2 simple object $m$, such that $c\otimes m =m\otimes c=m$ and \begin{equation}
m\otimes m=1\oplus c\oplus c^2\oplus c^3.
\end{equation}
The result of Tambara and Yamagami \cite{TambaraYamagami} implies that there are four symmetry categories with this fusion rule, distinguished by two sign choices.
These categories do not have a common name; let us temporarily call it $\mathrm{S}_{\pm\pm}$.
This completes the list of the integral symmetry category of total dimension 8.

Before moving on, we have two comments.
First,  every integral symmetry categories we saw so far, i.e.~ those symmetry categories for which dimensions of objects are integers, can be obtained by gauging a non-anomalous subgroup of a possibly anomalous group. This property fails when the total dimension is larger. Indeed, some of the Tambara-Yamagami categories we discuss next are integral but cannot be obtained by gauging a non-anomalous subgroup of a possibly anomalous group.

Second, the symmetry category $\mathrm{KP}$ is of a historical interest, since it is the category of representations of the first non-commutative non-cocommutative Hopf algebra that appeared in the literature, constructed by Kac and Paljutkin in 1966 \cite{KacPaljutkin}.
One way to construct a symmetry category is to pick a Hopf algebra $H$ and take the category of its representations.
When $H$ is commutative, the symmetry category is of the form $\cC(G)$, and when $H$ is cocommutative, the symmetry category is of the form $\Rep(G)$.
When we take the dual of a Hopf algebra, this naturally interchanges $\cC(G)$ and $\Rep(G)$.
Therefore considering Hopf algebras is a unified framework in which $\cC(G)$ and $\Rep(G)$ can be treated symmetrically.
That said, to treat the symmetries of two dimensional theories and their gauging, we need to deal with symmetry categories in general and we cannot stop at the level of the Hopf algebras.
The symmetry categories which are categories of representations of Hopf algebras can be characterized as symmetry categories which has $\Vec$ as a module category. 
But even the familiar $\cC(G,\alpha)$ with a nontrivial $\alpha$ does not have $\Vec$ as a module category!

\subsection{Tambara-Yamagami categories}
To construct a Tambara-Yamagami category, we start from an Abelian group $\cA$.
The simple objects of the category are elements $a\in \cA$ of dimension 1 together with an object $m$ of dimension $|\cA|$, with the commutative fusion rule \begin{equation}
a\otimes m = m,\qquad m\otimes m=\bigoplus_{a\in \cA} a.
\end{equation}
Tambara and Yamagami showed in \cite{TambaraYamagami} that any symmetry category with this fusion ring is given by the choice of a symmetric nondegenerate bicharacter $\chi: \cA\times \cA\to \UU(1)$
and the choice of the sign of $\tau=\pm 1/\sqrt{|\cA|}$. 
The nontrivial associators are given in terms of $\chi$ and $\tau$: \begin{align}
a_{a,m,b} &= \chi(a,b), \label{xxx}\\
a_{m,a,m} &= \bigoplus_{b} \chi(a,b) \id_b,\label{yyy}\\
a_{m,m,m} &= \tau (\chi(a,b)^{-1} )_{a,b} \in \Hom(\bigoplus_a m,\bigoplus_b m).\label{zzz}
\end{align}
Let us denote the resulting symmetry category by $\mathrm{TY}(\cA,\chi,\tau)$.

In our case where $\cA=\bZ_2\times \bZ_2$ generated by $a$ and $b$,  we just have two possible symmetric nondegenerate bicharacters $\chi$ up to the action of $\mathrm{SL}(2,\bZ_2)$. Explicitly, two such choices are specified by \begin{equation}
\chi(a,a)=1, \quad \chi(b,b)=-\chi(a,b)=\pm 1.
\end{equation} We denote the choices by $\chi_{\pm}$.
The choice of $\tau$ is $\tau=\pm1/2$.
Then we have the following correspondence: \begin{equation}
\begin{array}{c|cc}
& \chi & \tau \\
\hline
\Rep(D_8)& \chi_+ & +1/2 \\
\Rep(Q_8)& \chi_+ & -1/2 \\
\mathrm{KP} & \chi_- & +1/2 \\
\mathrm{TY} & \chi_- & -1/2
\end{array}.
\end{equation} 
Here we are slightly abusing the notation such that  $\mathrm{TY}$ alone stands for a specific symmetry category with total dimension 8, while $\mathrm{TY}(\cA,\chi,\tau)$ refers to a general construction.

Another Tambara-Yamagami category of total dimension 8 is based on $\cA=\bZ_4=\{1,c,c^2,c^3\}$. 
Any non-degenerate symmetric  bicharacter is of the form \begin{equation}
\chi_\pm(c^k,c^l)=(\pm i)^{kl}.
\end{equation} Together with the choice of the sign of $\tau$, we have four Tambara-Yamagami categories $\mathrm{S}_{\pm\pm}$ based on $\bZ_4$.
We note that the bicharacter $\chi_+$ is trivial on the subgroup $\bZ_2=\{1,a=c^2\}$.

It is known that  for all eight Tambara-Yamagami categories described above, we can gauge the subsymmetry $\bZ_2=\{1,a\}$ and obtain $\cC(D_8,\alpha)$ where $\alpha\in H^3(D_8,\UU(1))$  is chosen depending on the four cases. 
This fact is a specific instance of a general theorem determining when a Tambara-Yamagami category is obtained by gauging a subgroup of a possibly non-anomalous group \cite{MeirMusicantov}.
They showed that this occurs if and only if $\cA$ has a Lagrangian subgroup $H$ 
for $\chi$, i.e.~there is a subgroup $H\subset\cA$ such that i) the restriction of $\chi$ on $H$ is trivial and ii) $\cA/H \simeq \hat H$ via the pairing induced by $\chi$.
In the four cases above, $H$ is given by $\bZ_2=\{1,a\}$.

Let us explicitly describe below that gauging the subgroup $H$ of the Tambara-Yamagami symmetry category $\cC=\mathrm{TY}(\cA,\chi,\tau)$ produces a symmetry category of the form $\cC(G,\alpha)$.
The algebra object we use to gauge the system is $A=\bigoplus_{h\in H} h$.

We note that $\cA$ fits in the extension $0\to H\to \cA\to \hat H\to 0$.
We fix a specific section $s:\hat H\to \cA$ and denote by $\kappa$ the two-cocycle in $C^2(\hat H,H)$ characterizing this extension.
Then the symmetric nondegenerate bicharacter $\chi$ on $\cA$ defines a symmetric map
\begin{equation}
\chi: \hat H \times \hat H  \to \UU(1)
\end{equation} satisfying the condition \begin{equation}
\chi(\rho+\rho',\sigma)=\chi(\rho,\sigma)+\chi(\rho',\sigma)+\sigma(\kappa(\rho,\rho')).
\end{equation}

Simple $(A,A)$ bimodules turn out to be isomorphic to either of the following two types:
\begin{itemize}
\item $X_{\rho,\sigma}$ for $\rho,\sigma\in \hat H$. As an object in $\cC$, it is $\bigoplus_{h\in H} h\sigma$.  The right action of $A$ is trivial, and the left action of $A$ is given by $\rho(a) \id:   a\otimes h\sigma \to ah\sigma$.
\item $Y_{\rho,\sigma}$ for $\rho,\sigma\in \hat H$. As an object in $\cC$, it is just $m$.
The left action and the right action of $A$ is given by $\rho(a) \id: a\otimes m \to m$ and 
$\sigma(a)\id: m\otimes a\to m$.
\end{itemize}
The tensor product $\otimes_A$ in the category of bimodules, together with the projections $\proj$ \eqref{eq:projection} in the definition of $\otimes_A$ is then given as follows:
\begin{equation}
X_{\rho,\sigma}\otimes_A X_{\rho',\sigma'}=X_{\rho\rho',\sigma \sigma'}\label{1}
\end{equation} where the projections $\proj $ are trivial,
\begin{equation}
X_{\rho,\sigma} \otimes_A Y_{\rho',\sigma'}=Y_{\rho\rho',\sigma \sigma'}\label{2}
\end{equation} where the projections $\proj $ are given by $\rho'(a)\id:a\sigma\otimes m\to m$,
\begin{equation}
Y_{\rho,\sigma} \otimes_A X_{\rho',\sigma'}=Y_{\rho(\sigma')^{-1},\sigma (\rho')^{-1}}\label{3}
\end{equation} where the projections $\proj $ are given by $\sigma(a)\rho'(a)^{-1}\id:m\otimes a\sigma'\to m$,
and \begin{equation}
Y_{\rho,\sigma} \otimes_A Y_{\rho',\sigma'}=X_{\rho(\sigma')^{-1},\sigma (\rho')^{-1}}\label{4}
\end{equation} where the projections $\proj $ are given by \begin{equation}
\proj =\bigoplus_{a\in H}\sigma'(a)^{-1}\id_{a\sigma(\rho')^{-1}}: m\otimes m\to \bigoplus_{x\in\cA} x.
\end{equation}

The equations \eqref{1},\eqref{2},\eqref{3},\eqref{4} show that the simple objects $X_{\rho,\sigma}$ and $Y_{\rho,\sigma}$ form the group \begin{equation}
G=(\hat H\times \hat H)\rtimes \bZ_2
\end{equation} where the $\bZ_2$ acts by \begin{equation}
(\rho,\sigma)\mapsto (-\sigma,-\rho).
\end{equation}
The anomaly $\alpha$ can be computed using the projections given above and the associators \eqref{xxx}, \eqref{yyy} and \eqref{zzz} of the original category. 
We find \begin{align}
\alpha(X_{\rho,\sigma},X_{\rho',\sigma'},X_{\rho'',\sigma''})&=\rho''(\kappa(\sigma,\sigma')),\\
\alpha(X_{\rho,\sigma},X_{\rho',\sigma'},Y_{\rho'',\sigma''})&=\rho''(\kappa(\sigma,\sigma')),\\
\alpha(X_{\rho,\sigma},Y_{\rho',\sigma'},X_{\rho'',\sigma''})&=\chi(\sigma,\sigma''),\\
\alpha(Y_{\rho,\sigma},X_{\rho',\sigma'},X_{\rho'',\sigma''})&=(\rho'\rho''\sigma^{-1})(\kappa(\sigma',\sigma'')),\\
\alpha(X_{\rho,\sigma},Y_{\rho',\sigma'},Y_{\rho'',\sigma''})&=\sigma''(\kappa(\sigma,(\sigma')^{-1}\rho'')),\\
\alpha(Y_{\rho,\sigma},X_{\rho',\sigma'},Y_{\rho'',\sigma''})&=\chi(\sigma',\sigma(\rho')^{-1}(\rho'')^{-1}),\\
\alpha(Y_{\rho,\sigma},Y_{\rho',\sigma'},X_{\rho'',\sigma''})&=(\sigma'(\rho'')^{-1})(\kappa(\sigma(\rho')^{-1},\sigma'')),\\
\alpha(Y_{\rho,\sigma},Y_{\rho',\sigma'},Y_{\rho'',\sigma''})&=\mathrm{sgn}(\tau)\chi(\sigma(\rho')^{-1},\sigma'(\rho'')^{-1}).
\end{align}
As a check of the computation, we can directly confirm that these define a 3-cocycle on $G$.

In the eight cases $\Rep(D_8)$, $\Rep(Q_8)$, $\mathrm{KP}$, $\mathrm{TY}$ and $\mathrm{S}_{\pm\pm}$ we discussed above, we always have $H=\bZ_2$ and  the resulting group $G=(\bZ_2\times \bZ_2)\rtimes \bZ_2$ is  $D_8$. 
To see this,  regard $\bZ_2\times \bZ_2$ as the group of flipping the coordinates $x$ and $y$ of $\bR^2$ generated by \begin{equation}
(x,y)\mapsto (-x,y),\qquad (x,y)\mapsto (x,-y)
\end{equation} respectively,
 and $\bZ_2$ acting on $\bZ_2\times \bZ_2$ to be the exchange of $x$ and $y$ given by \begin{equation}
 (x,y)\mapsto (y,x).
\end{equation}
Dually, with a suitably chosen $\alpha$ on $D_8$ and gauging the $\bZ_2$ subgroup flipping the $x$ coordinate, we get the four symmetry categories given above.

\section{2d TFT with $\cC$ symmetry and their gauging}
\label{sec:tqft}
\subsection{2d TFTs without symmetry}

As a warm-up, let us recall the structure of 2d TFTs without any symmetry. 
We follow the exposition in \cite{Moore:2006dw} closely, see in particular their Appendix A. 

\begin{figure}
\[
\begin{array}{c|c|c|c}
\incc{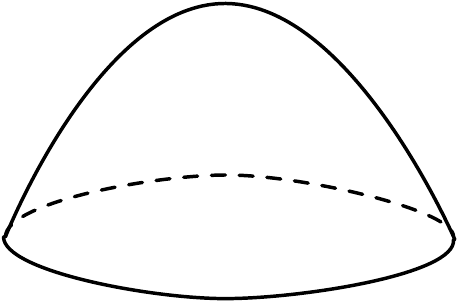} & \incc{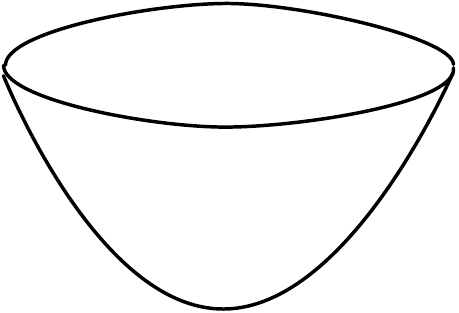}&\incc{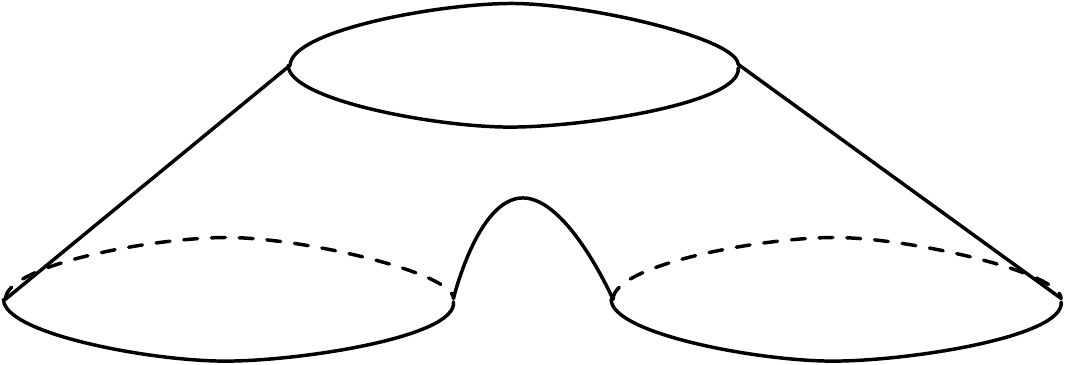}&\incc{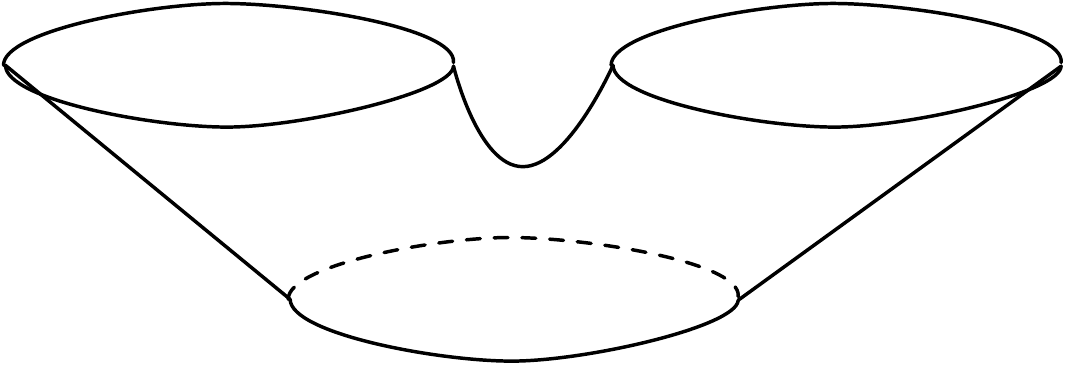}\\
\co{I}:V\to \bC& I:\bC\to V & M:V\otimes V\to V & \co{M}:V\to V\otimes V
\end{array}
\]
\caption{The basic building blocks of a 2d TFT.\label{fig:basic}}
\end{figure}

We start with a vector space $V$ of states on $S^1$ and one wants to define a consistent transition amplitude 
\begin{equation}
Z_\Sigma: V^{\otimes m}\to V^{\otimes n}
\end{equation}
 corresponding to a given topological surface $\Sigma$ 
with $m$ incoming circles and $n$ outgoing circles.
We need four basic maps $\co{I}$, $I$, $M$, $\co{M}$ corresponding to four basic geometries given in Fig.~\ref{fig:basic}.

\begin{figure}
\[
\begin{array}{c|c}
\incc{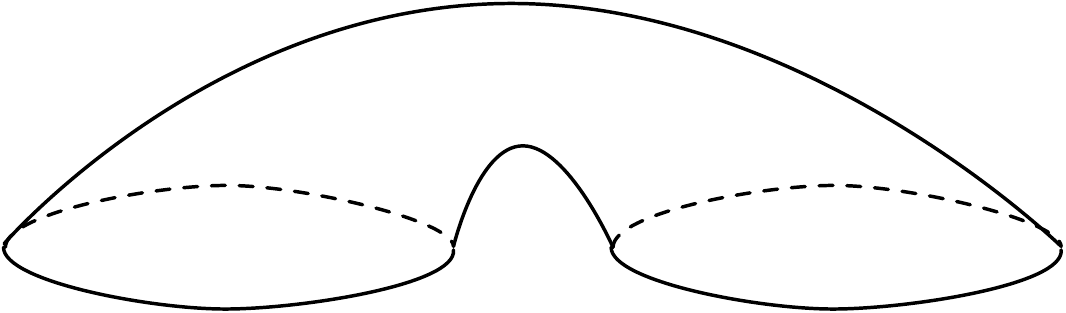} & \incc{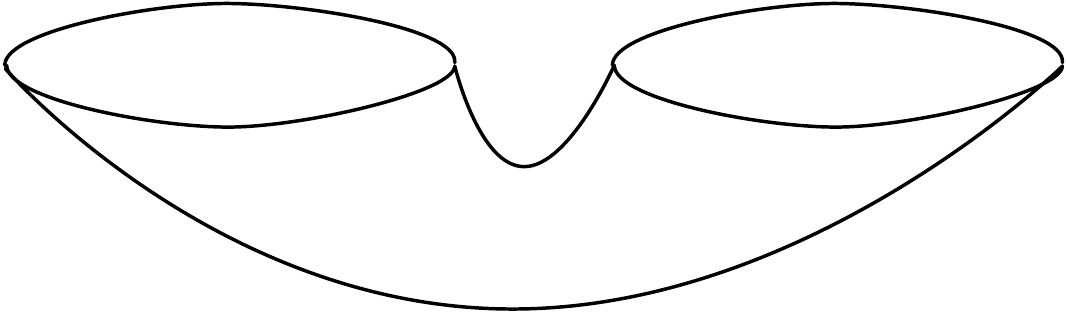}\\
\co{I}M:V\otimes V\to \bC& \co{M}I:\bC\to V\otimes V
\end{array}
\]
\caption{The pairing of $V$ with itself.\label{fig:pairing}}
\end{figure}

First, we construct maps $\co{I} M:V\otimes V\to \bC$ and $\co{M}I:\bC \to V\otimes V$ as in Fig.~\ref{fig:pairing}.
This inner product must be non-degenerate because it just corresponds to a cylinder geometry which pairs a state on one circle with a dual state on the other circle.
Using it, we can identify $V$ and $V^*$.
Then, $\co{I}$ is an adjoint of $I$ and $\co{M}$ is an adjoint of $M$.
Therefore, to every property involving $M$, we can write down a corresponding property involving $\co{M}$, and similarly for statements about $I$ and $\co{I}$.
This allows us to reduce the number of independent statements we need to write down roughly by half;
we do not repeat these adjoint statements below.

\begin{figure}
\[
\incc{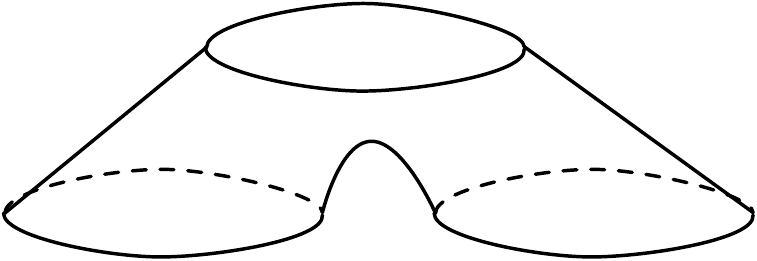}=\incc{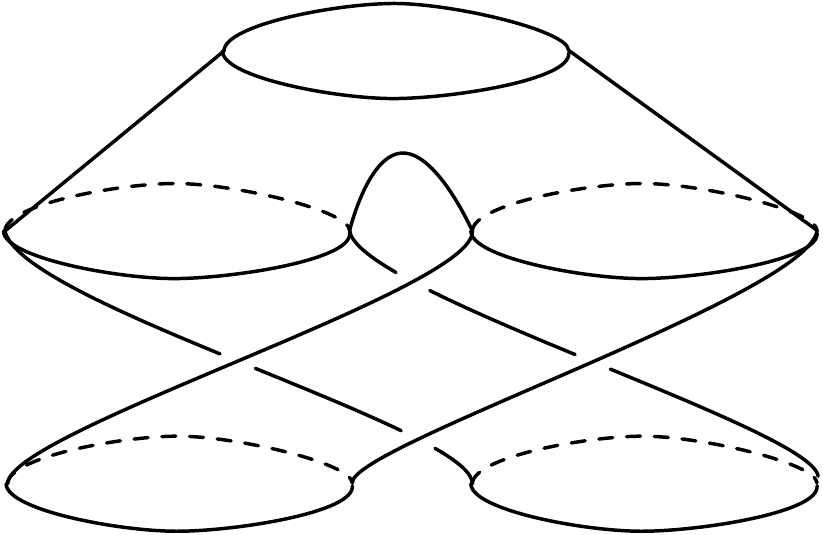}
\]
\caption{The product $M$ is commutative.\label{fig:comm}}
\end{figure}

\begin{figure}
\[
\incc{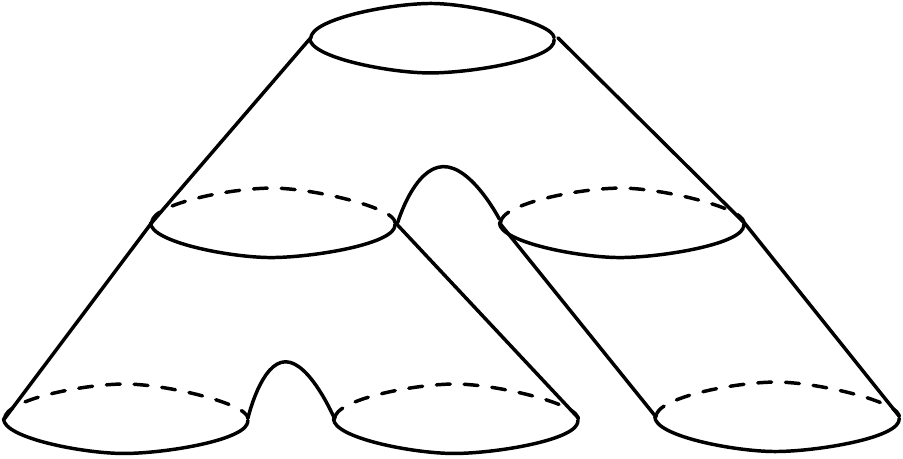}=\incc{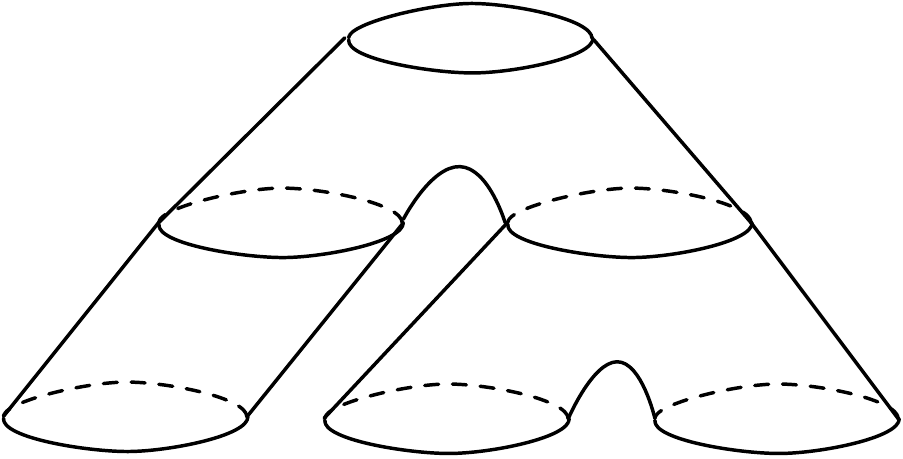}
\]
\caption{The product $M$ is associative.\label{fig:assoc}}
\end{figure}

\begin{figure}
\[
\incc{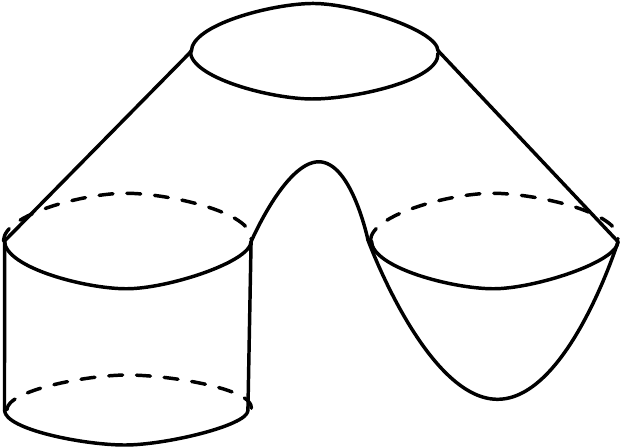}=\incc{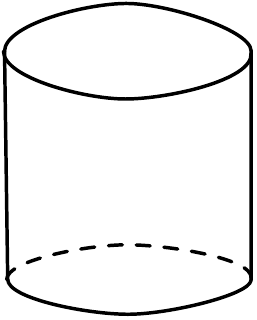}
\]
\caption{$I$ is a unit of the product $M$.\label{fig:unitT}}
\end{figure}

\begin{figure}
\[
\incc{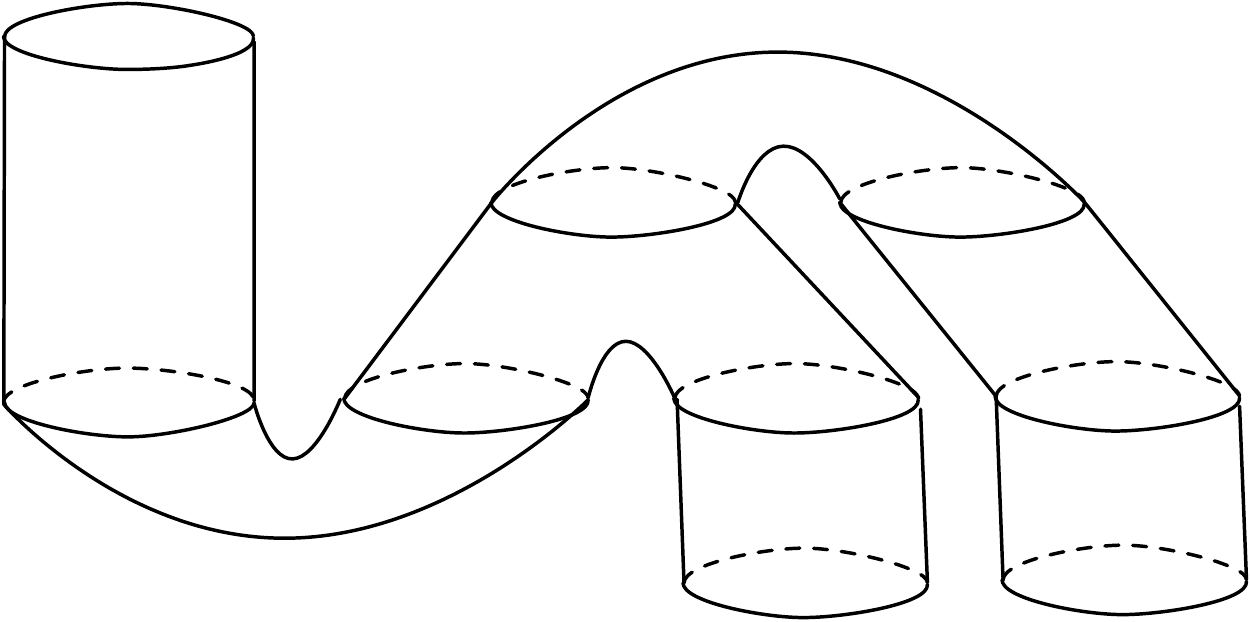}=\incc{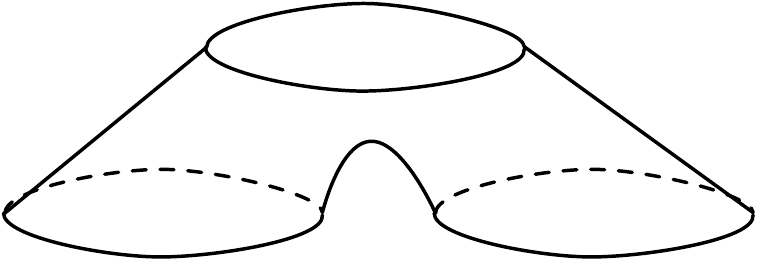}
\]
\caption{The product is invariant under exchanging an incoming circle and an outgoing circle.\label{fig:elephant}}
\end{figure}

We consider $M$ as giving a product on $V$.
There is no order on the two incoming circles of a pair of pants and hence the product is commutative, see Fig.~\ref{fig:comm}. 
We can also see that $M$ is associative from Fig.~\ref{fig:assoc}
and that $I$ is a unit of the multiplication $M$ from Fig.~\ref{fig:unitT}.
Also, by composing these inner products with the product, we see that the product is invariant under permuting three legs, see Fig.~\ref{fig:elephant}

After these preparations, let us associate \emph{a} map $Z_\Sigma:V^{\otimes m}\to V^{\otimes n}$ to a surface $\Sigma$ with $m$ incoming circles and $n$ outgoing circles.
We pick a time coordinate $t:\Sigma\to[0,1]$  such that at $t=0$ we start with $m$ initial circles and at $t=1$ we finish with $n$ final circles. As time goes from 0 to 1, the number of circles generically stay constant but can either increase or decrease by one unit at specific times $0=t_0<t_1<t_2<\cdots<t_p=1$. 
Cut $\Sigma$ once in each interval $(t_i,t_{i+1})$. This divides $\Sigma$ into $p$ pieces. 
The geometry of each piece contains some cylinders, which correspond to trivial transition amplitude, and exactly one non-trivial geometry out of the four non-trivial cases shown in Fig.~\ref{fig:basic}. This gives us \emph{an} expression for $Z_\Sigma$ in terms of the four maps $\co{I},I,M,\co{M}$.

However, one could choose a different time $t'$ which starts with same $m$ initial circles and ends with same $n$ final circles. In general, this would lead to a different cutting of $\Sigma$ and a different compositions of four maps $\co{I},I,M,\co{M}$.
We need to make sure that they agree.

We can continuously deform the time function $t$ to obtain the time function $t'$.
 The critical points $t_i$ will move under this deformation and will cross each other. 
It is also possible for two critical points to meet and annihilate each other or for two critical points to pop out of nowhere.
We therefore need to ensure that $Z_\Sigma$ remains invariant when $t_i$ and $t_{i+1}$ cross each other, and when two critical points are created or annihilated.
For this, we just need to ensure that the two-step composition from the cut between $t_{i-1}$ and $t_i$ to the cut between $t_{i+1}$ and $t_{i+2}$ remains invariant under these processes.

\begin{figure}
\[
\incc{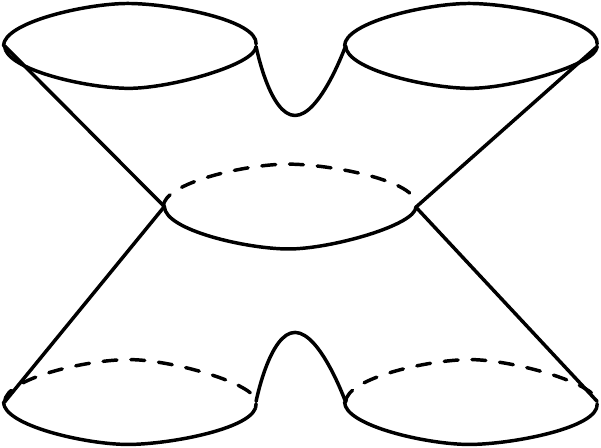}=\incc{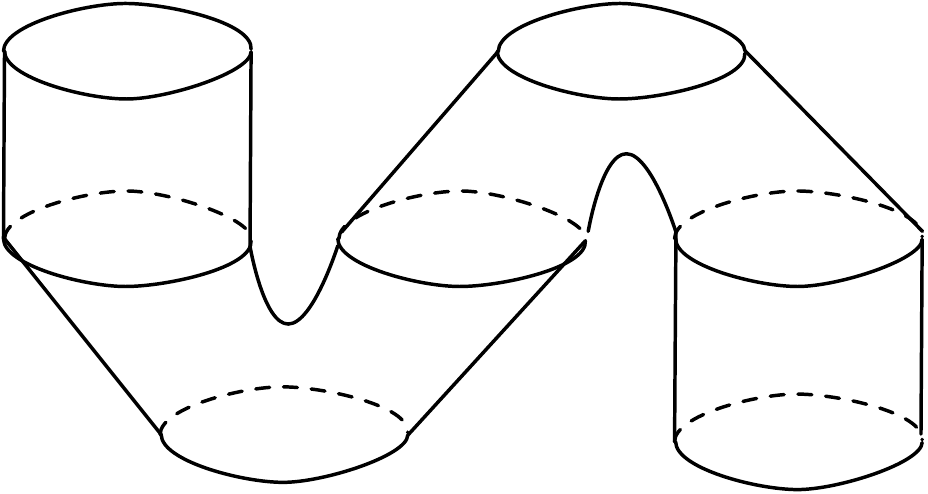}
\]
\caption{One possible topology change. \label{fig:st}}
\end{figure}

\begin{figure}
\[
\incc{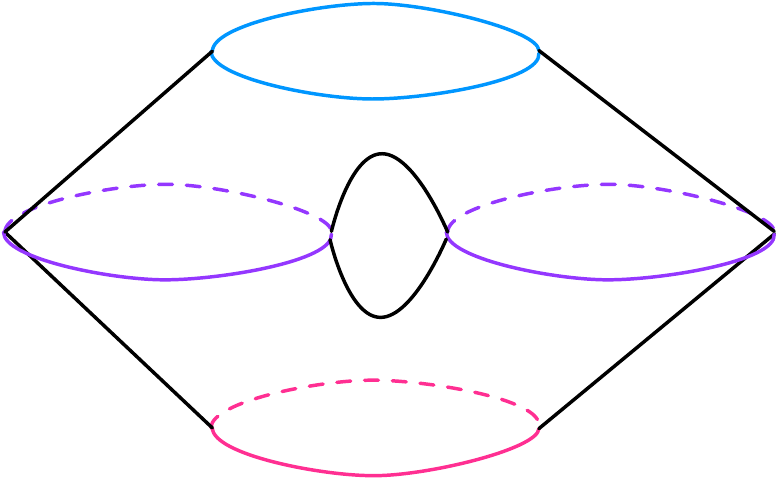}:\qquad \incc{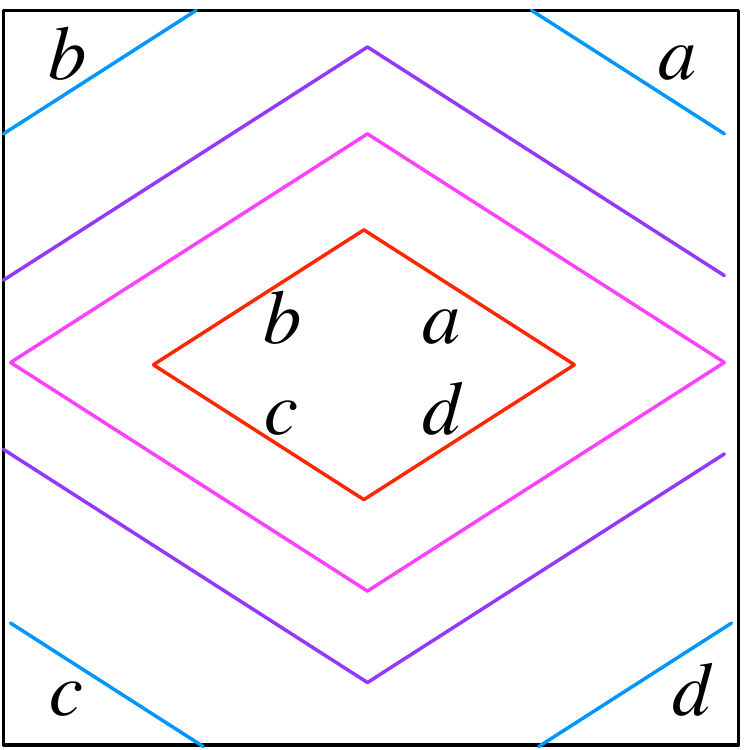} = \incc{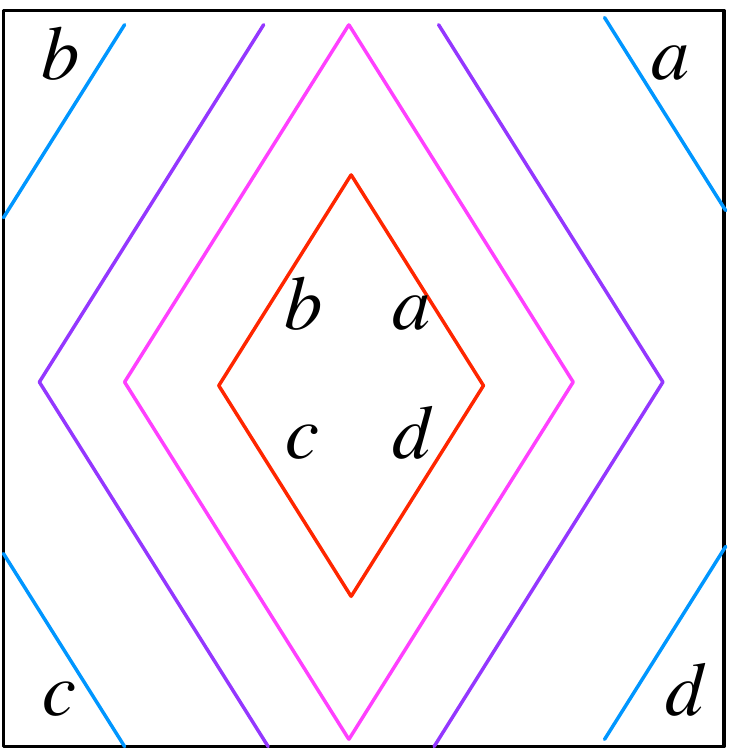}
\]
\caption{Another possible topology change concerns a torus with two holes. 
On the two figures on the right, the time flows from inside to the outside,
and the parallel edges of the boundary need to be identified to form a torus.
On one side, the intermediate two circles are along the A-cycle,
and on the other side, they are along the B-cycle.
 \label{fig:torus}}
\end{figure}

All possible types of the topology changes were enumerated carefully in Appendix A of \cite{Moore:2006dw}. 
The cases are the following and their adjoints:
\begin{itemize}
\item[1.] The creation or the annihilation of two critical points as shown in Fig.~\ref{fig:unitT}, or
\item[2a.] The exchange of two critical points as shown in  
 Fig.~\ref{fig:assoc}, which we already encountered, or
\item[2b.] the situation Fig.~\ref{fig:st} where  the number of intermediate circles changes from one to three, or
\item[2c.] the situation
 Fig.~\ref{fig:torus} where  the A-cycle and the B-cycle of a torus is exchanged. 
 In more detail, on one side, a circle consisting of segments $a,b,c,d$ in this order splits to two circles consisting of $a,b$ and $c,d$, which are now along the A-cycle. They then merge into a circle consisting of four segments with the order $b,a,d,c$.
On the other side, the two circles in the intermediate stage consists of segments $b,c$ and $d,a$, and are along the B-cycle. 
 \end{itemize}

The invariance of $Z_\Sigma$ under the change 1 is the unit property itself, and
the invariance under the change 2a is the associativity itself.
The invariance under the change 2b can be reduced to associativity by using the cyclic invariance of the product, shown in Fig.~\ref{fig:elephant}.
Finally, under the topology change 3b, the map $Z_\Sigma$ is trivially invariant.

In total, we have shown that a 2d TFT with no symmetry is completely defined by a vector space $V$ with the four maps $\co{I},I,M,\co{M}$ with the conditions described above.
Such a vector space is known as a \emph{commutative  Frobenius algebra} $V$.

\subsection{TFT with $\cC$ symmetry on a cylinder}
\label{sec:TFTcyl}
Let us now move on to the discussion of TFTs with symmetry given by a symmetry category $\cC$.
In this subsection we start with the simplest geometry, namely cylinders.
We already discussed basics in Sec.~\ref{sec:background}.
As mentioned there, we choose a \emph{base point} along each  constant-time cicle, and call its trajectory the \emph{auxiliary line}.

\begin{figure}
\[
\incb{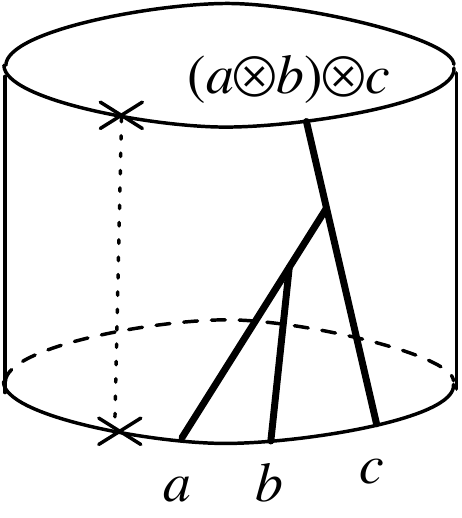}
\]
\caption{The Hilbert space of a circle with multiple line operators are identified with the Hilbert space of a circle with the fused line operator.\label{fig:mergedef}}
\end{figure}

\begin{figure}
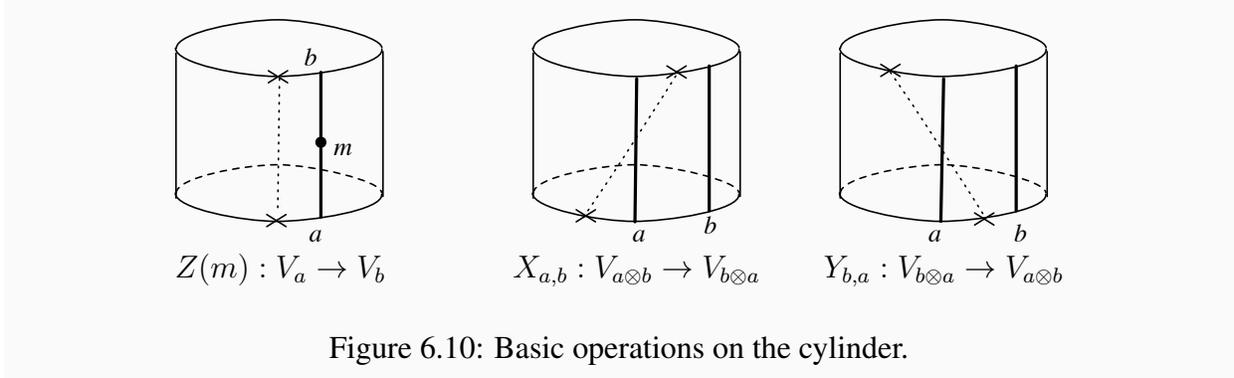

\[
\begin{array}{c@{\qquad\qquad}c@{\qquad}c}
\incb{Zm}&\incb{X} & \incb{Xinv} \\
Z(m):V_a\to V_b&X_{a,b}:V_{a\otimes b}\to V_{b\otimes a} & Y_{b,a}:V_{b\otimes a}\to V_{a\otimes b}
\end{array}
\]
\caption{Basic operations on the cylinder.\label{fig:basic-cylinder}}
\end{figure}

\paragraph{Basic ingredients:}
We first associate the Hibert space $V_a$ for a circle with a single insertion of a line labeled by $a\in \cC$. We require $V_{a\oplus b}=V_a\oplus V_b$.
We now associate a Hilbert space $V_{a,b,c,\ldots}$ for a circle with insertions of transverse lines $a$, $b$, $c$, \ldots\ 
by fusing them in a fixed particular order, starting from the closest line on the right of the base point and then toward the right : \begin{equation}
V_{a,b,c,\ldots}:=V_{(\cdots(( a\otimes b)\otimes c)\cdots )}.
\end{equation}
The case with three lines is shown in  Fig.~\ref{fig:mergedef}.

We have two basic operations we can perform on the cylinder, see Fig.~\ref{fig:basic-cylinder}.
One is to insert a morphism $m:a\to b$, which defines an operator $Z(m):V_a\to V_b$.
Another is to move the base point to the right and to the left, which defines morphisms $X_{a,b}:V_{a\otimes b}\to V_{b\otimes a}$ and $Y_{b,a}:V_{b\otimes a}\to V_{a\otimes b}$.

\paragraph{Assignment of a map to a given network:}
With these basic operations, we can assign \emph{a} map $V_{a,b,\ldots} \to V_{c,d,\ldots}$ for a cylinder equipped with an arbitrary network of lines and morphisms from the symmetry category $\cC$, where  an incoming circle have insertions $a$, $b$, \ldots\ 
and an outgoing circle have  insertions $c$, $d$, \ldots.

We choose a time function $t$ on it, and we call any time $t_i$ a \emph{critical point} when either of the following happens: i) there is an insertion of a morphism on a line, ii) there is a fusion of two lines $a,b$ into one line $a\otimes b$ or vice versa, and iii) a line crosses an auxiliary line.
 Note that we do not allow the auxiliary line to bend backward in time, as part of the definition.

We order $0=t_0<t_1<\cdots < t_{p-1} < t_p=1$ so that the incoming circle is at $t=0$
and the outgoing circle is at $t=1$.
Each critical point of type i) gives a factor of $Z(m)$,
that of type ii) gives a factor of $Z(\alpha)$ where $\alpha$ is an appropriate associator,
and that of type iii) gives a factor of $X$ or $Y$.
Then we define the map $V_{a,b,\cdots} \to V_{c,d,\cdots}$ associated with this time function $t$ to be the composition of factors corresponding to these critical points.

\begin{figure}
\begin{align*}
\text{If}& \quad\incb{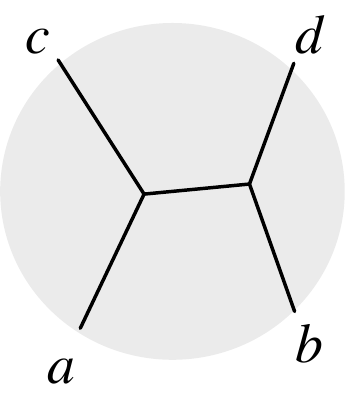}=\incb{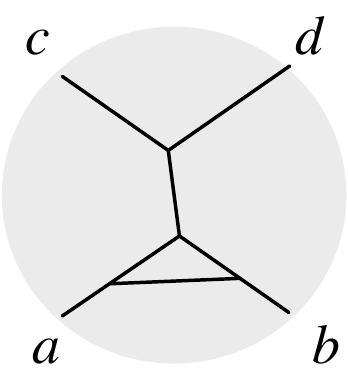}: a\otimes b\to c\otimes d\\
\text{then}& \quad\incb{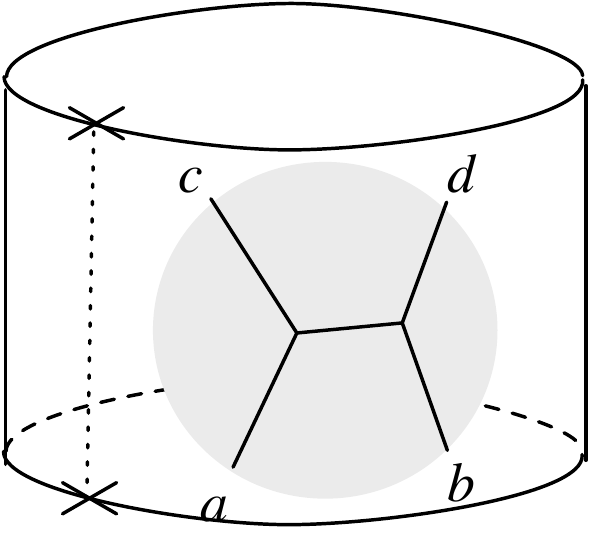}=\incb{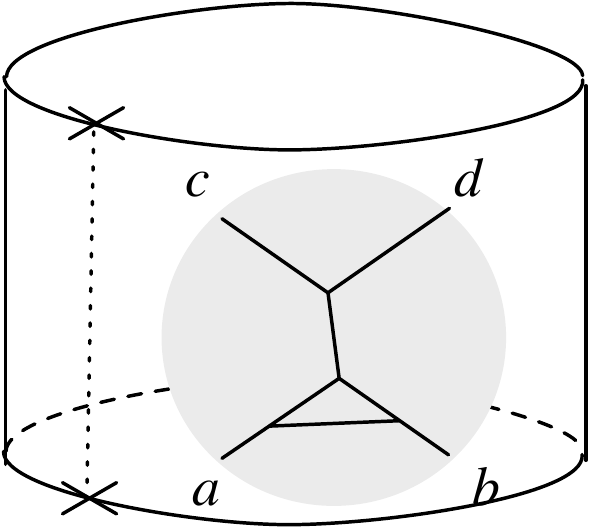}\\
\text{and} & \quad \incb{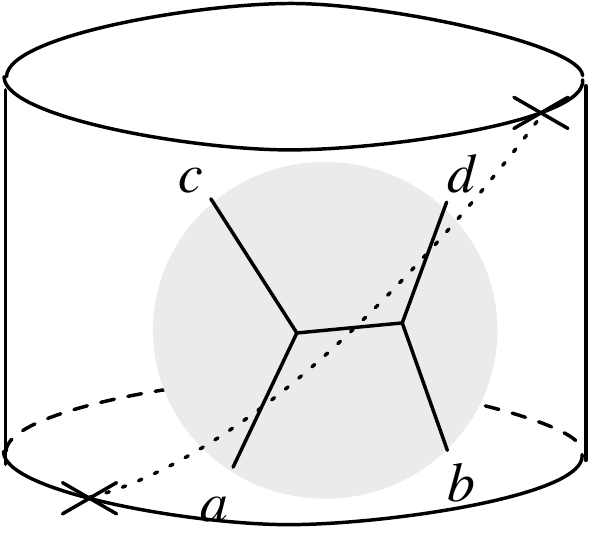}=\incb{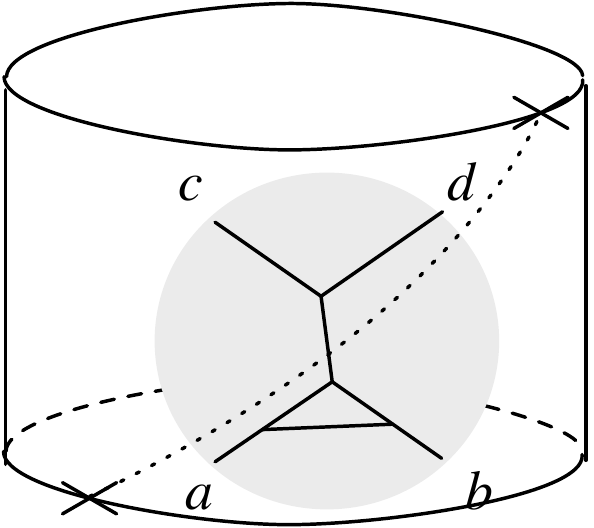}
\end{align*}
\caption{A local change in the network should not affect the map on the Hilbert space if the two subnetworks  give the same morphism.\label{fig:same}}
\end{figure}

\paragraph{Consistency of the assignment:}
We now need to show that this assignment is consistent. 
There are three types of changes under which the assignment needs to be constant, namely
\begin{itemize}
\item the change of the time function $t$,
\item the change of the positions of the auxiliary line, and
\item the change of the network in a disk region that does not change the morphism within it.
\end{itemize}
The third point might need some clarification. 
In the symmetry category $\cC$, a topologically different network can correspond to the same morphism.
Then we need to ensure that if we replace a subnetwork on a cylinder accordingly, the resulting map on the Hilbert space should also be the same, see Fig.~\ref{fig:same}
This is not just a change in the time function, therefore we need to guarantee the invariance separately.

The auxiliary line might cut though the subdiagram, as also shown in Fig.~\ref{fig:same},
but this does not have to be treated separately, since we can first move the auxiliary line outside of the disk region, assuming that it is shown that the auxiliary lines can be moved.

\begin{figure}
\[
\incb{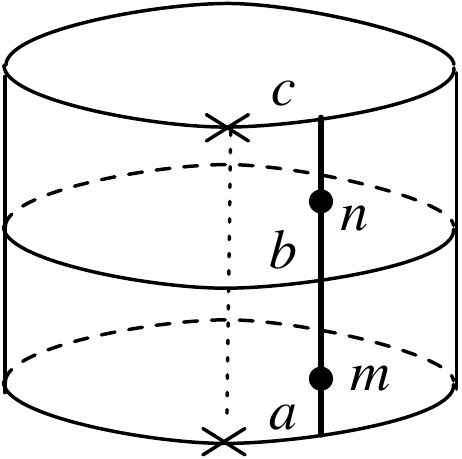}=\incb{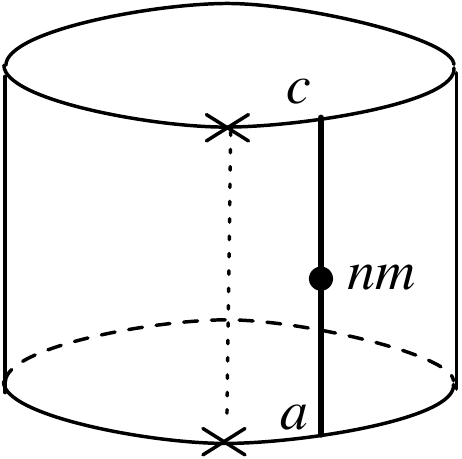}
\]
\caption{Two morphisms can be combined. \label{fig:fusemorph}}
\end{figure}

Then this third type of change can be just taken care of by assuming that 
we can fuse two local operators, leading to the following constraint, see Fig.~\ref{fig:fusemorph}:
\be
Z(n) Z(m)=Z(n\circ m).\label{Zfuse}
\ee

\begin{figure}
\[
\incb{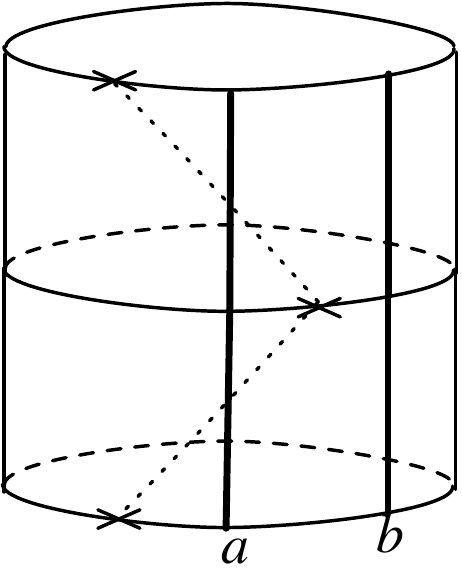}=\incb{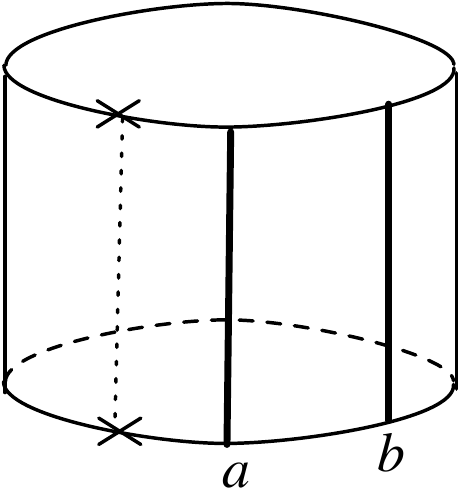}
\]
\caption{Moving the auxiliary line back and forth should do nothing. \label{fig:motion}}
\end{figure}

\begin{figure}
\[
\incb{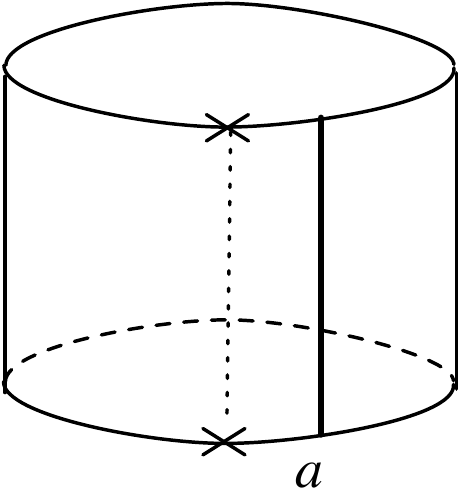}=\incb{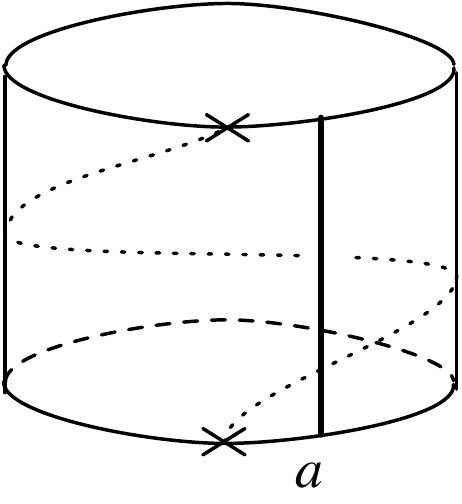}
\]
\caption{Winding the auxiliary line all the way around, represented by $X_{a,1}$, should not do anything. \label{fig:windaround}}
\end{figure}

Next, let us take care of the second type of change, where we move the auxiliary lines keeping the network and the time function fixed. 
First, moving the auxiliary line back and forth in succession should not do anything, so we have
\begin{equation}
X_{a,b}=Y_{b,a}^{-1},
\end{equation}
see Fig.~\ref{fig:motion}.
Rotating the base point all the way around should not do anything either, therefore we have \begin{equation}
X_{a,1}=\id,\label{wind}
\end{equation}
see Fig.~\ref{fig:windaround}.

\begin{figure}
\[
\incb{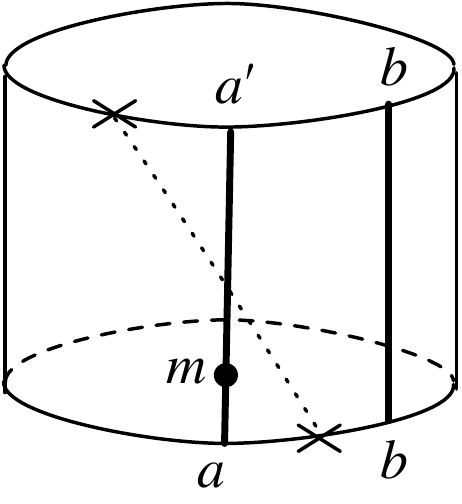}=\incb{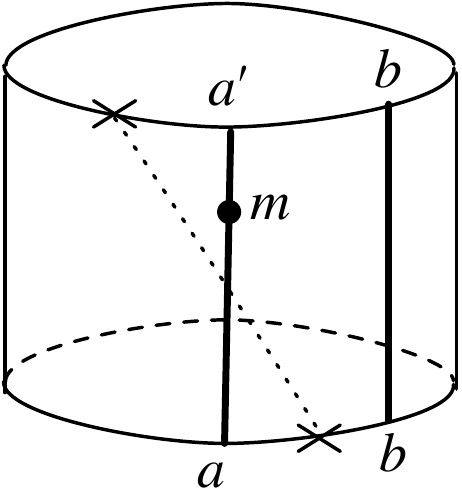};\qquad
\incb{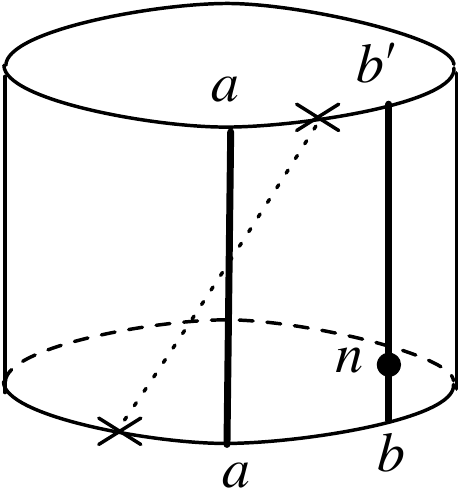}=\incb{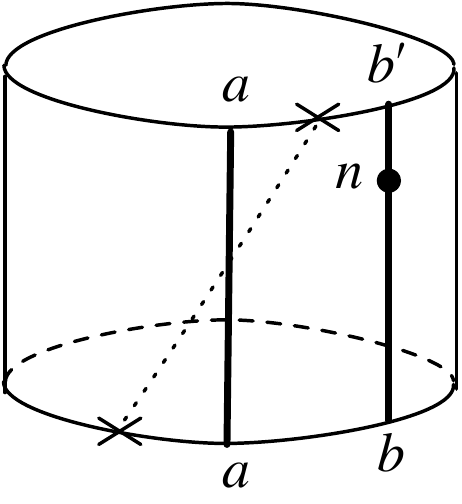}\qquad
\]
\caption{Morphisms can be crossed across the auxiliary line.\label{fig:movemorph}}
\end{figure}

\begin{figure}
\[
\incb{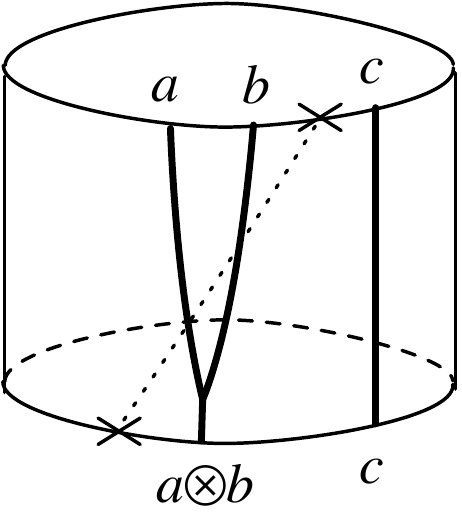}=\incb{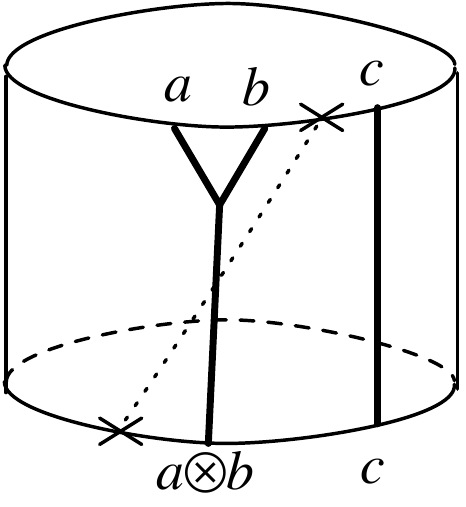}
\]
\caption{Crossing the fused line over the auxiliary line should be the same with crossing two lines separately.\label{fig:foobar}}
\end{figure}

Then we should be able to move the morphisms across the auxiliary line, leading to two relations, as illustrated in Fig.~\ref{fig:movemorph}:
\begin{align}
X_{a',b} Z(m\otimes 1)&=Z(1\otimes m) X_{a,b} \label{actaux},\\
X_{a,b'} Z(1\otimes n)&=Z(n\otimes 1) X_{a,b} \label{auxact}
\end{align} for $m:a\to a'$ and $n:b\to b'$.
We can also fuse two lines before crossing the auxiliary line, see Fig.~\ref{fig:foobar}.
This leads to the constraint
\begin{equation}
X_{b,c\otimes a} Z(\alpha_{b,c,a}) X_{a,b\otimes c} Z(\alpha_{a,b,c})=Z(\alpha_{c,a,b}{}^{-1}) X_{a\otimes b,c}. \label{auxaux}
\end{equation}

Finally, on the cylinder, the change in the time function itself does not do much, and possible changes are already all covered.
Thus, we see that to define a consistent TFT with $\cC$ symmetry on a cylinder, we need the data of an  additive functor $Z:\cC\to\Vec$ with morphisms  $X_{a,b}:V_{a\otimes b}\simeq V_{b\otimes a}$  satisfying (\ref{wind}),  (\ref{actaux}), (\ref{auxact}) and (\ref{auxaux}).

\paragraph{Generalized associators on the cylinder:}
The relations so far guarantees that we can always move the base point and change the order of the tensoring of lines in a consistent manner. .
For example, the relation \eqref{auxaux} means that there is a single well-defined isomorphism between $V_{(a\otimes b)\otimes c}$ and $V_{(c\otimes a)\otimes b}$.
We introduce a notation \begin{equation}
\cA_{(a\otimes b)\otimes c \to (c\otimes a)\otimes b} : V_{(a\otimes b)\otimes c} \to V_{(c\otimes a)\otimes b}
\end{equation} for it, and call it a generalized associator on the cylinder.
We similarly introduce generalized associators for an arbitrary motion of the base point and an arbitrary rearrangement of parentheses. 
Each such generalized associator have multiple distinct-looking expressions in terms of sequences of $Z(\alpha)$, $X$ and $X^{-1}$, but they give rise to the same isomorphism.

\subsection{TFT with $\cC$ symmetry on a general geometry}

\begin{figure}
\[
\begin{array}{c|c|c|c}
\incc{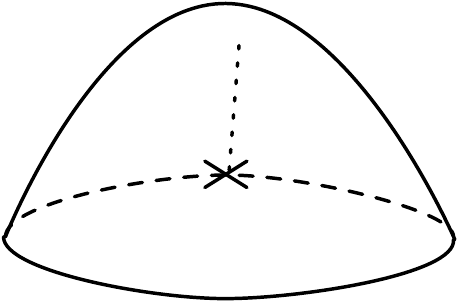} & \incc{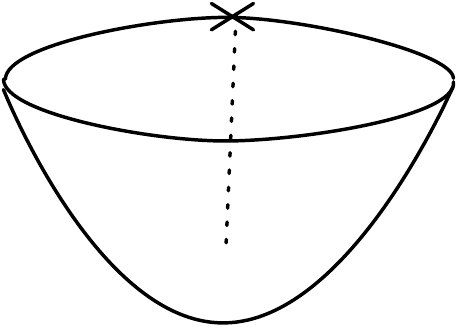}&\incc{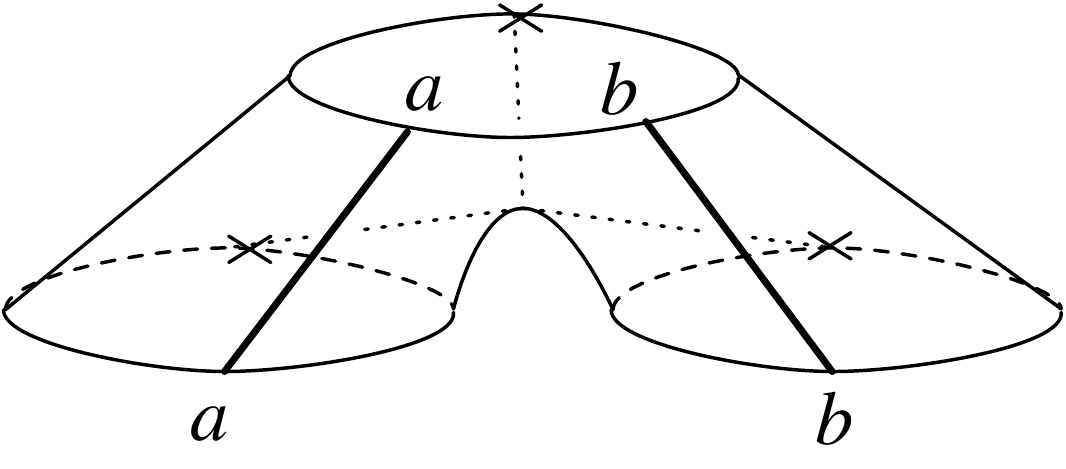}&\incc{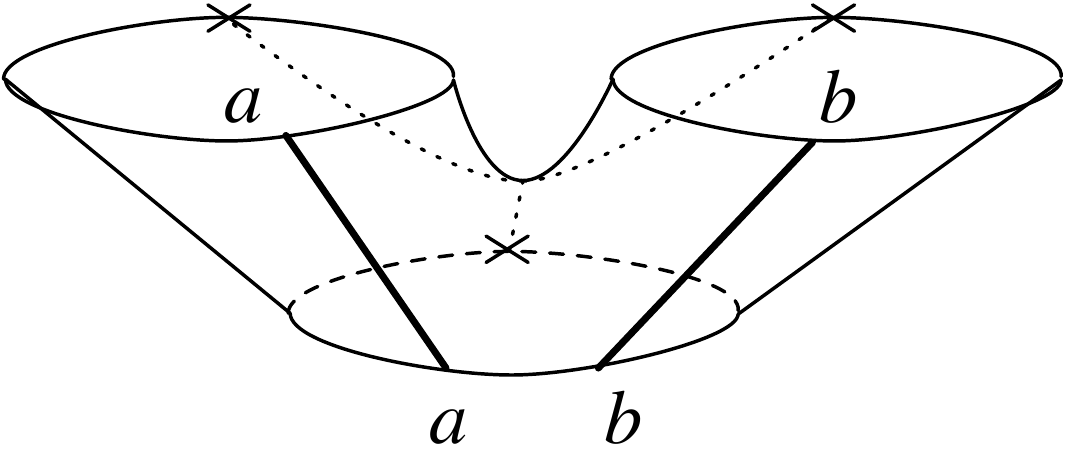}\\
\co{I}:V_1\to \bC& I:\bC\to V_1 & M_{a,b}:V_a\otimes V_b\to V_{a\otimes b} & \co{M}_{a,b}:V_{a\otimes b}\to V_{a}\otimes V_b
\end{array}
\]
\caption{The basic building blocks of a 2d TFT with $\cC$ symmetry.\label{fig:Cbasic}}
\end{figure}

\paragraph{Basic data:}
Let us discuss now the TFT with $\cC$ symmetry on a general geometry.
The four basic geometries are given in Fig.~\ref{fig:Cbasic}.
For a pair of pants,
we need to join the two auxiliary lines coming from each leg into a single auxiliary line. We take the point where this happens to coincide with the critical point where two circles join to form a single circle. 
In what follows, we will refer to the initial two legs of a pair of pants as the initial legs and the final leg as the product leg.

We can now associate to any geometry $\Sigma$ with $m$ initial legs and $n$ final legs with an arbitrarily complicated network of lines and morphisms from $\cC$ a linear map as follows.
We first choose a time function $t:\Sigma\to [0,1]$.
We call a time value $t_i$ critical when any of the following happens: i) the topology of the constant time slice change, ii) there is a morphism, or iii) a line crosses a auxiliary line.
We order the critical times so that $0=t_0<t_1<t_2<\cdots<t_p=1$. 
We cut $\Sigma$ once in each interval $(t_i,t_{i+1})$, and associate to each critical time $t_i$ 
one of the basic linear maps.
We then compose them.
We now need to guarantee that this assignment is consistent. 

\begin{figure}
\[
\begin{array}{c|c}
\incc{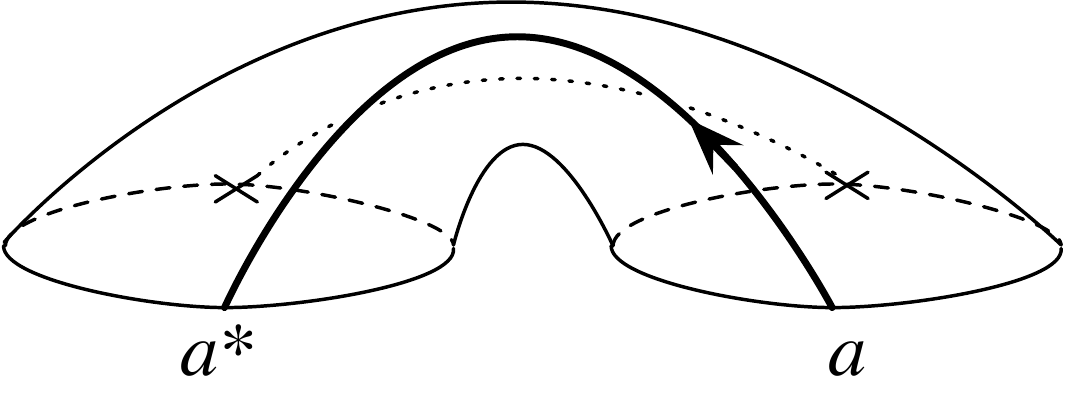} & \incc{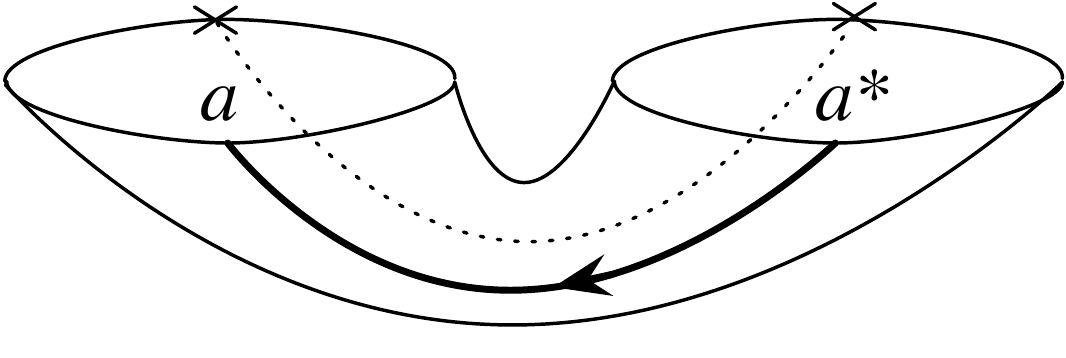}
\end{array}
\]
\caption{The pairing of $V_a$ and $V_{a^*}$.\label{fig:Xpairing}}
\end{figure}

\paragraph{Basic consistency conditions:}
Let us first enumerate basic consistency conditions. 
First, we  define the pairing of $V_a$ and $V_{a^*}$ as in Fig.~\ref{fig:Xpairing}:
\begin{align}
\co{I}Z(\ev_a^L)M_{a^*,a}&: V_{a^*}\otimes V_{a} \to \bC,\\
\co{M}_{a,a^*} Z(\coev_a^L) I &: \bC \to V_{a} \otimes V_{a^*}.
\end{align}
Then we require that \begin{equation}
\text{\emph{this pairing is non-degenerate and can be used to identify $(V_a)^* \simeq V_{a^*}$.}}
\label{Xpairing}
\end{equation}
Under this pairing, the product $M_{a,b}$ and the coproduct $\co{M}_{b^*,a^*}$ are adjoint, etc.
This again allows us to reduce the number of cases need to be mentioned below roughly by half.

Before proceeding, we note that we used $\ev_a^L$, $\coev_a^R$ to define the pairing.
We can also use $\ev_{a^*}^R$, $\coev_{a^*}^R$ to define a slightly different pairing. 
Exactly which pairing to be used in each situation can be determined by fully assigning orientations to every line involved in the diagram. 
Below, we assume that every line carries an upward orientation, unless otherwise marked in the figure.

\begin{figure}
\[
\incc{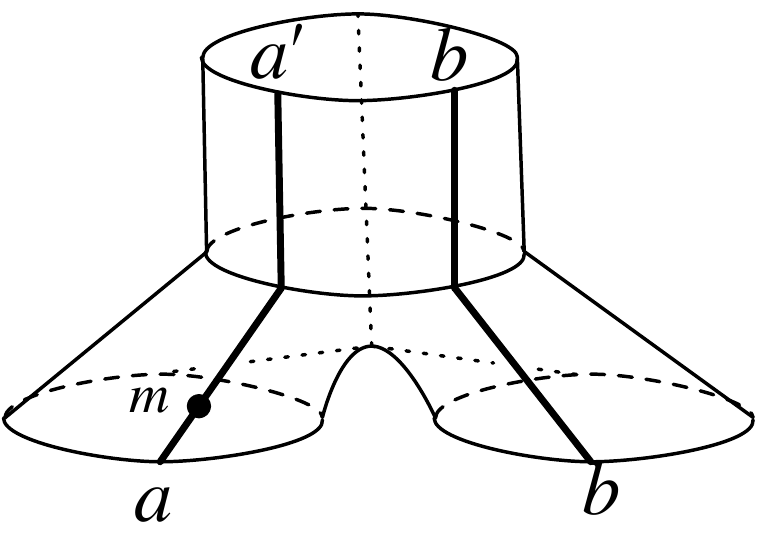}=\incc{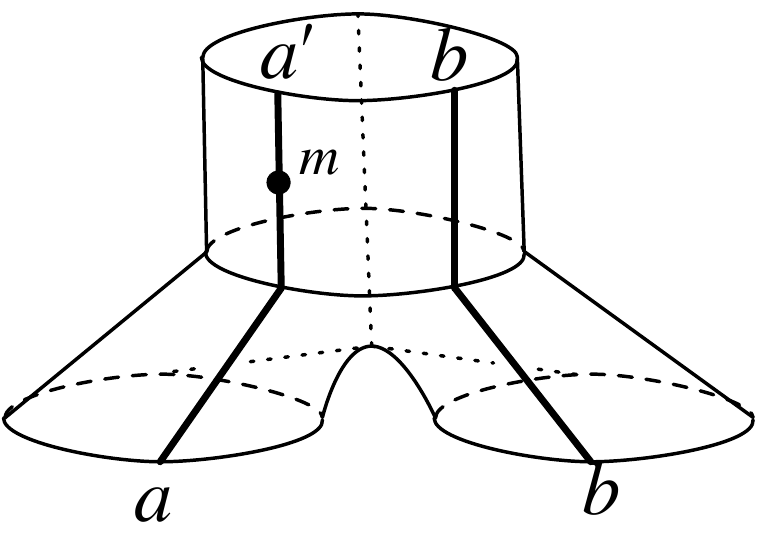}
\]
\caption{A morphism can be moved across the product.\label{fig:morphmove}}
\end{figure}

\begin{figure}
\[
\incc{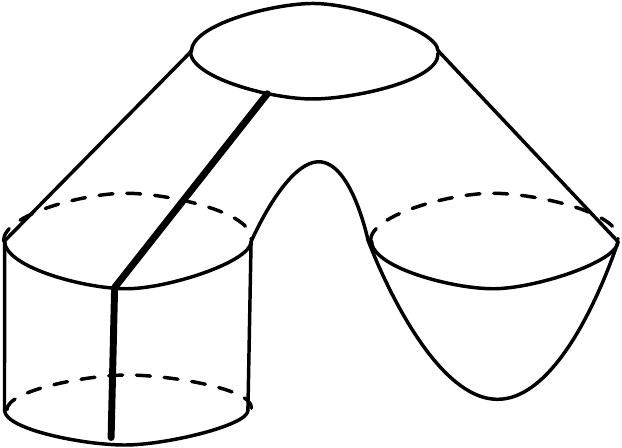}=\incc{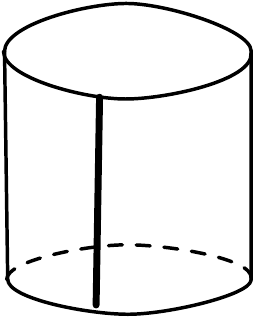}
\]
\caption{$I$ is a unit of the product $M_{a,1}$.\label{fig:XunitT}}
\end{figure}

\begin{figure}
\[
\incc{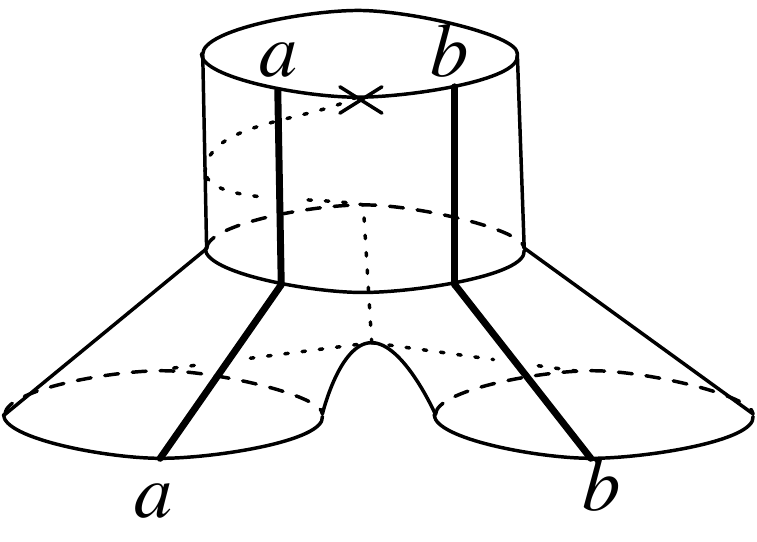}=\incc{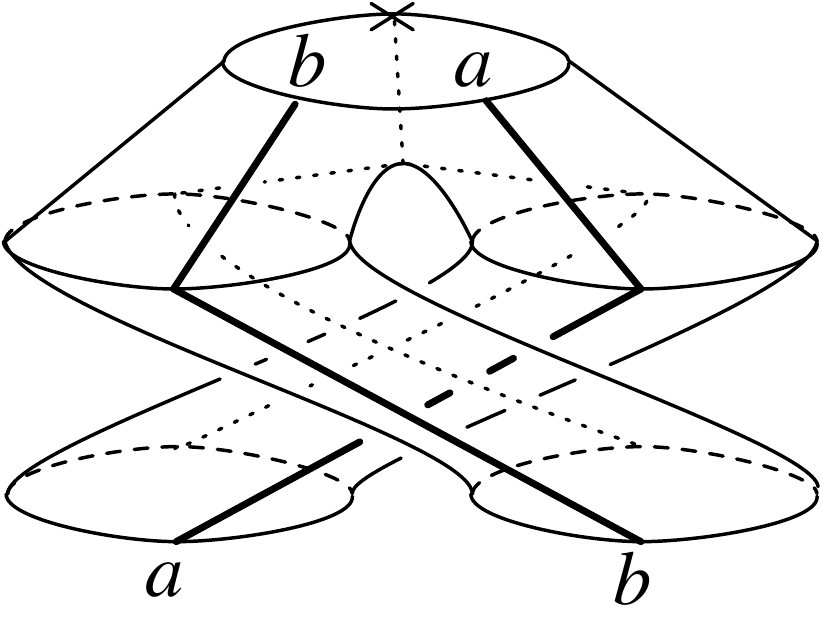}
\]
\caption{The product $M$ is twisted commutative.\label{fig:Xcomm}}
\end{figure}

\begin{figure}
\[
\incc{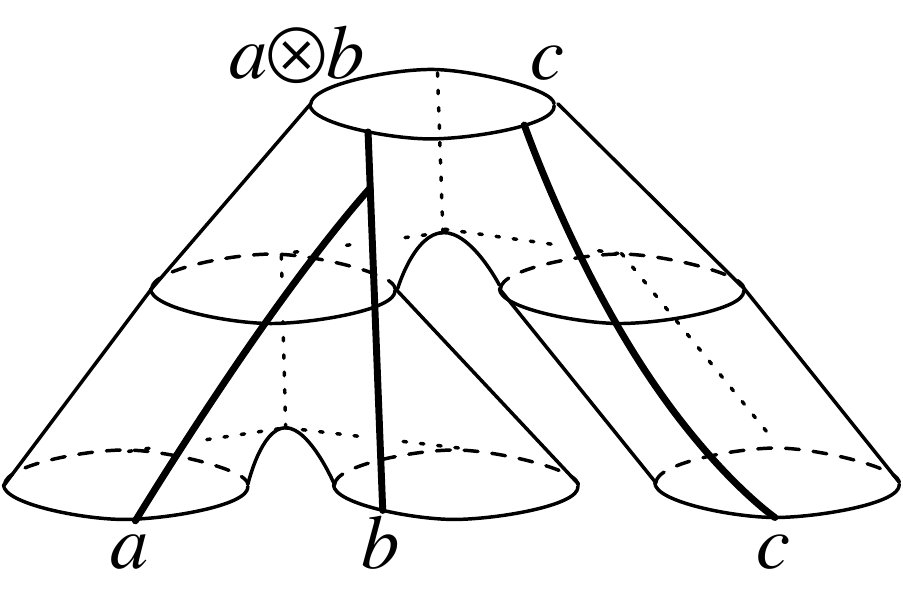} \stackrel{\text{up to assoc.}}{=} \incc{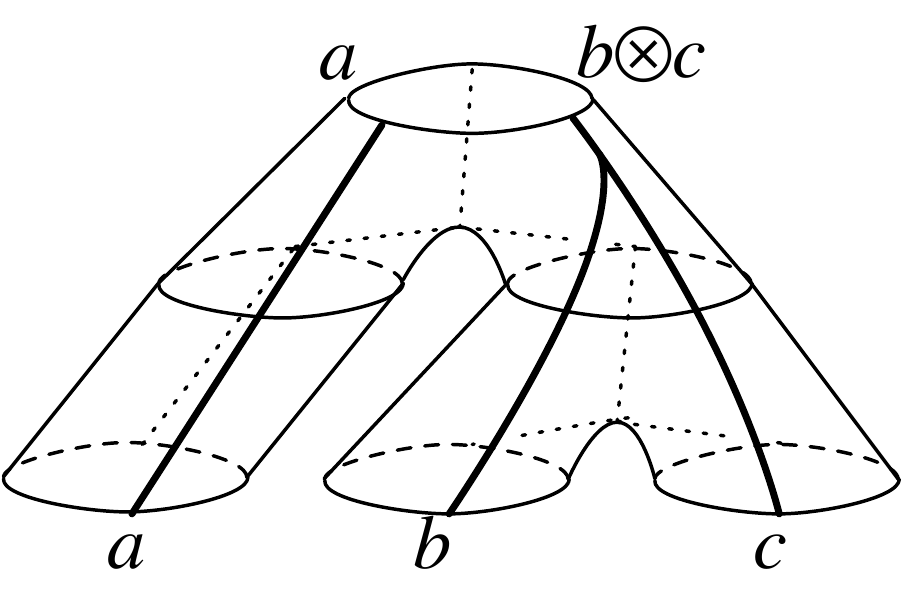}
\]
\caption{The product $M$ is associative up to the associator.\label{fig:Xassoc}}
\end{figure}

Second, a morphism can be moved across the product, see Fig.~\ref{fig:morphmove}: \begin{equation}
M_{a,b} (Z(m)\otimes \id_{V_b}) = Z(m\otimes \id_b) M_{a,b}.\label{morphmove}
\end{equation}
Third, the map $I$ defined by the bowl geometry gives the unit, see Fig.~\ref{fig:XunitT}: 
\begin{equation}
M_{a,1} (v\otimes I) = v, \qquad v\in V_a. \label{XunitT}
\end{equation}
Fourth, it is twisted commutative: \begin{equation}
X_{a,b} M_{a,b} ( v\otimes w) = M_{b,a} (w\otimes v) ,\qquad v\in V_a,\quad w\in V_b
\label{Xcomm}
\end{equation} as illustrated in Fig.~\ref{fig:Xcomm}.
Fifth, it is associative up to the associator: \begin{equation}
Z(\alpha_{a,b,c})  M_{a\otimes b,c} (M_{a,b}\otimes \id_c) = 
M_{a,b\otimes c} (\id_a\otimes M_{b,c}),
\label{Xassoc}
\end{equation}
as shown in Fig.~\ref{fig:Xassoc}.

\begin{figure}
\[
\incc{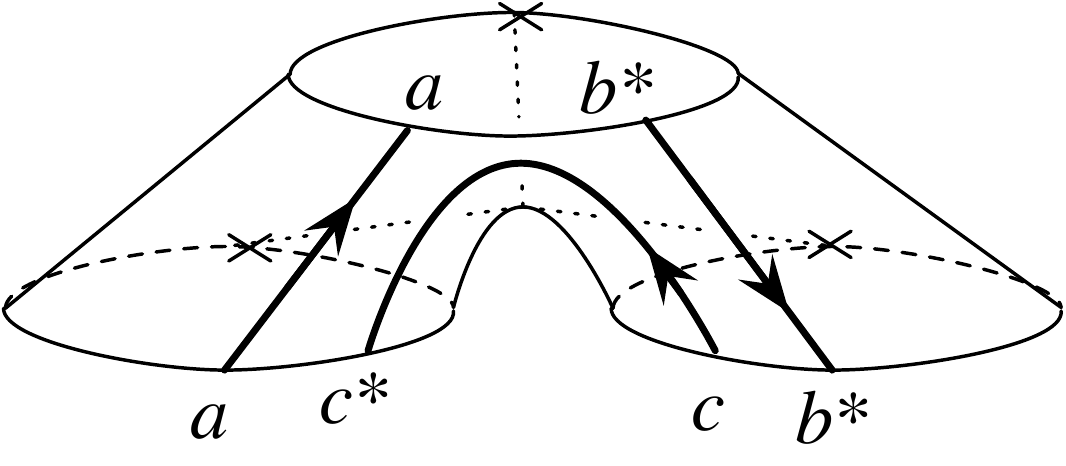}=\incc{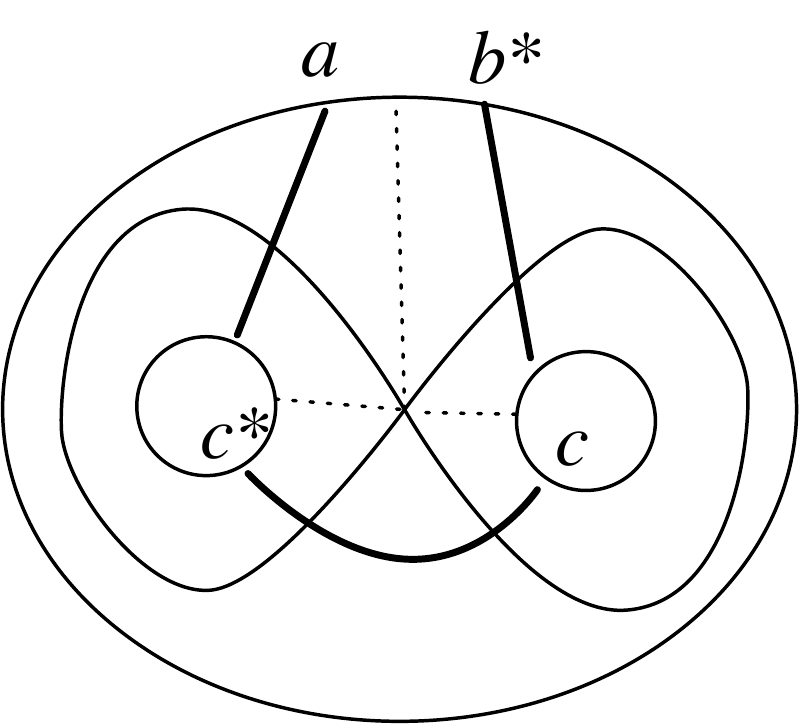}
\]
\caption{A slightly generalized version of the product operation. In the figure on the right, the time flows from inside to the outside.\label{fig:genpro}}
\end{figure}

Sixth, we want to formulate that the product is  symmetric under the cyclic permutation of three circles. 
To do this we first introduce a slightly generalized form of the product shown in Fig.~\ref{fig:genpro}:
\begin{equation}
M_{(a\otimes c^*,c\otimes b^* )\to a\otimes b^*}: V_{a\otimes c^*} \otimes V_{c\otimes b^*} \to V_{a\otimes b^*}
\end{equation} given by \begin{equation}
M_{(a\otimes c^*,c\otimes b^*) \to a\otimes b^*} 
= Z(\id_a \otimes \ev_{c}^L \otimes \id_{b^*} )
\cA_{(a\otimes c^*)\otimes (c \otimes b^*) \to a\otimes (c^*\otimes c) \otimes b^*} M_{a\otimes c^*,c\otimes b^*}\label{gen1}
\end{equation}
where $\ev_{c}^L: c^*\otimes c\to 1$ is the evaluation morphism and
$\cA$ is the generalized associator introduced in the last subsection.

\begin{figure}
\[
\incc{gen2}=\incc{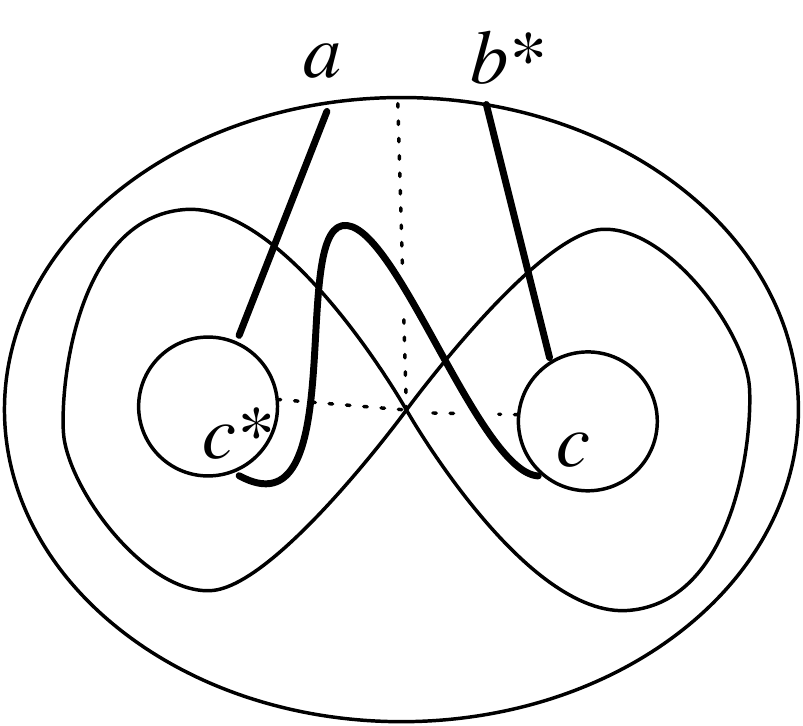}
\]
\caption{An alternative definition of the generalized product. The time flows from the inside to the outside.\label{fig:alt}}
\end{figure}

The generalized product has an alternative definition as given in Fig.~\ref{fig:alt}, where the line $c$ crosses three auxiliary lines. This gives an alternative expression\begin{multline}
M_{(a\otimes c^*,c\otimes b^*) \to a\otimes b^*} 
=\\
Z(\id_{a\otimes b^*} \otimes \ev_c^R)
\cA_{(c^*\otimes a)\otimes( b^* \otimes c) \to (a\otimes b^*) \otimes(c \otimes c^*) }
M_{c^*\otimes a, b^* \otimes c} (X_{a,c^*}\otimes X_{c,b^*})\label{gen2}
\end{multline}
and we demand \begin{equation}
\text{\emph{the right hand sides of the equations \eqref{gen1} and \eqref{gen2} are the same.}}
\label{mess}
\end{equation}

\begin{figure}
\[
\incc{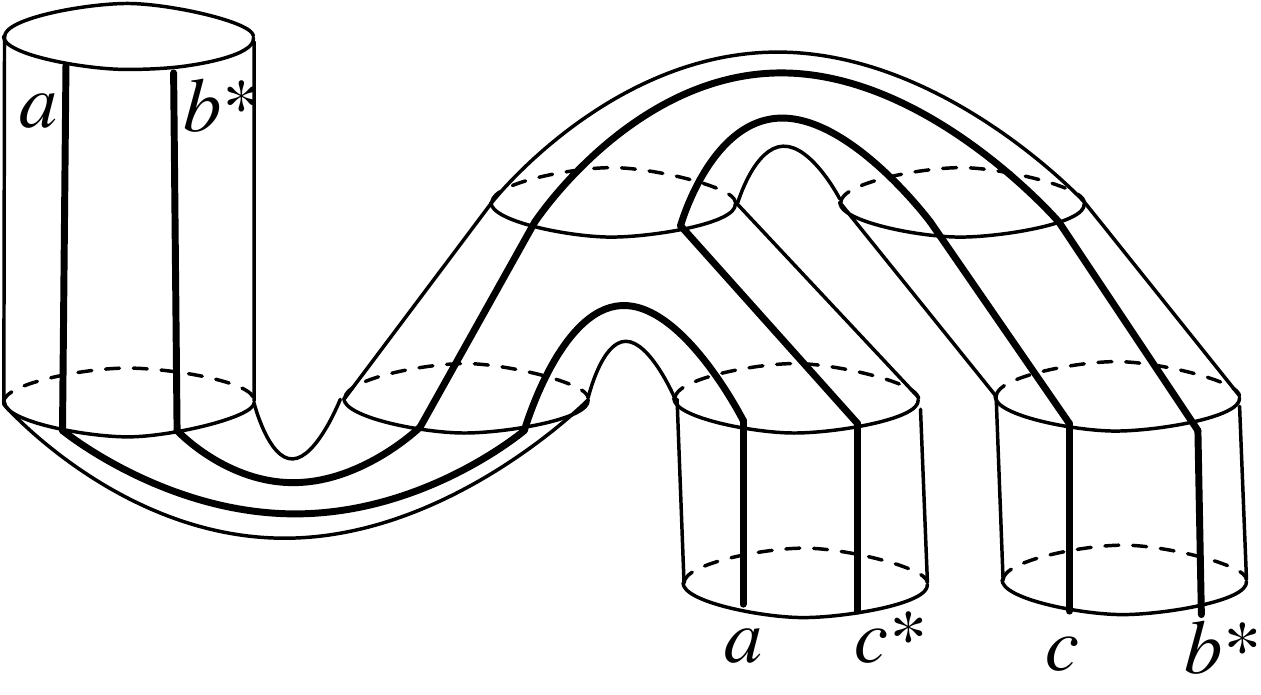}=\incc{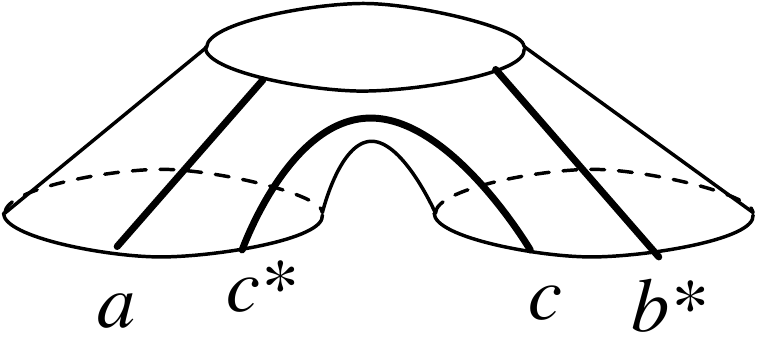}
\]
\caption{The product is invariant under exchanging an incoming circle and an outgoing circle.\label{fig:Xelephant}}
\end{figure}

We can now formulate the cyclic symmetry of the product: \begin{equation}
\text{\emph{$M_{(a\otimes c^*,c\otimes b^*) \to a\otimes b^*}$ 
and $M_{(c\otimes b^*, b\otimes a^*) \to c\otimes a^*}$ 
are related by the inner products,}}\label{Xelephant}
\end{equation} see Fig.~\ref{fig:Xelephant}.
We can in fact derive this relation for general $a$, $b$, $c$ just from the subcase when $c=1$ and the relations already mentioned.
We keep the general case for cosmetic reasons, since it looks more symmetric.

\begin{figure}
\[
\begin{array}{c@{\qquad}c}
\incc{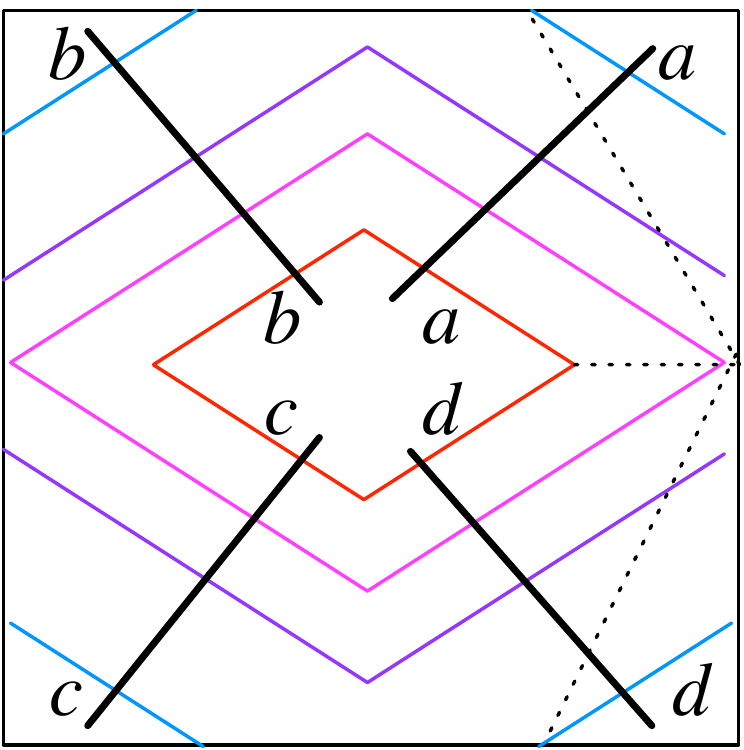} & \incc{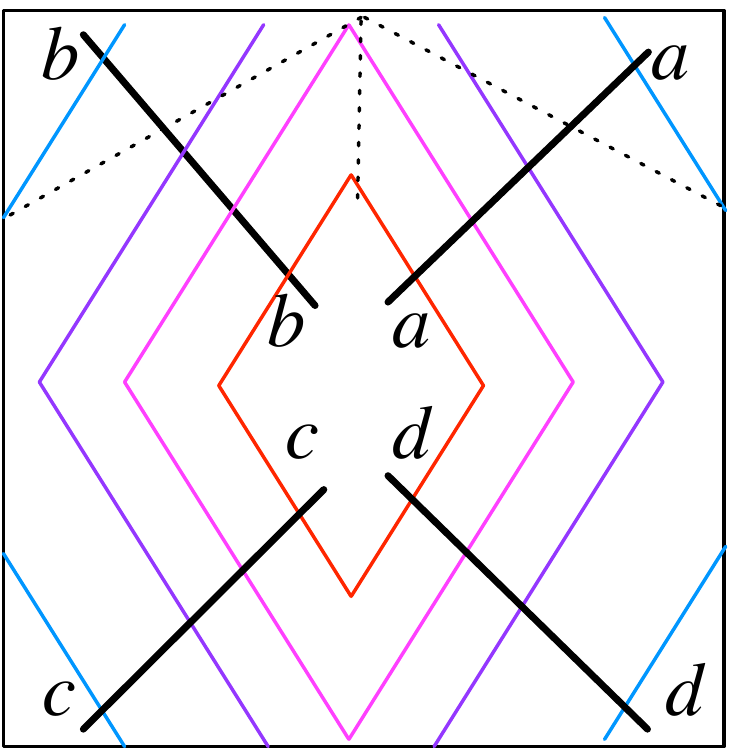}\\
V_{(a\otimes b)\otimes (c\otimes d)}\to V_{(b\otimes a)\otimes (d\otimes c)} & 
V_{(b\otimes c)\otimes (d\otimes a)}\to V_{(c\otimes b)\otimes (a\otimes d)} 
\end{array}
\]
\caption{Two ways a circle splits into two and then rejoins. They should be equal up to the action of $X$ and the associators.\label{fig:Xtorus}}
\end{figure}

Seventh, we need a consistency on the torus. An incoming circle consisting of four segments with lines $a$, $b$, $c$, $d$ can first split into two circles with two segments $a$, $b$ and $c$, $d$ each and then rejoins to form a circle with four segments in the order $b$, $a$, $d$, $c$;
another way this happens is that the two intermediate circles have segments $b$, $c$ and $d$, $a$,
see Fig.~\ref{fig:Xtorus}.
They each determine maps 
$(M\Delta)_{a,b;c,d}:V_{(a\otimes b)\otimes (c\otimes d)}\to V_{(b\otimes a)\otimes (d\otimes c)}$
and
$(M\Delta)_{b,c;d,a}:V_{(b\otimes c)\otimes (d\otimes a)}\to V_{(c\otimes b)\otimes (a\otimes d)} $ given by 
\begin{align}
(M\Delta)_{a,b;c,d} & :=
M_{b\otimes a,d\otimes c}
(X_{a,b}\otimes X_{c,d})
\Delta_{a\otimes b, c\otimes d},  \\
(M\Delta)_{b,c;d,a} & :=
M_{c\otimes b,a\otimes d}
(X_{b,c}\otimes X_{d,a})
\Delta_{b\otimes c, d\otimes a}. 
\end{align}
We then demand that they are equal up to the generalized associators: \begin{equation}
\cA_{(b\otimes a)\otimes (d\otimes c)\to (c\otimes b)\otimes (a\otimes d)}(M\Delta)_{a,b;c,d} 
=
(M\Delta)_{b,c;d,a}
\cA_{(a\otimes b)\otimes (c\otimes d) \to (b\otimes c)\otimes (d\otimes a)}.
\label{Xtorus}
\end{equation}

\paragraph{Consistency in the general case:}

We finally finished writing down basic moves. 
Now we can analyze the general moves. We again have three cases:
\begin{itemize}
\item the change of the time function $t$ on the surface $\Sigma$,
\item the change of the positions of the auxiliary line, and
\item the change of the network in a disk region that does not change the morphism.
\end{itemize}

Let us start by discussing the third case.
This is in fact automatic once the first two cases are taken care of, since any disk region can be put into a cylinder under a topological change, 
and then the auxiliary line can be moved off away from it. 
Then all we have to assume is that $Z(m)$ fuses appropriately, \eqref{Zfuse}.

The change in the position of the auxiliary line can happen in the following three ways:
\begin{itemize}
\item The auxiliary line can move within a single cylinder. This was already discussed in the last subsection.
\item When a circle with line $a$ and a circle with line $b$ join to form a circle, the order of $a$, $b$ and the base point $x$ in the product leg can either be $x$, $a$, $b$ or $a$, $b$, $x$.
The invariance under this is the twisted commutativity \eqref{Xcomm}.
\item  A line  $c\in \cC$ can cross the trivalent vertex of the auxiliary line. 
This move changes the number of the intersection of the line $c$ with the three auxiliary lines in one of the two ways, $0\leftrightarrow 3$ or $1\leftrightarrow 2$.
One example of the move $0\leftrightarrow 3$ is the equality \eqref{mess} of the two definitions \eqref{gen1} and \eqref{gen2} of the generalized product.
The move $1\leftrightarrow 2$ can be deduced by combining the twisted commutativity.
\end{itemize}

\begin{figure}
\[
\incc{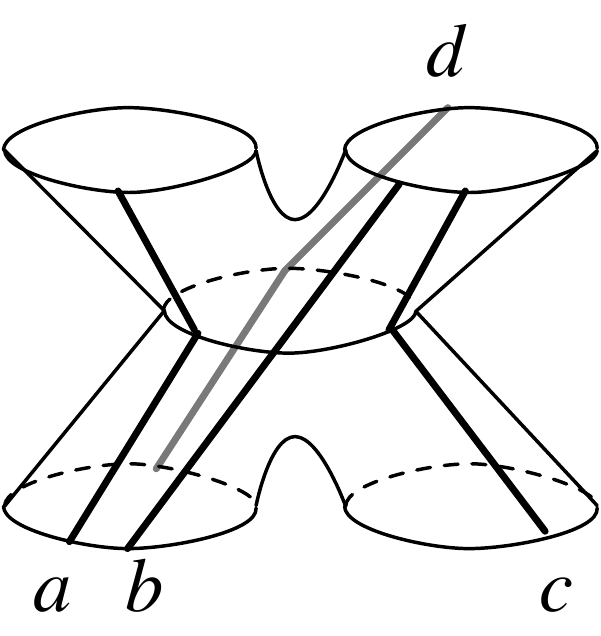}=\incc{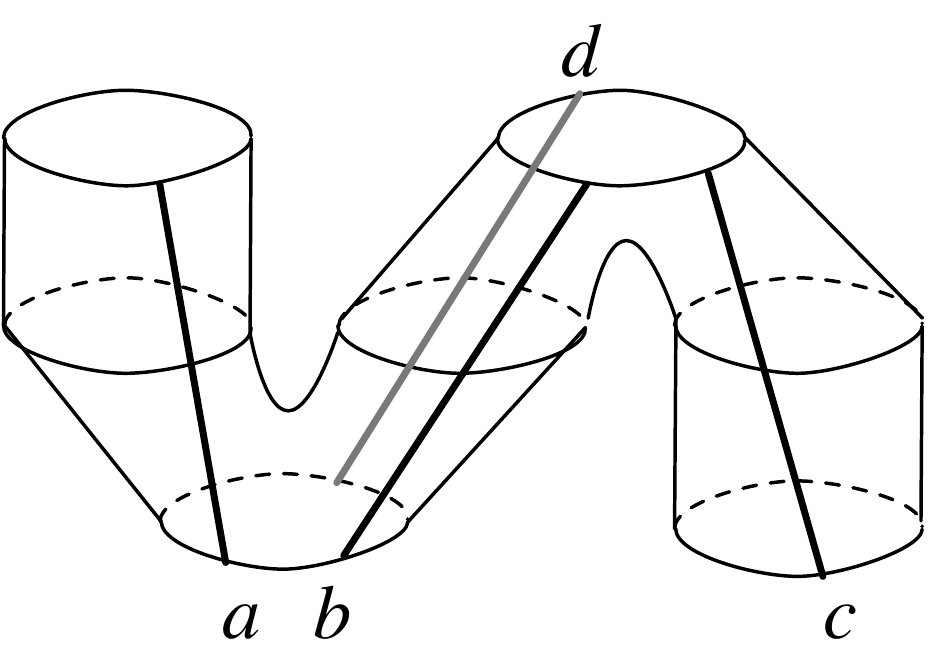}
\]
\caption{One possible topology change. The line $d$ is on the back side of the figures.\label{fig:Xst}}
\end{figure}

Finally we need to take care of the changes in the time function $t$.
One possible change is that 
a morphism and a product can happen in two different orders. The invariance under this move is \eqref{morphmove}.
Then there are topological changes in the cutting of the surface, which again comes in the following varieties:
\begin{itemize}
\item[1.] The creation or the annihilation of two critical points does Fig.~\ref{fig:XunitT}.
The consistency under this change is the unit property \eqref{XunitT}.
\item[2a.] The exchange of two critical points does 
 Fig.~\ref{fig:Xassoc}.
 The consistency under this change is the associativity \eqref{Xassoc}.
\item[2b.] The number of intermediate circles changes from one to three. One example is drawn in Fig.~\ref{fig:Xst}. 
The consistency under this change can be reduced to the cyclic symmetry of the generalized product \eqref{Xelephant}.
\item[2c.] How the torus is decomposed is changed as in 
 Fig.~\ref{fig:Xtorus}, for which we assigned a basic relation \eqref{Xtorus}.
 \end{itemize}

Summarizing, a TFT with $\cC$ symmetry is captured by the data $(V;Z,X;\co{I},I,M,\co{M})$
satisfying the various relations listed above. Namely, on the cylinder, we have \eqref{Zfuse}, \eqref{wind}, \eqref{actaux}, \eqref{auxact}, \eqref{auxaux}, and on the general geometry, we have in addition \eqref{Xpairing}, \eqref{morphmove}, \eqref{XunitT}, \eqref{Xcomm}, \eqref{Xassoc}, \eqref{mess}, \eqref{Xelephant}, and \eqref{Xtorus}, and finally, diagrams turned upside down correspond to adjoint linear maps.

\subsection{Gauged TFT with the dual symmetry}
Now we would like to discuss the definition of the TFT $T/A$ gauged by an algebra object $A$ in terms of the ungauged TFT $T$. 
We start from the data $(V;Z,X;\co{I},I,M,\co{M})$ for the original theory $T$.

The Hilbert space of the gauged theory was introduced in Sec.~\ref{sec:Hil}. See that section for some necessary background. 
We use the action by $A$ on $V_p$, depicted in Fig.~\ref{fig:proj}. This is given by 
\be
P:=U_{p,A}(\co{x}_R,x_L):=Z(x_L) X_{p,A} Z(\co{x}_R)
\ee
where $x_L:A\otimes p\to  p$ and $\co{x}_R:p \to p\otimes A$ are the morphisms defining the $A$-bimodule structure on $p$.
As we already discussed, $P$ turns out to be a projector, and we define $W_p$ to be the projection $P V_p$.
Now, we define the data $(W;\tilde Z,\tilde X,\co{\tilde I},\tilde I,\tilde M,\co{\tilde M})$ for $T/A$ in terms of corresponding data for $T$. 

\paragraph{The morphism map $\tilde Z$ :}
We define the new $\tilde Z$ to be the restriction of the old $Z$. 
We need to check that if the initial state lay in $W_p\subset V_p $ then the final state also necessarily lies in $W_q$, for a bimodule morphism $p\to q$.
This can be checked by gluing a cylinder on top of the final state which corresponds to the action of $A$. 
The wrapped $A$ line can then be taken across any bimodule morphism until it wraps the $p$ line at the start of the cobordism. 
But then the wrapped $A$ line has no effect and can be removed because the initial state we started with is invariant under the action of $A$.

\paragraph{The base point-change map $\tilde X$ :} 
We now want to define the new $\tilde X_{p,q}: W_{p\otimes_A q}\to W_{q\otimes_A p}$. We use $\proj$ and $\coproj$ (see Sec.~\ref{sec:bimod}) to define $\tilde X_{p,q}=Z(\proj) X_{p,q} Z(\coproj)$. This is well defined because a wrapped $A$ line at the end of the cobordism can be moved to an $A$ line propagating between $a$ and $b$ at the start of the cobordism which can be removed because of the definition of $\otimes_A$.

\paragraph{The unit and counit maps $\tilde I $ and $\co{\tilde I}$ :} 
Now the new unit morphism $\tilde I $ would be a map from $\bC$ to $W_A\subset V_A$. We define it as $\tilde I=Z(u)I$ where $u:1\to A$ is the unit morphism in the definition of $A$. Similarly, $\tilde {\co{I}}=\co{I} Z(v)$ where $v:A\to 1$ is the co-unit morphism in the definition of $A$. These are well-defined as can be shown by manipulations similar to those we are now going to perform for the definition of $\tilde M$.

\paragraph{The product and coproduct maps $\tilde M$ and $ \co{\tilde M}$ :}
\begin{figure}
\centering
\begin{tikzpicture}[line width=1pt]
\begin{scope}[every node/.style={sloped,allow upside down}]
\coordinate (C1) at (0,0);
\coordinate (mid1) at ($(C1)-(0,2)$);
\coordinate (A1) at ($(mid1)+(-1.2,-2)$);
\coordinate (B1) at ($(mid1)+(1.2,-2)$);
\coordinate (U) at ($(C1)-(0,0.75)$);
\coordinate (V) at ($(C1)+(0,-1.5)$);
\coordinate (Co) at ($(mid1)-(6,5)$);
\coordinate (Co') at ($(mid1)+(6,-5)$);

\arrowpath{(mid1)}{(C1)}{0.5};
\arrowpath{(A1)}{(mid1)}{0.5};
\arrowpath{(B1)}{(mid1)}{0.5};

\draw[fill=black] (A1) circle(1.5pt);
\draw[fill=black] (B1) circle(1.5pt);
\draw[fill=black] (C1) circle(1.5pt);
\draw (U) .. controls (Co) and (Co') .. (V);

\node[above] at (C1) {\scriptsize{$p\otimes_A q$}};
\node[below] at (A1) {\scriptsize{$p$}};
\node[below] at (B1) {\scriptsize{$q$}};
\node at ($(mid1)+(3,0)$) {$=_1$};

\coordinate (C2) at ($(C1)+(6,0)$);
\coordinate (mid2) at ($(C2)-(0,2)$);
\coordinate (A2) at ($(mid2)+(-1.2,-2)$);
\coordinate (B2) at ($(mid2)+(1.2,-2)$);
\coordinate (U2) at ($(mid2)!0.33!(A2)$);
\coordinate (V2) at ($(mid2)!0.33!(B2)$);
\coordinate (Co2) at ($(mid2)-(6,4)$);
\coordinate (Co'2) at ($(mid2)+(6,-4)$);

\arrowpath{(mid2)}{(C2)}{0.5};
\arrowpath{(A2)}{(mid2)}{0.5};
\arrowpath{(B2)}{(mid2)}{0.5};

\draw[fill=black] (A2) circle(1.5pt);
\draw[fill=black] (B2) circle(1.5pt);
\draw[fill=black] (C2) circle(1.5pt);
\draw (U2) .. controls (Co2) and (Co'2) .. (V2);

\node[above] at (C2) {\scriptsize{$p\otimes_A q$}};
\node[below] at (A2) {\scriptsize{$p$}};
\node[below] at (B2) {\scriptsize{$q$}};

\coordinate (C3) at ($(C1)-(0,7)$);
\coordinate (mid3) at ($(C3)-(0,2)$);
\coordinate (A3) at ($(mid3)+(-1.2,-2)$);
\coordinate (B3) at ($(mid3)+(1.2,-2)$);
\coordinate (U3) at ($(mid3)!0.33!(A3)$);
\coordinate (U3') at ($(mid3)!0.6!(B3)$);
\coordinate (V3) at ($(mid3)!0.33!(B3)$);
\coordinate (V3') at ($(mid3)!0.8!(B3)$);
\coordinate (Co3) at ($(mid3)-(6,4)$);
\coordinate (Co'3) at ($(mid3)+(6,-4)$);
\coordinate (Co3') at ($(mid3)-(1.5,3)$);
\coordinate (Co'3') at ($(mid3)+(3.5,-3)$);

\arrowpath{(mid3)}{(C3)}{0.5};
\arrowpath{(A3)}{(mid3)}{0.5};
\arrowpath{(B3)}{(mid3)}{0.5};

\draw[fill=black] (A3) circle(1.5pt);
\draw[fill=black] (B3) circle(1.5pt);
\draw[fill=black] (C3) circle(1.5pt);
\draw (U3) .. controls (Co3) and (Co'3) .. (V3);
\draw (U3') .. controls (Co3') and (Co'3') .. (V3');

\node at ($(mid3)-(3,0)$) {$=_2$};
\node[above] at (C3) {\scriptsize{$p\otimes_A q$}};
\node[below] at (A3) {\scriptsize{$p$}};
\node[below] at (B3) {\scriptsize{$q$}};

\coordinate (C4) at ($(C3)+(6,0)$);
\coordinate (mid4) at ($(C4)-(0,2)$);
\coordinate (A4) at ($(mid4)+(-1.2,-2)$);
\coordinate (B4) at ($(mid4)+(1.2,-2)$);
\coordinate (U4) at ($(mid4)!0.33!(A4)$);
\coordinate (U4') at ($(mid4)!0.7!(B4)$);
\coordinate (V4) at ($(mid4)!0.33!(B4)$);
\coordinate (V4') at ($(B4)-(1,1.15)$);
\coordinate (Co4) at ($(mid4)-(6,4)$);
\coordinate (Co'4) at ($(mid4)+(6,-4)$);
\coordinate (Co4') at ($(mid4)-(1.5,3)$);
\coordinate (Co'4') at ($(mid4)+(3.5,-3)$);

\arrowpath{(mid4)}{(C4)}{0.5};
\arrowpath{(A4)}{(mid4)}{0.5};
\arrowpath{(B4)}{(mid4)}{0.5};

\draw[fill=black] (A4) circle(1.5pt);
\draw[fill=black] (B4) circle(1.5pt);
\draw[fill=black] (C4) circle(1.5pt);
\draw (U4) .. controls (Co4) and (Co'4) .. (V4);
\draw (U4')--(V4');

\node at ($(mid4)-(3,0)$) {$=_3$};
\node[above] at (C4) {\scriptsize{$p\otimes_A q$}};
\node[below] at (A4) {\scriptsize{$p$}};
\node[below] at (B4) {\scriptsize{$q$}};

\coordinate (C5) at ($(C3)-(0,7)$);
\coordinate (mid5) at ($(C5)-(0,2)$);
\coordinate (A5) at ($(mid5)+(-1.2,-2)$);
\coordinate (B5) at ($(mid5)+(1.2,-2)$);
\coordinate (U5) at ($(mid5)!0.33!(A5)$);
\coordinate (U5') at ($(mid5)!0.6!(B5)$);
\coordinate (V5) at ($(mid5)!0.33!(B5)$);
\coordinate (V5') at ($(mid5)!0.8!(B5)$);
\coordinate (Co5) at ($(mid5)-(6,4)$);
\coordinate (Co'5) at ($(mid5)+(6,-4)$);
\coordinate (Co5') at ($(mid5)-(1.5,3)$);
\coordinate (Co'5') at ($(mid5)+(0,-3)$);

\arrowpath{(mid5)}{(C5)}{0.5};
\arrowpath{(A5)}{(mid5)}{0.5};
\arrowpath{(B5)}{(mid5)}{0.5};

\draw[fill=black] (A5) circle(1.5pt);
\draw[fill=black] (B5) circle(1.5pt);
\draw[fill=black] (C5) circle(1.5pt);
\draw (U5) .. controls (Co5) and (Co'5') .. (U5');
\draw (V5') .. controls (Co5') and (Co'5) .. (V5);

\node at ($(mid5)-(3,0)$) {$=_4$};
\node[above] at (C5) {\scriptsize{$p\otimes_A q$}};
\node[below] at (A5) {\scriptsize{$p$}};
\node[below] at (B5) {\scriptsize{$q$}};

\coordinate (C6) at ($(C5)+(6,0)$);
\coordinate (mid6) at ($(C6)-(0,2)$);
\coordinate (A6) at ($(mid6)+(-1.2,-2)$);
\coordinate (B6) at ($(mid6)+(1.2,-2)$);
\coordinate (U6) at ($(mid6)!0.33!(A6)$);
\coordinate (U6') at ($(mid6)!0.6!(A6)$);
\coordinate (V6) at ($(mid6)!0.33!(B6)$);
\coordinate (V6') at ($(mid6)!0.8!(B6)$);
\coordinate (Co6) at ($(mid6)-(5,4)$);
\coordinate (Co'6) at ($(mid6)+(6,-4)$);
\coordinate (Co6') at ($(mid6)-(1.5,3)$);
\coordinate (Co'6') at ($(mid6)+(1.5,-3)$);

\arrowpath{(mid6)}{(C6)}{0.5};
\arrowpath{(A6)}{(mid6)}{0.5};
\arrowpath{(B6)}{(mid6)}{0.5};

\draw[fill=black] (A6) circle(1.5pt);
\draw[fill=black] (B6) circle(1.5pt);
\draw[fill=black] (C6) circle(1.5pt);
\draw (U6) .. controls (Co6) and (Co'6') .. (U6');
\draw (V6') .. controls (Co6') and (Co'6) .. (V6);

\node at ($(mid6)-(3,0)$) {$=_5$};
\node[above] at (C6) {\scriptsize{$p\otimes_A q$}};
\node[below] at (A6) {\scriptsize{$p$}};
\node[below] at (B6) {\scriptsize{$q$}};
\end{scope}
\end{tikzpicture}
\caption{The new product $\tilde M$ and its well-definedness.
For details of the manipulation, see the main text.}\label{fig:pro}
\end{figure}
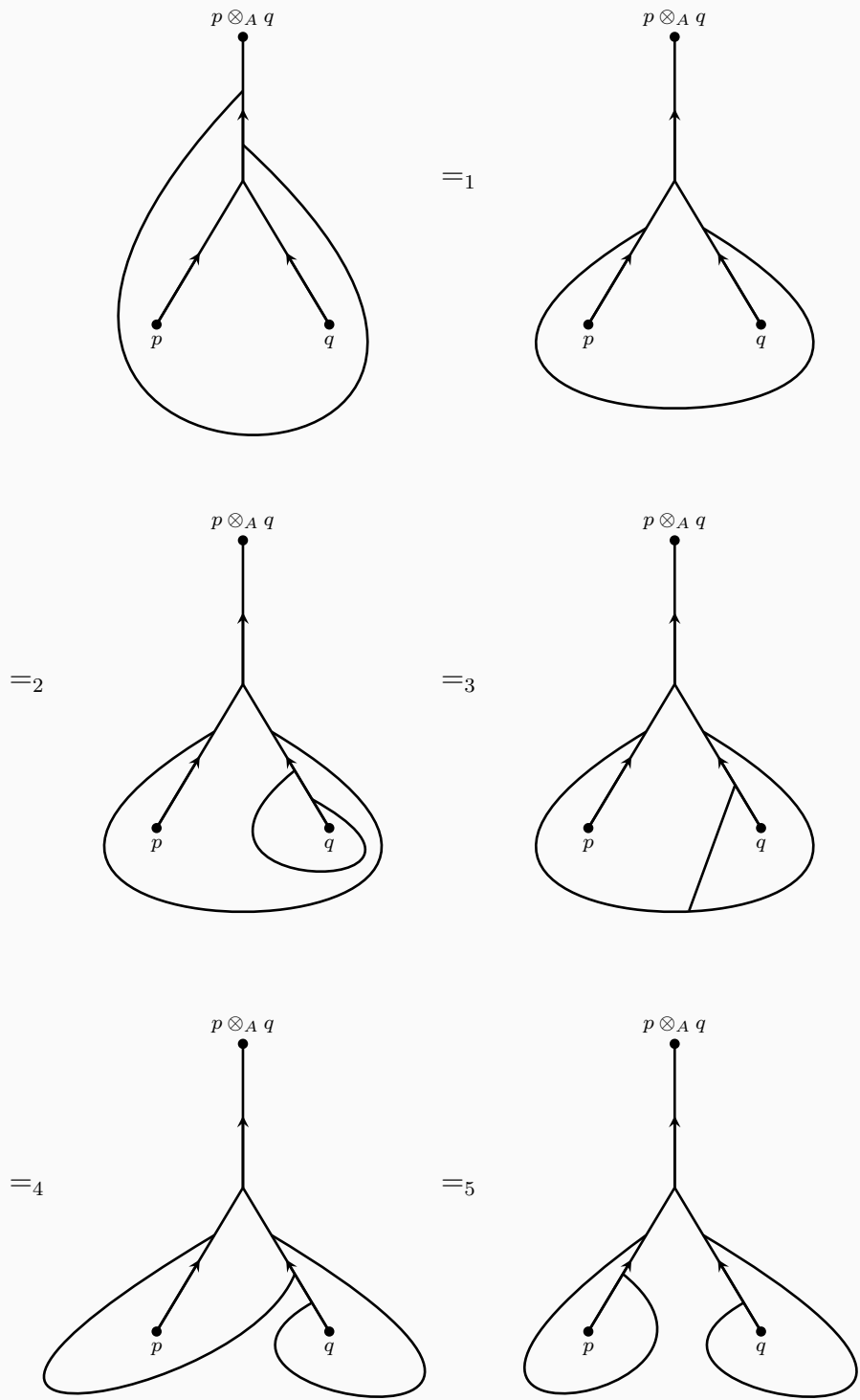

The new $\tilde M$ is defined analogously as $\tilde M_{p,q}=Z(\proj)M_{p,q}$. 
This $\tilde M$ can be shown to be well defined as a map $W_p\otimes W_q \to W_{p\otimes_A q}$ by a series of manipulations using a lot of properties of $A$. See Figure \ref{fig:pro}.
We represent a pair of pants as a 3-punctured plane for ease of illustration. The lower punctures correspond to input legs and the upper puncture corresponds to the product leg of the pair of pants.
Unlabeled lines correspond to $A$. To explain various manipulations, let us refer to manipulations involving $=_i$ as ``step $i$". 
\begin{itemize}
\item Step 1 just tells us that the left action of $A$ on $p\otimes_A q$ is defined by the left action on $p$, and the right action on $p\otimes_A q$ is defined by the right action on $q$.
\item  In step 2, we introduce an $A$ line wrapping the leg carrying $q$. We can do that because the input state in that leg is invariant under the action of $A$.
\item  In step 3, we first use the fact that $q$ is an $A$-bimodule and then use the fact that $q$ is a right $A$-module.
\item  In step 4, we use the fact that $q$ is a left module for $A$.
\item Finally, in step 5, we first use the fact that $q$ is a bimodule, then the fact that the tensor product is $\otimes_A$, and then the fact that $p$ is a bimodule.
\item   Ultimately, we can simply remove both the $A$ lines because the input states are invariant under the action of $A$.
\end{itemize}
We define $\co{\tilde M}$ as the adjoint of $\tilde M$.

To complete the definition of $T/A$, we have to check that the operations defined above satisfy the various conditions that we described in the last section. Most of them just concern some trivial topological manipulations of lines and are manifestly satisfied. Some others, such as  \eqref{XunitT}, can be checked using manipulations similar to the ones we have been doing in this sub-section. Yet some others, like the complicated relation (\ref{mess}) and (\ref{Xtorus})), require us to simplify a lot of $\proj$ and $\coproj$, but this can be done. This completes the definition of $T/A$.

\section{Conclusions}
\label{sec:conclusions}
In this paper we reviewed the notion of unitary fusion categories, or symmetry categories as we prefer to call them, and how they formalize the generalized notion of finite symmetries of a two-dimensional system.
We studied various explicit examples of such symmetry categories, some of which are related rather directly to finite groups and some of which are not.
We then studied how a symmetry category can be gauged and be re-gauged back.
We also defined 2d topological quantum field theories admitting a symmetry given by a symmetry category.
Many questions remain. Here we mention just two.

The first is how to generalize the constructions discussed in this paper to higher dimensions.
In a sense, this is a merger of the generalized symmetry in two dimensions in the sense of this paper,
and of the generalized symmetry in the sense of \cite{Gaiotto:2014kfa}.
That there should be something that combines both is clear: 
even in general spacetime dimension $d$, the 0-form symmetries can be any non-Abelian group $G$, possibly with an anomaly specified by $H^d(G,\UU(1))$ in the bosonic case and by subtler objects in the fermionic case.
Then the $(d-2)$-form symmetry needs to be extended at least to allow $\Rep(G)$.
When $d=3$, it seems that the notion of 1-form symmetries needs to be extended at least to include modular tensor categories, with an action of the 0-form symmetry group $G$ with an anomaly.
What should be the notion in $d=4$ and higher?

The second is to actually construct two-dimensional systems $T$ for a given symmetry category $\cC$.
For any group $G$ without an anomaly, there is the trivial theory where the Hilbert space is always one-dimensional. How about the other cases?
We can roughly classify symmetry categories $\cC$ as follows, depending on the simplest possible theories $T$ that have $\cC$ as a symmetry:
\begin{enumerate}
\item The simplest $\cC$-symmetric theories have one-dimensional Hilbert space. 
These would be  $\cC$-symmetry protected topological (SPT) phases.
\item The simplest $\cC$-symmetric theories have finite-dimensional Hilbert space. 
These would be   $\cC$-symmetry enriched topological (SET) phases.
\item The simplest $\cC$-symmetric theories have infinite-dimensional Hilbert space. 
Taking the low energy limit, these would be  $\cC$-symmetric conformal field theories (CFTs).
\item There is no $\cC$-symmetric theory.
\end{enumerate} 
Clearly this classification forms a hierarchy,
and it would be nice if we have  a uniform construction that tells easily which stage of the above classification a given symmetry category $\cC$ belongs to.
There is a recent paper in this general direction \cite{Buican:2017rxc}, 
where a construction of 2d theory starting from any given symmetry category $\cC$ was discussed.
We hope to see more developments in the future.

Actually, there are various symmetry categories $\cC$ constructed in the subfactor theory, e.g.~what is called the Haagerup  fusion category,  for which no $\cC$-symmetric theory is known. 
If a theory symmetric under the Haagerup fusion category could be constructed,  it would be considered as a huge breakthrough.

\section*{Acknowledgements}
The authors would like to thank Mikhail Kapranov for bringing the crucial reference \cite{EGNO} to the authors' attention. LB is also grateful to Kevin Costello, Davide Gaiotto and Theo Johnson-Freyd for helpful discussions.
The authors thank Nils Carqueville and Eric Sharpe for comments on, and  Marcel Bischoff, Kentaro Hori and Shu-Heng Shao for pointing out erroneous claims in an earlier version of the manuscript.
The work of LB  is partially supported by the Perimeter Institute for Theoretical Physics. Research at Perimeter Institute is supported by the Government of Canada through Industry Canada and by the Province of Ontario through the Ministry of Economic Development and Innovation.
The work of YT is partially supported in part 
by JSPS KAKENHI Grant-in-Aid (Wakate-A), No.17H04837
and JSPS KAKENHI Grant-in-Aid (Kiban-S), No.16H06335
and  by WPI Initiative, MEXT, Japan at IPMU, the University of Tokyo.

\appendix
\section{Group cohomology}
\label{sec:cohomology}
In this appendix, we collect various standard facts about group cohomology.
\paragraph{Definition:}
Given a finite group $G$ and its module $A$, we define $n$-cochains $C^n(G,A)$ as functions $G^n\to A$.
The differential is given by \begin{multline}
df(g_1,\ldots,g_{n+1})=g_1 f(g_2,\ldots,g_{n+1})\\
+\sum_{i=1}^n (-1)^{i}f(g_1,\ldots,g_i g_{i+1},\ldots,g_{n+1}) + (-1)^{n+1} f(g_1,\ldots, g_n).
\end{multline}
The differential squares to zero: $d^2=0$.
Then we define the group cohomology $H^i(G,A)$ as the cohomology of this differential.
Explicitly, the first few differentials are given by \begin{align}
df(g,h)&=gf(h)-f(gh)+f(g) , \\
df(a,b,c)&=af(b,c)-f(ab,c)+f(a,bc)-f(a,b),\\
df(x,y,z,w)&=xf(y,z,w)-f(xy,z,w)+f(x,yz,w)-f(x,y,zw)+f(x,y,z).
\end{align}

\paragraph{Some points on notation:}
It does not lead to any loss of generality if we assume that every cochain/cocycle/coboundary is \emph{normalized}, i.e.~it is zero whenever at least one of $g_i=1$. See e.g.~\cite{Chen:2011pg}. We have assumed throughout the paper that every cochain is normalized. We are often interested in $H^i(G,\UU(1))$ for $i=2,3$ where the action of $G$ on $\UU(1)$ is taken to be trivial. Henceforth, we will assume the trivial action whenever we write $\UU(1)$. It is also convenient to treat $\UU(1)$ elements as phases and in this case the $+$ sign in above definitions should be replaced by the usual multiplication of phases. For instance we have,
\be
df(g,h)=\frac{f(h)f(g)}{f(gh)}.
\ee
We have used the product notation throughout the paper in the context of group cohomology valued in $\UU(1)$.

\paragraph{Pull-back:}
Recall that given a map $M_1\to M_2$ between two manifolds, one can \emph{pull-back} $n$-forms on $M_2$ to $n$-forms on $M_1$. The analogous statement in group cohomology is that given a group homomorphism $G\to G'$, we obtain a module homomorphism $H^i(G',A)\to H^i(G,A)$. Explicitly, let $h:G\to G'$, then $\tilde h:H^i(G',A)\to H^i(G,A)$ is given by
\be
\tilde h(\alpha)(g_1,\cdots,g_i)=\alpha\left(h(g_1),\cdots,h(g_i)\right).
\ee

\paragraph{Cup product:}
There is an operation called \emph{cup product} $C^i(G,A)\times C^j(G,A)\to C^{i+j}(G,A)$ when $A$ is a ring. If $\alpha\in H^i(G,A)$ and $\beta\in H^j(G,A)$, then the cup product is defined as
\be
(\alpha\cup\beta)(g_1,\cdots,g_{i+j})=\alpha(g_1,\cdots,g_i)\beta(g_{i+1},\cdots,g_{i+j}).
\ee
It can be easily checked that this product descends to a product on cohomologies.

\paragraph{One-dimensional representations of $G$:}
Let us ask what is the meaning of $H^1(G,\UU(1))$. The 1-cochains are maps from $G$ to $\UU(1)$ and imposing the cocycle condition turns them into group homomorphisms. Hence $H^1(G,\UU(1))$ is the group formed by one-dimensional representations of $G$. In particular, when $G$ is a finite Abelian group, then $H^1(G,\UU(1))\simeq \hat G$ is the dual group.

\paragraph{Projective representations of $G$:}
Now, let us ask what is the meaning of $\epsilon\in H^2(G,\UU(1))$. We want to interpret the $\epsilon(g_1,g_2)$ as the phases defining a projective representation of $G$. The cocyle condition reads
\be
\epsilon(g_1,g_2)\epsilon(g_1g_2,g_3)=\epsilon(g_2,g_3)\epsilon(g_1,g_2g_3)
\ee
which is the associativity condition such phases are required to satisfy. Such a cocycle can be shifted by a coboundary of the form
\be
d\beta(g_1,g_2)=\frac{\beta(g_1g_2)}{\beta(g_1)\beta(g_2)}
\ee
which corresponds to rephasing of the group action on the projective representation. Thus, we see that $H^2(G,\UU(1))$ classifies the phases encountered in projective representations upto rephasing. The usual representations correspond to the trivial element of $H^2(G,\UU(1))$.

\paragraph{Crossed products and extensions of groups:}
Consider an Abelian group $H$ and a (possibly non-Abelian) group $K$. Consider an action of $K$ on $H$ and use it to define $H^2(K,H)$. An element $\kappa\in H^2(K,H)$ can be used to define a group extension $G$ of $K$ by $H$, that is there is a short exact sequence
\be
0\to H\to G\to K\to 0
\ee
and $G$ is called the $\kappa$-cross product of $K$ and $H$ and it is written as $G=H\rtimes_\kappa K$.
Explicitly, the group $G$ as a set is the direct product $ H\times K$ with the group multiplication given as follows
\be
(h_1,k_1)(h_2,k_2)=(h_1+(k_1\triangleright h_2)+\kappa(k_1,k_2),k_1 k_2)\label{crossed}.
\ee
Here, $\triangleright$ denotes the action of $K$ on $H$ via inner automorphism in $G$. The reader can verify that the associativity of the group multiplication is ensured by the cocycle condition on $\kappa$. Shifting $\kappa$ by a coboundary changes $G$ upto isomorphism. Hence, group extensions of $K$ by an Abelian group $H$ are classified by a group action of $K$ on $H$ along with an element in $H^2(K,H)$ defined using the group action.

\paragraph{Bicharacters on $G$:}
Consider an Abelian group $G$ and a trivial module $A$ of $G$. Given a cohomology element in $H^2(G,A)$ represented by a cocycle $\epsilon(g,h)$, one can form $\alpha(g,h)=\epsilon(g,h)-\epsilon(h,g)$ which is an antisymmetric function on $G$. This is indeed well defined because adding a coboundary to $\epsilon$ doesn't change $\alpha$. When $A$ is $\UU(1)$ this $\alpha$ is a bicharacter, and there's a bijection between an antisymmetric bicharacter on $G$ and $H^2(G,\UU(1))$.

\paragraph{A useful isomorphism:}
Recall the familiar statement that any closed $n$-form is exact \emph{locally}. In group cohomology, the analogous statement is that
\be
H^i(G,\bR)=1,
\ee
that is $H^i(G,\bR)$ is trivial.
We can use this to obtain $H^{i+1}(G,\bZ)\simeq H^i(G,\UU(1))$. 
This follows from the long exact sequence associated to the sequence $0\to \bZ \to \bR \to \UU(1)\to 0$.

\paragraph{Explicit group cohomologies:}
\begin{itemize}
\item $H^*(\bZ_n,\bZ)=\bZ[x_2]/(nx_2)$.
\item $H^2((\bZ_n)^k,\UU(1)) = (\bZ_n)^{k(k-1)/2}$. 
\item $H^3((\bZ_n)^k,\UU(1)) = (\bZ_n)^{k+k(k-1)/2+k(k-1)(k-2)/2}$, with generators given by \begin{align}
\alpha^{(i)}(a,b,c)&=e^{2\pi i a_i(b_i+c_i-\vev{b_i+c_i})/n^2},\\
\alpha^{(i,j)}(a,b,c)&=e^{2\pi i a_i(b_j+c_j-\vev{b_j+c_j})/n^2},\\
\alpha^{(i,j,k)}(a,b,c)&=e^{2\pi i a_ib_jc_k/n}
\end{align} where  $a,b,c=\{0,1,\ldots,n-1\}$ and $\vev{a}$ is the mod $n$  function to $\{0,\ldots,n-1\}$.
In particular, $H^3(\bZ_n,\UU(1))=\bZ_n$ and  its generator has the cocycle \begin{equation}
\alpha(a,b,c)=e^{2\pi i a(b+c-\vev{b+c})/n^2}.
\end{equation} 
\item For $D_m$ the dihedral group with $m$ elements we have \cite{Handel}, 
\begin{equation}
H^*(D_m,\bZ)=\bZ[a_2,b_2,c_3,d_4]/(2a_2,2b_2,2c_3,md_4,(b_2)^2+a_2b_2+(m^2/4)d_4).
\end{equation} 
In particular, $H^2(D_{2n+1},\UU(1))=0$, $H^3(D_{2n+1},\UU(1))=\bZ_{4n+2}$; $H^2(D_{2n},\UU(1))=\bZ_2\times \bZ_2$, $H^3(D_{2n},\UU(1))=\bZ_2\times \bZ_2 \times \bZ_{2n}$.
The explicit generators can be found in \cite{deWildPropitius:1995cf}.
\item For $Q_8$, the quaternion group, we have \cite{HayamiSanada} \begin{equation}
H^*(Q_8,\bZ)=\bZ[A_2,B_2,C_4]/(2A_2,2B_2,8C_4,A_2^2,B_2^2,A_2B_2-4C_4).
\end{equation}
\end{itemize}

\bibliographystyle{ytphys}
\bibliography{ref}

\end{document}